\documentclass[10pt,journal,compsoc,letterpaper]{IEEEtran}

\ifCLASSOPTIONcompsoc
  \usepackage[nocompress]{cite}
\else
  \usepackage{cite}
\fi

\usepackage[pdftex]{graphicx}

\usepackage[cmex10]{amsmath}
\usepackage{amsfonts}
\usepackage{array}
\usepackage{fixltx2e}

\usepackage{color}
\usepackage{url}
\usepackage{booktabs}
\usepackage{comment}
\usepackage{multirow}

\usepackage{flushend}

\usepackage[table]{xcolor}

\newcommand{\squishlist}{ 
   \begin{list}{$\bullet$}
    { \setlength{\itemsep}{0pt}      \setlength{\parsep}{3pt} 
      \setlength{\topsep}{3pt}       \setlength{\partopsep}{0pt}
      \setlength{\leftmargin}{1.5em} \setlength{\labelwidth}{1em}
      \setlength{\labelsep}{0.5em} } }

\newcommand{\squishend}{
  \end{list}  }

\newcommand{\approach}{ENTRUST}

\usepackage{fancyhdr}
\begin{document}

\title{Engineering Trustworthy Self-Adaptive Software with Dynamic Assurance Cases}
\author{Radu~Calinescu,
			Danny~Weyns,
			 Simos~Gerasimou,
			 M.~Usman~Iftikhar,
			 Ibrahim~Habli,
			 and~Tim Kelly
\IEEEcompsocitemizethanks{\IEEEcompsocthanksitem R.~Calinescu, S.~Gerasimou, I.~Habli and T.~Kelly are with the Department of Computer Science at the University of York, UK.
\IEEEcompsocthanksitem D.~Weyns is with the Department of Computer Science of the Katholieke Universiteit Leuven, Belgium.
\IEEEcompsocthanksitem M.~U.~Iftikhar is with the Department of Computer Science at  Linnaeus University, Sweden.}}

\IEEEtitleabstractindextext{
\begin{abstract}
Building on concepts drawn from control theory, \emph{self-adaptive software} handles environmental and internal uncertainties by dynamically adjusting its architecture and parameters 
in response to events such as workload changes and component failures. Self-adaptive software is increasingly expected to meet strict functional and non-functional requirements in applications from areas as diverse as manufacturing, healthcare and finance. To address this need, we introduce a methodology for the systematic ENgineering of TRUstworthy Self-adaptive sofTware (\approach). \approach\ uses a combination of (1)~design-time and runtime modelling and verification, and (2)~industry-adopted assurance processes to develop trustworthy self-adaptive software \emph{and} assurance cases arguing the suitability of the software for its intended application. To evaluate the effectiveness of our methodology, we present a tool-supported instance of \approach\ and its use to develop proof-of-concept self-adaptive software for embedded and service-based systems from the oceanic monitoring and e-finance domains, respectively. The experimental results show that \approach\ can be used to engineer self-adaptive software systems in different application domains and to generate dynamic assurance cases for these systems. 
\end{abstract}

}

\maketitle


\thispagestyle{fancy}
\section{Introduction}

Software systems are regularly used in applications characterised by uncertain environments, evolving requirements and unexpected failures. The correct operation of these applications depends on the ability of software to adapt to change, through the dynamic reconfiguration of its parameters or architecture. When events such as variations in workload, changes in the required throughput or component failures are observed, alternative adaptation options are analysed, and a suitable new software configuration may be selected and applied. 

\begin{figure}
\centering
\includegraphics[width=8.3cm]{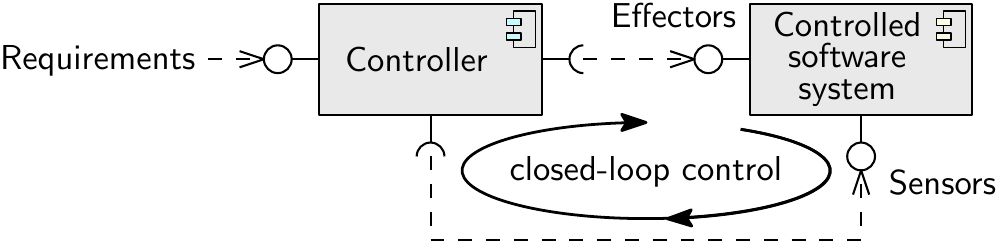}
\caption{Closed-loop control is used to automate software adaptation
\label{fig:closed-loop-control}}
\end{figure}

As software adaptation is often too complex or too costly to be performed by human operators, its automation has been the subject of intense research. Using concepts borrowed from the control of discrete-event systems \cite{ramadge-wonham-1989}, this research proposes the extension of software systems with \emph{closed-loop control}. As shown in Fig.~\ref{fig:closed-loop-control}, the paradigm involves using an external software \emph{controller} to monitor the system and to adapt its architecture or configuration after environmental and internal changes.
Inspired by the autonomic computing manifesto \cite{horn2001,KephartC2003} and by pioneering work on self-adaptive software \cite{KarsaiS1999,OreizyGTHJMQRW1999}, this research has been very successful. 
Over the past decade, numerous research projects proposed architectures \cite{white2004architectural,GarlanSASS2004,Kramer2007:FOSE} and frameworks \cite{elkhodary2010fusion,weyns2012forms,tesauro2004multi,CalinescuGKMT2011} for the engineering of \emph{self-adaptive systems}. Extensive surveys of this research and its applications are available in \cite{HuebscherM08,PsaierDustdar2011,SalehieT2009}.

In this paper, we are concerned with the use of self-adaptive software in systems with strict functional and non-functional requirements. A growing number of systems are expected to fit this description in the near future. Service-based telehealth systems are envisaged to use self-adaptation to cope with service failures and workload variations \cite{CalinescuGKMT2011,Epifani2009:ICSE,Weyns:2015:TAS:2821357.2821373}, avoiding harm to patients. Autonomous robots used in applications ranging from manufacturing \cite{7353608,D'Ippolito:2014:HBP:2568225.2568264} to oceanic monitoring \cite{CGB2015,Gerasimou2014:SEAMS} will need to rely on self-adaptive software for completing their missions safely and effectively, without damage to, or loss of, expensive equipment. Employing self-adaptive software in these applications is very challenging, as it requires assurances about the correct operation of the software in scenarios affected by uncertainty.

Assurance has become a major concern for self-adaptive software only recently \cite{CamaraLGL2013,Cheng2014,delemos_et_al:DR:2014:4508,seamsRoadmap2013}. Accordingly, the research in the area is limited, and often confined to providing evidence that individual aspects of the self-adaptive software are correct (e.g.\ the software platform used to execute the controller, the controller functions, or the runtime adaptation decisions).  However, such evidence is only one component of the established industry process for the assurance of software-based systems \cite{BloomfieldB2010,LittlewoodW07,alarp1998}. In real-world applications, assuring a software system requires the provision of an \emph{assurance case}, which standards such as \cite{DefenceStandard00-56} define as 

\begin{quote}
  ``\emph{a structured argument, supported by a body of evidence, that provides a compelling, comprehensible and valid case that a system is safe for a given application in a given environment}''.
\end{quote}

Our work addresses this discrepancy between the state of practice and the current research on assurances for self-adaptive software.  To this end, we introduce a generic methodology for the joint development of trustworthy self-adaptive software systems \emph{and} their associated assurance cases. Our methodology for the ENgineering of TRUstworthy Self-ad\-apt\-ive sofTware (\approach) is underpinned by a combination of (1) design-time and runtime modelling and verification, and (2) an industry-adopted standard for the formalisation of assurance arguments \cite{gsn-2011,spriggs-2012}. 

\approach\ uses design-time modelling, verification and synthesis of assurance evidence for the 
control aspects of a self-adaptive system that are engineered before the system is deployed. These design-time activities support the initial controller enactment and the generation of a partial assurance case for the self-adaptive system. The dynamic selection of a system configuration (i.e., architecture and parameters) during the initial deployment and after internal and environmental changes involves further modelling and verification, and the synthesis of the additional assurance evidence required to complete the assurance case. These activities are fully automated and carried out at runtime.

The \approach\ methodology is not prescriptive about the modelling, verification and assurance evidence generation methods used in its design-time and runtime stages. This generality exploits the fact that the body of evidence underpinning an assurance case can combine verification evidence from activities including formal verification, testing and simulation. As such, our methodology is applicable to a broad range of application domains, software engineering paradigms and verification methods.

\approach\ supports the systematic engineering and assurance of self-adaptive systems. In line with other research on self-adaptive systems (see e.g. \cite{SalehieT2009,Weyns2012}), we assume that the controlled software system from Figure~\ref{fig:closed-loop-control} already exists, and we focus on its enhancement with self-adaptation capabilities through the addition of a high-level monitor-analyse-plan-execute (MAPE) control loop. The components of the controlled software system may already support low-level, real-time adaptation to localised changes. For instance, the self-adaptive embedded system used as a running example in Section~\ref{sec:tool-supported} is a controlled unmanned vehicle that employs built-in low-level control to maintain the speed selected by its high-level \approach\ controller. Mature approaches from the areas of robust control of discrete-event systems (e.g. \cite{ramadge-wonham-1989,DBLP:journals/tac/LahijanianAB15,Tabuada2009,DBLP:journals/scl/ZamaniWM13}) and real-time systems (e.g. \cite{Krishna2001,Lu2002}) already exist for the engineering of such low-level control, which is outside the scope of \approach. 
Likewise, established assurance processes are available for the non-self-adaptive aspects of software systems (e.g.\ \cite{BishopBloomfield1998,BloomfieldB2010,Hawkins201355,hawkins2013principles,Rushby:Cases15}). We do not duplicate this work. Using these processes to construct assurance arguments for the correct design, development and operation of the controlled software system, and for the derivation, validity, completeness and formalisation of the requirements from Fig.~\ref{fig:closed-loop-control} is outside the scope of our paper. 
Thus, \approach\ focuses on the correct engineering of the controller and on the correct operation of self-adaptive system, assuming that the controlled system and its requirements are both correct.

The main contributions of our paper are:
\squishlist
\item[1)] The first end-to-end methodology 
for (a)~engineering self-adaptive software systems with assurance evidence for the controller platform, its functions and the adaptation decisions; and (b)~devising assurance cases whose assurance arguments bring together this evidence.
\item[2)] A novel assurance argument pattern for self-adaptive systems, expressed in the Goal Structuring Notation (GSN) standard~\cite{gsn-2011} that is widely used for assurance case development in industry \cite{spriggs-2012}.
\item[3)] An instantiation of our methodology whose stages are supported by the established modelling and verification tools UPPAAL \cite{behrmann2006uppaal} and PRISM \cite{KNP11}. This methodology instance extends and integrates for the first time our previously separate strands of work on developing formally verified control loops \cite{Iftikhar14}, runtime probabilistic model checking   \cite{Calinescu2012:CACM} and dynamic safety cases \cite{DHP2015}.
\squishend
These contributions are evaluated using two case studies with different characteristics and goals, and belonging to different application domains.

The remainder of the paper is organised as follows. In Section~\ref{sec:preliminaries}, we provide background information on assurance cases, GSN and assurance argument patterns. Section~\ref{sec:systems} introduces two proof-of-concept self-adaptive systems that we use to illustrate our methodology, which is described in Section~\ref{sec:methodology}, and to illustrate and evaluate its tool-supported instance, which is presented in Section~\ref{sec:tool-supported}.  
Section~\ref{sec:evaluation} presents our evaluation results, which show that the methodology can be used for the effective engineering of self-adaptive systems from different domains and for the generation of dynamic assurance cases for these systems.
In Section~\ref{sec:related}, we overview the existing approaches to providing assurances for self-adaptive software systems, and we compare them to \approach. Finally, Section~\ref{sec:conclusion} concludes the paper with a discussion and a summary of future work directions.

\section{Preliminaries}
\label{sec:preliminaries}

This section provides background information on assurance cases, introducing the assurance-related terminology and concepts used in the rest of the paper. We start by defining assurance cases and their components in Section~\ref{subsec:assurance-cases}. Next, we introduce a commonly used notation for the specification of assurance cases in Section~\ref{subsect:gsn}. Finally, we introduce the concept of an assurance argument pattern in Section~\ref{subsect:assurance-paterns}.

\subsection{Assurance Cases \label{subsec:assurance-cases}}

An \emph{assurance case}\footnote{Assurance cases developed for safety-critical systems are also called \emph{safety cases}. In this work, we are concerned with any self-adaptive software systems that must meet strict requirements, and therefore we talk about assurance cases and assurance arguments.} is a report that supports a specific \emph{claim} about the requirements of a system  \cite{BishopBloomfield1998}. As an example, the assurance case in \cite{nefab-2011} provides documented assurance that the ``\textsl{implementation and operation of North European Functional Airspace Block (NEFAB) is acceptably safe according to ICAO, EC and EUROCONTROL safety requirements}.'' The documented assurance within an assurance case comprises (1)~\emph{evidence} and (2)~structured \emph{arguments} that link the evidence to the claim \cite{BishopBloomfield1998}, possibly through intermediate claims. 

Assurance cases are becoming mandatory for software systems used in safety-critical and mission-critical applications \cite{alarp1998,LittlewoodW07,BloomfieldB2010}. They are used in domains ranging from nuclear energy \cite{ONR-2013} and medical devices \cite{DHHS-2014} to air traffic control \cite{Eurocontrol2006} and defence \cite{DefenceStandard00-56}. A growing number of assurance cases from these and other domains are openly available (e.g., \cite{nefab-2011,Virginia-2014}). 

The development of assurance cases comprises processes carried out at all stages of the system life cycle \cite{alarp1998}. Requirements analysis evidence and design evidence demonstrate that system reliability, safety, maintainability, etc. are considered in the early stages of the life cycle. Implementation, validation and verification evidence are then generated as the system is developed. Finally, evidence collected at runtime is used to update assurance cases during system maintenance. 

As aptly described in \cite{alarp1998}, the assurance case must be ``\textsl{a living, cradle-to-grave document.}'' This is particularly true for self-adaptive software systems. For these systems, existing evidence needs to be continuously combined with new \emph{adaptation evidence}, i.e., evidence that the system will continue to operate safely after self-adaptation activities.

\subsection{Goal Structuring Notation \label{subsect:gsn}}

\begin{figure}
\centering
\includegraphics[width=\hsize]{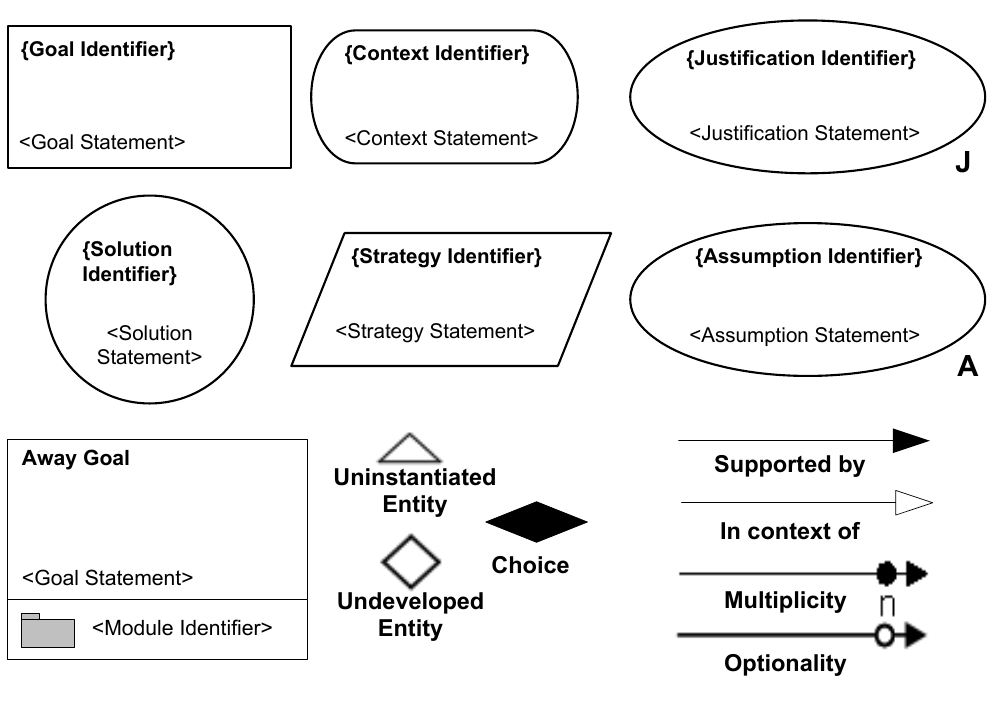}

\vspace*{-4mm}
\caption{Core GSN elements
\label{fig:gsn}}

\vspace*{-3mm}
\end{figure}

The assurance cases for self-adaptive systems introduced later in the paper are devised in the \emph{Goal Structuring Notation} (GSN) \cite{Kelly04thegoal}, a community standard \cite{gsn-2011} widely used for assurance case development in industry \cite{spriggs-2012}. The main GSN elements (Fig.~\ref{fig:gsn}) can be used to construct an argument by showing how an assurance claim (represented in GSN by a \emph{goal}) is broken down into sub-claims (also represented by GSN \emph{goals}), until eventually it can be supported by GSN \emph{solutions} (i.e., assurance evidence from verification, testing, etc.). 
\emph{Strategies} are used to partition the argument and describe the nature of the inference that exists between a goal and its supporting goal(s). The rationale (\emph{assumptions} and \emph{justifications}) for individual elements of the argument can be captured, along with the \emph{context} (e.g.\ to describe the operational environment) in which the claims are stated.

In a GSN diagram, claims are linked to strategies, sub-claims and ultimately to solutions using `\emph{supported by}' connectives, which are rendered as lines with a solid arrowhead and declare inferential or evidential relationships. `Supported by' connectives may be decorated with their multiplicity or marked as optional. The `\emph{in context of}' connective, rendered as a line with a hollow arrowhead, declares a contextual relationship between a goal or strategy on the one hand and a context, assumption or justification on the other hand.

Large or complex sections of the assurance argument can be organised into modules by means of GSN \emph{away goals} referenced in the main argument and defined separately. 
Finally, GSN entities can be marked as \emph{uninstantiated} to indicate that they are placeholders that need to be replaced with a concrete instantiation, and GSN goals can be marked as \emph{undeveloped} to indicate that they need to be further developed into sub-goals, strategies and solutions. 

\begin{figure}
\centering
\includegraphics[width=7cm]{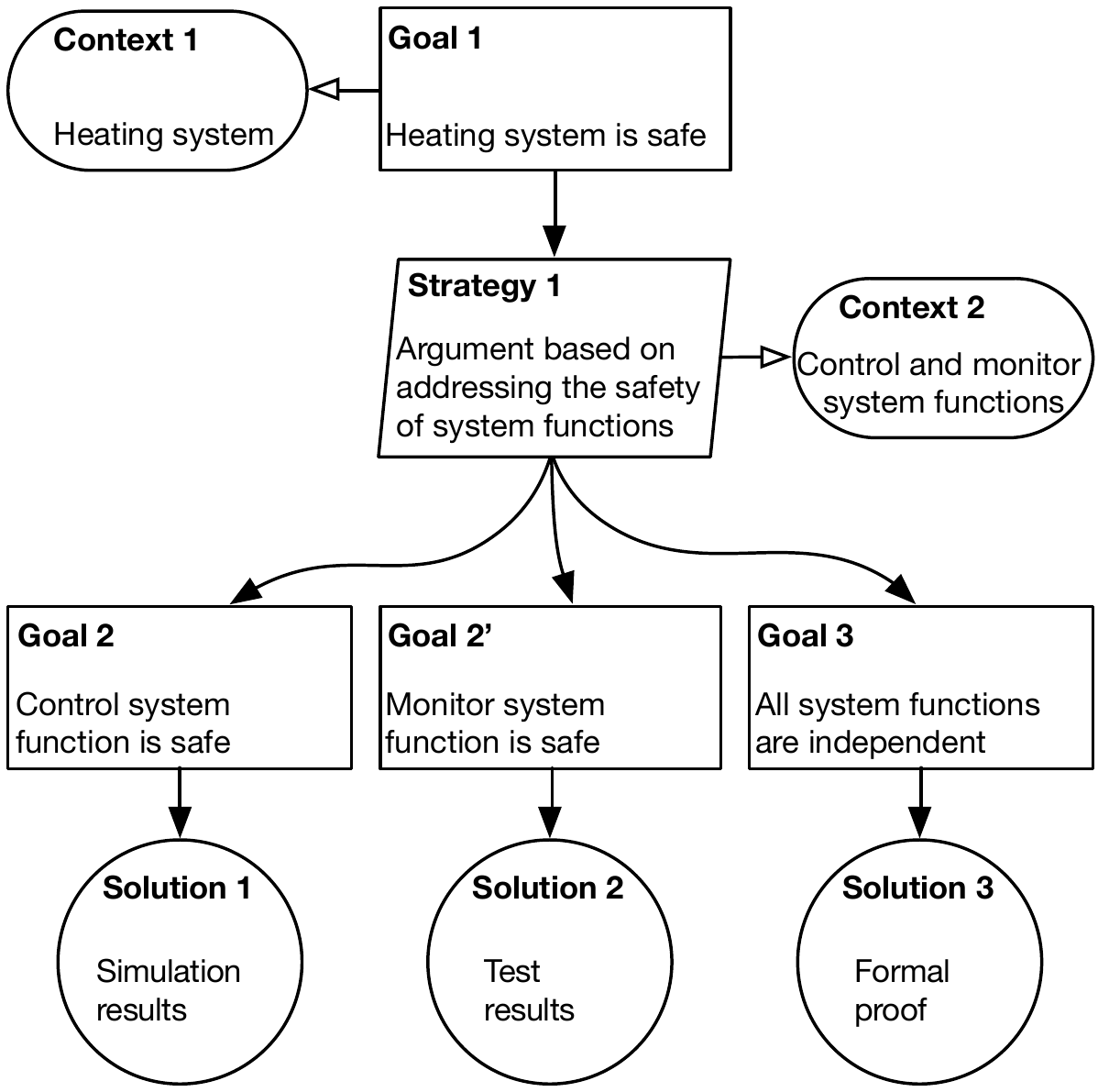}
\caption{Example of a GSN assurance argument
\label{fig:safety-case}}

\vspace*{-3mm}
\end{figure}

As an example, Figure~\ref{fig:safety-case} shows a simple GSN assurance argument for the software part of a heating system. Its root goal (\textbf{Goal~1}) claims that the system is safe at all times. This claim is partitioned into sub-claims using a strategy (\textbf{Strategy~1}) that addresses the safety of the two system functions (i.e.\ control and monitoring) separately through sub-claims \textbf{Goal~2} (for the control system) and \textbf{Goal~2'} (for the monitor system), and includes sub-claim \textbf{Goal 3} that the two functions are independent. The three sub-claims are supported by three solutions comprising assurance evidence from simulation, testing and formal proof, respectively.

\subsection{Assurance Argument Patterns \label{subsect:assurance-paterns}}

To reduce the significant effort required to develop assurance cases, in our previous work on software assurance \cite{DBLP:conf/safecomp/HawkinsHK13,hawkins2013principles} we collaborated to the creation of a catalog of reusable GSN \emph{assurance argument patterns}  \cite{hawkins2011using}. Each pattern considers the contribution made by the software to system hazards for a particular class of systems and scenarios. The GSN elements of a pattern that are generic to the entire class are fully developed and instantiated, whereas the entities that are specific to each system and scenario within the class are left undeveloped and/or uninstantiated. 

\begin{figure}
\centering
\includegraphics[width=7cm]{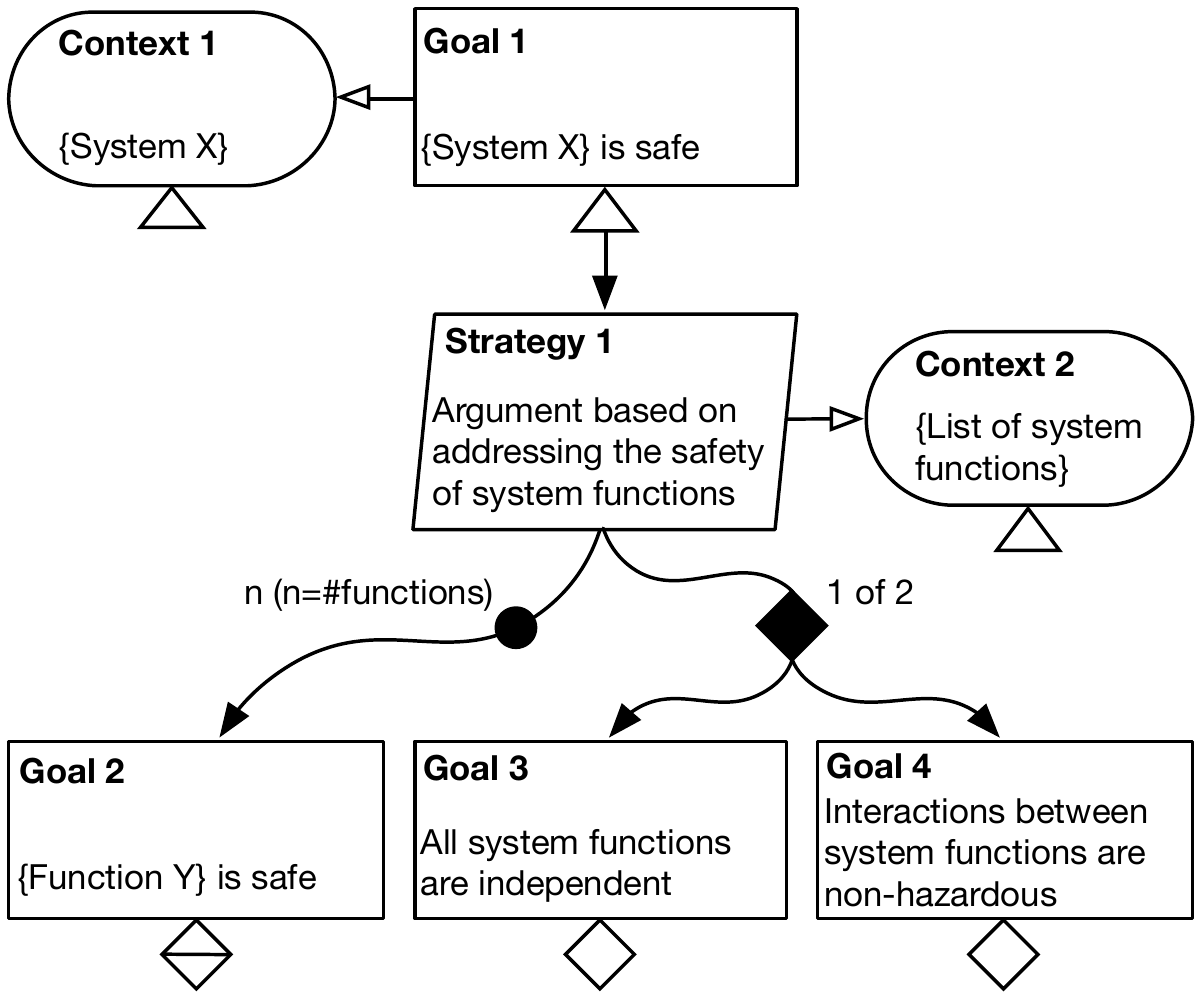}
\caption{Example of a GSN assurance argument pattern 
\label{fig:safety-case-pattern}}

\vspace*{-3mm}
\end{figure}

As an example, Fig.~\ref{fig:safety-case-pattern} depicts an assurance argument pattern that is instantiated by the GSN assurance argument from Fig.~\ref{fig:safety-case}. The elements surrounded by curly brackets `$\{$' and `$\}$' in the pattern must be \emph{instantiated} for each assurance argument based on the pattern, as further indicated by the triangular `uninstantiated' symbol under the GSN entities that contain them. \textbf{Goal 2} is marked with both this `uninstantiated' symbol (because it contains elements in curly brackets) and a diamond-shaped `undeveloped' symbol (because, like for the `choice' sub-claims \textbf{Goal~3} and \textbf{Goal~4}, additional GSN entities must be added underneath to complete the assurance argument); the two symbols are rendered overlapping under \textbf{Goal 2}.

In this paper, we devise a new assurance argument pattern, which is applicable to self-adaptive software systems.
\section{\mbox{Self-adaptive System Examples}}
\label{sec:systems}

This section introduces two examples of self-adaptive software systems that we will use to illustrate our the generic \approach\ methodology in Section~\ref{sec:methodology}, and to illustrate and evaluate its tool-supported instantiation in Section~\ref{sec:tool-supported}.

\subsection{Unmanned Underwater Vehicle (UUV) System}
\label{sec:running}

The first system is a self-adaptive UUV embedded system adapted from~\cite{Gerasimou2014:SEAMS}. UUVs are increasingly used in a wide range of oceanographic and military tasks, including oceanic surveillance (e.g., to monitor pollution levels and ecosystems), undersea mapping and mine detection. Limitations due to their operating environment (e.g., impossibility  to maintain UUV-operator communication during missions and  unexpected changes) require that UUV systems are self-adaptive. These systems are often mission critical (e.g., when used for mine detection) or business critical (e.g., they carry expensive equipment that should not be lost).

The self-adaptive system we use consists of a UUV deployed to carry out a data gathering mission. The UUV is equipped with $n \geq 1$ on-board sensors that can measure the same characteristic of the ocean environment (e.g., water current, salinity or temperature). When used, the sensors take measurements with different, variable rates $r_1$, $r_2$, \ldots, $r_n$. The probability that each sensor produces measurements that are sufficiently accurate for the purpose of the mission depends on the UUV speed $\mathit{sp}$, and is given by  $p_1$, $p_2$, \ldots, $p_n$. 
For each measurement taken, a different amount of energy is consumed, given by $e_1$, $e_2$, \ldots, $e_n$.  
Finally, the $n$ sensors can be switched on and off individually (e.g., to save battery power when not required), but these operations consume an amount of energy given by $e^\mathrm{on}_1$, $e^\mathrm{on}_2$, \ldots, $e^\mathrm{on}_n$ and $e^\mathrm{off}_1$, $e^\mathrm{off}_2$, \ldots, $e^\mathrm{off}_n$, respectively.
The UUV  must adapt to changes in the sensor measurement rates $r_1$, $r_2$, \ldots, $r_n$ and to sensor failures by dynamically adjusting:
\squishlist
\item[(a)] the UUV speed $sp$
\item[(b)] the sensor configuration $x_1$, $x_2$, \ldots, $x_n$ (where $x_i=1$ if the $i$-th sensor is on and $x_i=0$ otherwise)
\squishend
in order to meet the quality-of-service requirements below:

\squishlist
	\item[\textbf{R1 (throughput)}:] The UUV should take at least 20 measurements of sufficient accuracy for every 10~metres of mission distance.
	\item[\textbf{R2 (resource usage)}:] The energy consumption of the sensors should not exceed 120 Joules per 10 surveyed metres.
	\item[\textbf{R3 (cost)}:] If requirements R1 and R2 are satisfied by multiple configurations, the UUV should use one of these configurations that minimises the cost function
\begin{equation}
\label{eq:cost1}
cost = w_1 E + w_2 \mathit{sp}^{-1},
\end{equation}
where $E$ is the energy used by the sensors to survey a 10m mission distance, and $w_1, w_2 \!>\!0$ are weights that reflect the relative importance of carrying out the mission with reduced battery usage and completing the mission faster.\footnote{Cost (or \emph{utility}) functions that employ weights to combine several performance, reliability, resource use and other quality attributes of software---accounting for differences in attribute value ranges and relative importance---are extensively used in self-adaptive software systems (e.g.\ \cite{elkhodary2010fusion,GarlanSASS2004,CalinescuGKMT2011,SalehieT2009,Walsh-etal-2004}).}
	\item[\textbf{R4 (safety)}:] If a configuration that meets requirements \textbf{R1}--\textbf{R3} is not identified within 2~seconds after a sensor rate change, the UUV speed must be reduced to 0m/s. This ensures that the UUV does not advance more than the distance it can cover at its maximum speed within 2~seconds without taking appropriate measurements, and waits until the controller identifies a suitable configuration (e.g., after the UUV sensors recover) or new instructions are provided by a human operator.
\squishend

\subsection{Foreign Exchange Trading System \label{sect:FX-case-study}}

\begin{figure}
\centering
\includegraphics[trim= 5mm 5mm 5mm 9mm, clip, width=8.25cm, height=6cm]{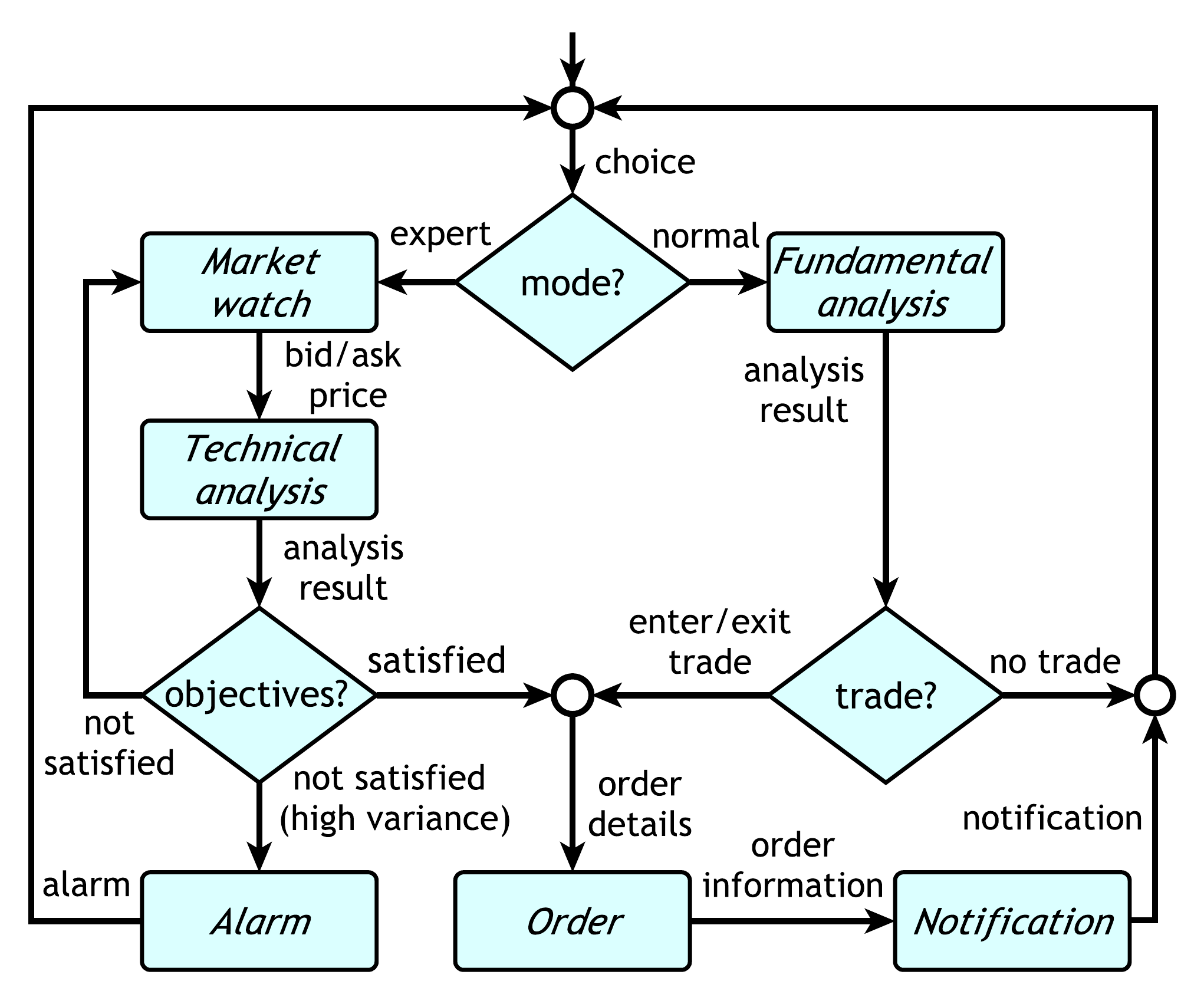}

\vspace*{-2mm}
\caption{\textcolor{black}{Foreign exchange trading (FX) workflow}}
\label{fig:workflow}

\vspace*{-2mm}
\end{figure}

Our second system is a service-based system from the area of foreign exchange trading, taken from our recent work in \cite{gerasimouCB2015}. This system, which we anonymise as FX for confidentiality reasons, is used by an European foreign exchange brokerage company. The FX system implements the workflow shown in Fig.~\ref{fig:workflow} and described below.

An FX customer (called a trader) can use the system in two operation modes. In the \textit{expert}  mode, FX executes a loop that analyses market activity, identifies patterns that satisfy the trader's objectives, and automatically carries out trades. Thus, the \emph{Market watch} service extracts real-time exchange rates (bid/ask price) of selected currency pairs. This data is used by a \emph{Technical analysis} service that evaluates the current trading conditions, predicts future price movement, and decides if the trader's objectives are: (i)~``satisfied" (causing the invocation of an \emph{Order} service to carry out a trade); (ii)~``unsatisfied" (resulting in a new \emph{Market watch} invocation); or (iii)~``unsatisfied with high variance" (triggering an \emph{Alarm} service invocation to notify the trader about discrepancies/opportunities not covered by the trading objectives). In the \emph{normal} mode, FX assesses the economic outlook of a country using a \emph{Fundamental analysis} service that collects, analyses and evaluates information such as news reports, economic data and political events, and provides an assessment on the country's currency. If satisfied with this assessment, the trader can use the \emph{Order} service to sell or buy currency, in which case a \emph{Notification} service confirms the completion of the trade. We assume that the FX system has to dynamically select third-party implementations for each service from Fig.~\ref{fig:workflow}, in order to meet the following system requirements: 
\squishlist
	\item[\textbf{R1 (reliability)}:] Workflow executions must complete successfully with probability at least 0.9.
	\item[\textbf{R2 (response time)}:] The total service response time per workflow execution must be at most 5s.
	\item [\textbf{R3 (cost)}:] If requirements R1 and R2 are satisfied by multiple configurations, the FX system should use one of these configurations that minimises the cost function:
	\begin{equation}
	\label{eq:cost2}
	\mathit{cost} = w_1 \mathit{price} + w_2 \mathit{time},
	\end{equation}
	where $\mathit{price}$ and $\mathit{time}$ represent the total price of the services invoked by a workflow execution and the response time for a workflow execution, respectively, and $w_1,w_2>0$ are weights that encode the desired trade-off between price and response time.
	\item[\textbf{R4 (safety)}:] If a configuration that ensures requirements \textbf{R1}--\textbf{R3} cannot be identified within 2s after a change in service characteristics is signalled by the sensors of the self-adaptive FX system, the \emph{Order} service invocation is bypassed, so that the FX system does not carry out any trade that might be based on incorrect or stale data.
\squishend
Note that requirements R1--R3 express two constraints and an optimisation criterion that are qualitatively different from those specified by the requirements from our first case study (cf.\ Section~\ref{sec:running}). Nevertheless, our tool-supported instance of the \approach\ methodology enabled the development of the self-adaptive FX system as described in Section~\ref{sec:sbs}.


\begin{figure*}
\centering
\includegraphics[width=0.9\hsize]{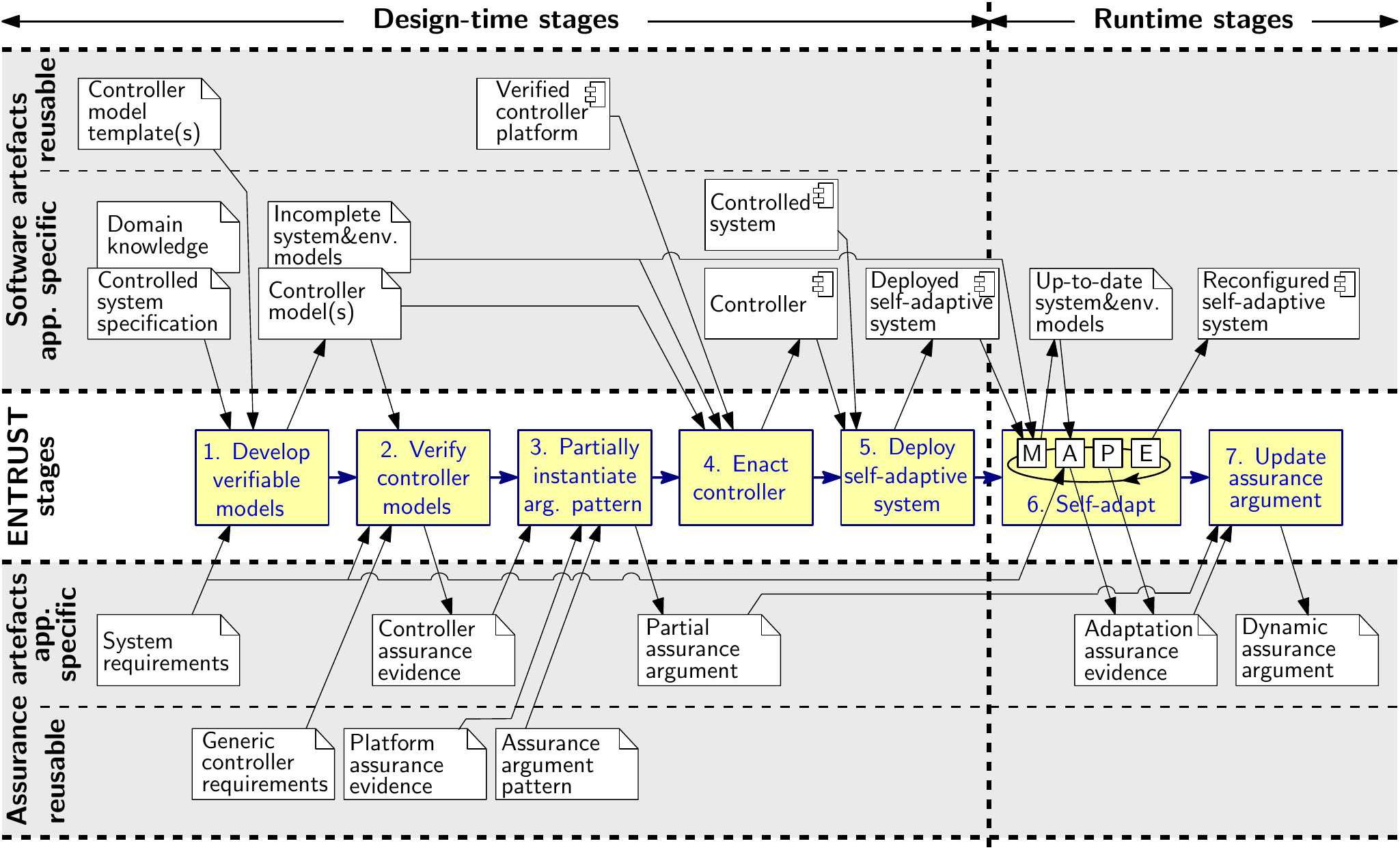}

\vspace*{-1.5mm}
\caption{Stages and key artefacts of the \approach\ methodology. In line with the two principles underpinning the methodology, its first  stage involves the development of verifiable models for the controller, controlled system and environment of the self-adaptive system used throughout the remaining stages, and multiple stages reuse application-independent software and assurance artefacts.
\label{fig:methodology}}

\vspace*{-1.5mm}
\end{figure*}

\section{The \approach\ Methodology}
\label{sec:methodology}

The \approach\ methodology supports the systematic engineering and assurance of self-adaptive systems based on monitor-analyse-plan-execute (MAPE) control loops. This is by far the most common type of control loop used to devise self-adaptive software systems \cite{Brun:2009:ESS:1573856.1573860,HuebscherM08,seamsRoadmap2013,delemos_et_al:DR:2014:4508,SalehieT2009,elkhodary2010fusion,MaciasEscriva20137267,Krupitzer2015184}. The engineering of self-adaptive systems based on essentially different control techniques, such as the control theoretical paradigm proposed in \cite{Filieri14}, is not supported by our methodology.

\approach\ comprises the tool-supported design-time stages and the automated runtime stages shown in Figure~\ref{fig:methodology}, and is underpinned by two key principles:
\squishlist
\item[1)] \emph{Model-driven engineering is essential for developing trustworthy self-adaptive systems and their assurance cases.} As emphasised in the previous section, model-based analysis, simulation, testing and formal verification---at design time and during reconfiguration---represent the main sources of assurance evidence for self-adaptive software. As such, both the design-time and the runtime stages of our methodology are model driven. Models of the structure and behaviour of the functional components, controller and environment are the basis for the engineering and assurance of \approach\ self-adaptive systems. 
\item[2)] \emph{Reuse of application-independent software and assurance artefacts significantly reduces the effort and expertise required to develop trustworthy self-adaptive systems.} Assembling an assurance case for a software system is a costly process that requires considerable effort and expertise. Therefore, the reuse of both software and assurance artefacts is essential for \approach. In particular, the reuse of application-independent controller components and of templates for developing application-specific controller elements also enables the reuse of assurance evidence that these software artefacts are trustworthy. 
\squishend

The \approach\ stages and their exploitation of these two principles are described in the remainder of this section.

\subsection{Design-time \approach\ Stages}

\subsubsection{Stage 1: Development of Verifiable Models} 

In \approach, the engineering of a self-adaptive system with the architecture from Figure~\ref{fig:closed-loop-control} starts with the development of models for: 
\squishlist
\item[1)] The controller of the self-adaptive system;
\item[2)] The relevant aspects of the controlled software system and its environment.
\squishend
A combination of structural and behavioural models may be produced, depending on the evidence needed to assemble the assurance case for the self-adaptive system under development. \approach\ is not prescriptive in this respect. However, we require that these models are \emph{verifiable}, i.e., that they can be used in conjunction with methods such as model checking or simulation, to obtain evidence that the controller and the self-adaptive system meet their requirements. As an example, finite state transition models may be produced for the controllers of our UUV and FX systems from Section~\ref{sec:systems}, enabling the use of model checking to verify that these controllers are deadlock free.

The verifiable models are application-specific. As illustrated in Figure~\ref{fig:methodology}, their development requires \emph{domain knowledge},\footnote{The \approach\ software and assurance artefacts that appear in \emph{italics} in the text are also shown in Figure~\ref{fig:methodology}.} is based on a \emph{controlled system specification}, and is informed by the \emph{system requirements}. As in other areas of software engineering, we envisage that tool-supported methods will typically be used to obtain these models. However, their manual development or fully automated synthesis are not precluded by \approach. 

In line with the ``reuse of artefacts'' principle, \approach\ exploits the fact that the controllers of self-adaptive systems implement the established MAPE workflow, and uses application-independent \emph{controller model template(s)} to devise the \emph{controller model(s)}. These templates model the generic aspects of the MAPE workflow and contain placeholders for the application-specific elements of an \approach\ controller. 

Given the environmental and internal uncertainty that characterises self-adaptive systems, only \emph{incomplete system and environment models} can be produced in this \approach\ stage. These incomplete models may include unknown or estimated parameters, nondeterminism (i.e., alternative options whose likelihoods are unknown), parts that are missing, or some combination of all of these. For example, parametric Markov chains may be devised to enable the runtime analysis of the requirements for our UVV and FX systems detailed in Sections~\ref{sec:running} and~\ref{sect:FX-case-study}, respectively, by means of probabilistic model checking or simulation.

\subsubsection{Stage 2: Verification of Controller Models} 

The main role of the second \approach\ stage is to produce \emph{controller assurance evidence}, i.e., compelling evidence that a controller based on the controller model(s) from Stage~1 will satisfy a set of \emph{generic controller requirements}. These are requirements that must be satisfied in any self-adaptive system (e.g., deadlock freeness) and are predefined in a format compatible with that of the controller model templates and with the method that will be used to verify the controller models. For example, if labelled transition systems are used to model the controller and model checking to establish its correctness as in \cite{DBLP:conf/icse/DIppolitoBPU11,D'Ippolito:2014:HBP:2568225.2568264}, these generic controller requirements can be predefined as temporal logic formulae. 

The controller assurance evidence may additionally include evidence that some of the system requirements are satisfied. Thus, it may be possible to show that---despite the uncertainty characteristic to any self-adaptive system---application-specific failsafe operating modes (e.g.\ those specified by requirements R4 of our UUV and FX systems from Section~\ref{sec:systems}) are always reachable. 

The assurance evidence generated in this stage of the methodology may be obtained using a range of methods that include formal verification, theorem proving and simulation. The methods that can be used depend on the types of models produced in the previous \approach\ stage, and on the generic controller requirements and system requirements for which assurance is sought. The availability of tool support in the form of model checkers, theorem provers, SMT solvers, domain-specific simulators, etc.\ will influence the choice of these methods.

Preparing the design-time models, i.e., developing verifiable models and verifying the controller models, comes with a cost. However, by using tool-supported methods and exploiting reusable application-independent software, this cost can significantly be reduced and does not affect the usability of \approach\ compared to other related approaches. Related approaches that only provide a fraction of the assurances that \approach\ achieves (as detailed when we discuss related work in Section~\ref{sec:related}) operate with design-time models that require a comparable effort to specify the models and provide the controller assurance evidence.

\subsubsection{Stage 3: Partial Instantiation of Assurance Argument Pattern}

This \approach\ stage uses the controller assurance evidence from Stage~2 to support the partial instantiation of a generic \emph{assurance argument pattern} for self-adaptive software. As explained in Section~\ref{subsect:assurance-paterns}, this pattern is an incomplete assurance argument containing placeholders for the system-specific assurance evidence. A subset of the placeholders correspond to the controller assurance evidence obtained in Stage~2, and are therefore instantiated using this evidence. The result is a \emph{partial assurance argument}, which still contains placeholders for the assurance evidence that cannot be obtained until the uncertainties associated with the self-adaptive system are resolved at runtime. 

For example, the partial assurance argument for our UUV and FX systems should contain evidence that their controllers are deadlock free and that their failsafe requirements R4 are always satisfied. These requirements can be verified at design time. In contrast, requirements R1--R3 for the two systems cannot be verified until runtime, when the controller acquires information about the measurement rates of the UUV sensors and the third-party services available for the FX operations, respectively. Assurance evidence that requirements R1--R3 are satisfied can only be obtained at runtime.

In addition to the two types of placeholders, the assurance argument pattern used as input for this stage includes assurance evidence that is application independent. In particular, it includes evidence about the correct operation of the \emph{verified controller platform}, i.e.\ the software that implements application-independent controller functionality used to execute the \approach\ controllers. 
This \emph{platform assurance evidence} is reusable across self-adaptive systems.

\subsubsection{Stage 4: Enactment of the Controller}

This \approach\ stage assembles the \emph{controller} of the self-adaptive system. The process involves integrating the verified controller platform with the application-specific controller elements, and with the sensors and effectors that interface the controller with the controlled software system from Figure~\ref{fig:closed-loop-control}. 

The application-specific controller elements must be devised from the verified controller models, by using a trusted model-driven engineering method. This can be done using \emph{model-to-text transformation}, a method that employs a trusted \emph{model compiler} to generate a low-level executable representation of the controller models. Alternatively, the \approach\ verified controller platform may include a trusted virtual machine\footnote{Throughout the paper, the term ``virtual machine'' refers to a software component capable to interpret and execute controller models, much like a Java virtual machine executes Java code.} able to directly interpret and run the controller models. The second, \emph{model interpretation} method \cite{Spi00b}, has the advantage that it eliminates the need to generate controller code and to provide additional assurances for it.

\subsubsection{Stage 5: Deployment of the Self-Adaptive System \label{sect:deployment-stage}}

In the last design-time stage, the integrated controller and \emph{controlled components} of the self-adaptive system are installed, preconfigured and activated by means of an application-specific process. The pre-configuration is responsible for setting the deployment-specific parameters and architectural aspects of the system. For example, the pre-configuration of the UUV system from Section~\ref{sec:running} involves selecting the initial speed and active sensor set for the UUV, whereas for the FX system from Section~\ref{sect:FX-case-study} it involves choosing initial third-party implementations for each FX service.

The \emph{deployed self-adaptive system} will be fully configured and a complete assurance argument will be available only after the first execution of the MAPE control loop. This execution is typically triggered by the system activation, to ensure that the newly deployed self-adaptive system takes into account the current state of its environment as described next. 

\subsection{Runtime \approach\ Stages}

\subsubsection{Stage 6: Self-adaptation \label{sssec:self-adaptation}}

In this \approach\ stage, the deployed self-adaptive system is dynamically adjusting its parameters and architecture in line with observed internal and environmental changes. To this end, the controller executes a typical MAPE loop that monitors the system and its environment, using the information obtained in this way to resolve the ``unknowns'' from the incomplete system and environment models. The resulting \emph{up-to-date system and environment models} enable the MAPE loop to analyse the system compliance with its requirements after changes, and to plan and execute suitable reconfigurations if necessary.

Whenever the MAPE loop produces a \emph{reconfigured self-adaptive system}, its analysis and planning steps generate \emph{adaptation assurance evidence} confirming the correctness of the analysis results and of the reconfiguration plan devised on the basis of these results. This assurance evidence is a by-product of analysis and planning methods that may include runtime verification, simulation and runtime model checking. Irrespective of the methods that produce it, the adaptation assurance evidence is essential for the development of a complete assurance argument in the next \approach\ stage.

\subsubsection{Stage 7: Synthesis of Dynamic Assurance Argument}

The final \approach\ stage uses the adaptation correctness evidence produced by the MAPE loop to fill in the placeholders from the partial assurance argument, and to devise the complete assurance case for the reconfigured self-adaptive system. For example, runtime evidence that requirements R1--R3 of the UUV and FX systems from Section~\ref{sec:systems} are met will be used to complete the remaining placeholders from their partial assurance arguments.
Thus, an \approach\ assurance case is underpinned by a \emph{dynamic assurance argument} that is updated after each reconfiguration of the system parameters and architecture. This assurance case captures both the full assurance argument and the evidence that justifies the active configuration of the self-adaptive system.

The \approach\ assurance case versions generated for every system reconfiguration have two key uses. First, they allow decision makers and auditors to understand and assess the present and past versions of the assurance case. Second, they allow human operators to endorse major reconfiguration plans in human-supervised self-adaptive systems. This type of self-adaptive systems is of particular interest in domains where human supervision represents an important risk mitigation factor or may be required by regulations. As an example, UK Civil Aviation Authority regulations \cite{CAP-722} permit self-adaptation in certain functions (e.g., power management, flight management and collision avoidance) of unmanned aircraft of no more than 20 kg provided that the aircraft operates within the visual line of sight of a human operator.

\section{Tool-Supported Instance of \approach}
\label{sec:tool-supported}

This section presents an instance of \approach\ in which the stages described in Section~\ref{sec:methodology} are supported by the modelling and verification tools UPPAAL \cite{behrmann2006uppaal} and PRISM \cite{KNP11}. We start with an overview of this tool-supported \approach\ instance in Section~\ref{subsect:overview}, followed by a description of each of its stages in Section~\ref{subsect:stage-description}. The UUV self-adaptive system introduced in Section~\ref{sec:running} is used as a running example throughout these stage descriptions.  We conclude with an end-to-end illustration of how  the \approach\ instance can be used to develop the FX self-adaptive system in Section~\ref{sec:sbs}.

\subsection{Overview \label{subsect:overview}}

The \approach\ methodology can be used with different combinations of modelling, verification and controller enactment methods, which may employ different self-adaptive system architectures and types of assurance evidence. This section presents a tool-supported instance of \approach\ that uses one such combination of methods. We developed this instance of the methodology with the aim to validate \approach\ and to ease its adoption.  

\begin{figure}
\centering
\includegraphics[width=0.8\hsize]{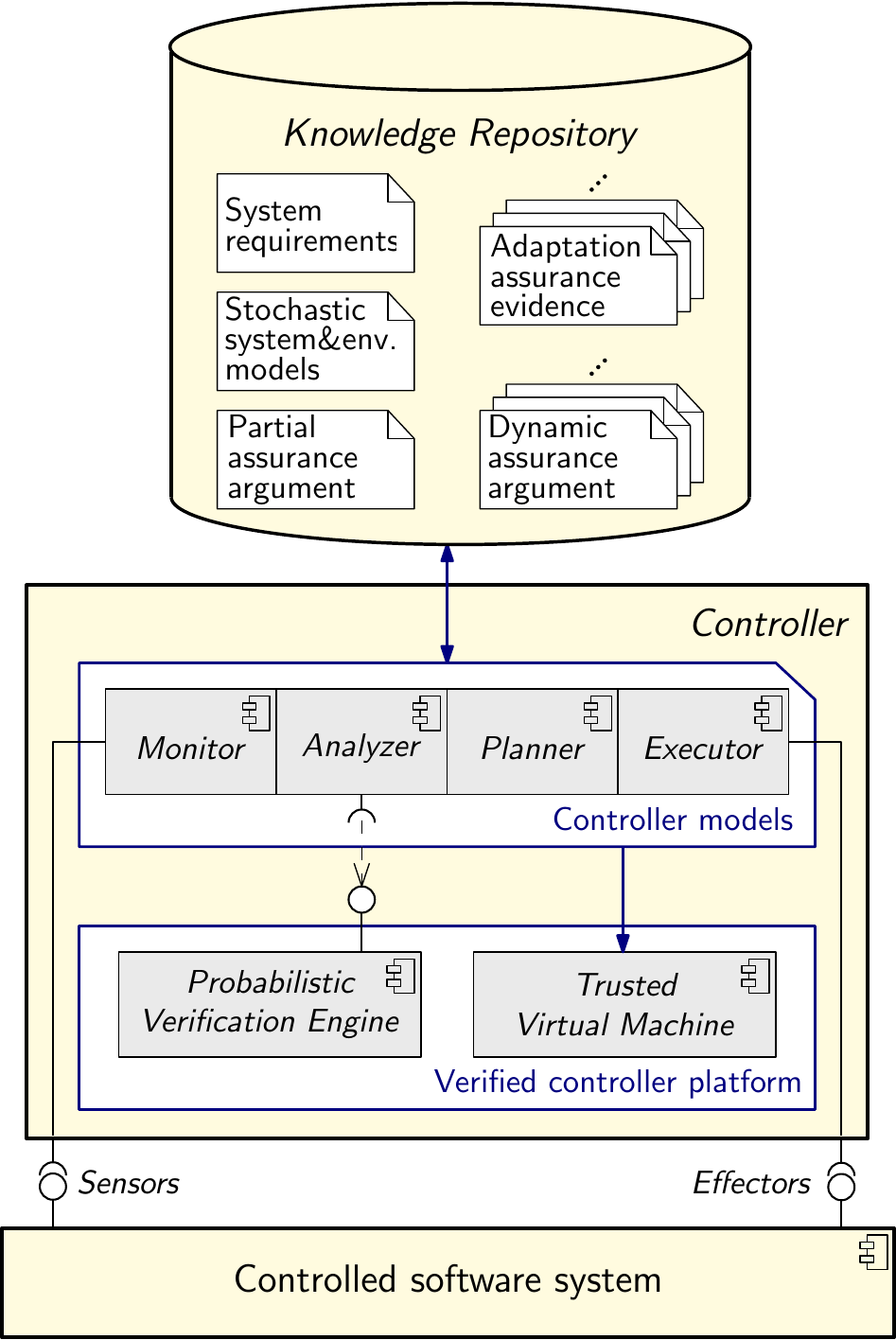}
\caption{Architecture of an \approach\ self-adaptive system
\label{fig:architecture}}
\end{figure}

Our \approach\ instance supports the engineering of self-adaptive systems with the architecture shown in Fig.~\ref{fig:architecture}. The reusable verified controller platform at the core of this architecture comprises:
\squishlist
\item[1)] A \emph{Trusted Virtual Machine} that directly interprets and executes models of the four steps from the MAPE control loop\footnote{Hence the controller models are depicted as software components in Figure~\ref{fig:architecture}.} (i.e., the \approach\ controller models).
\item[2)] A \emph{Probabilistic Verification Engine} that is used to verify stochastic models of the controlled system and its environment during the analysis step of the MAPE loop. 
\squishend
Using the \emph{Trusted Virtual Machine} for controller model interpretation eliminates the need for a model-to-text transformation of the controller models into executable code, which is a complex, error-prone operation. Not having to devise this transformation and to provide assurance evidence for it are major benefits of our \approach\ instance. Although we still need assurance evidence for the virtual machine, this was obtained when we developed and verified the virtual machine,\footnote{This assurance evidence is in the form of a comprehensive test suite and a report describing its successful execution by the virtual machine, both of which are available on our \approach\ project website at \url{https://www-users.cs.york.ac.uk/simos/ENTRUST/}.}  and is part of the reusable \emph{platform assurance evidence} for the \approach\ instance.

\begin{table*}
\renewcommand{\arraystretch}{1.2}
	\centering
		\caption{Stages of the tool-supported instance of the \approach\ methodology \label{table:stage-summary}}
		
		\vspace*{-3mm}
		\begin{small}
		\begin{tabular}{p{0.6cm} p{2.1cm} p{10.8cm} p{2.9cm}} %
			\toprule
			\textbf{Stage} & \textbf{Type} & \textbf{Description} & \textbf{Supporting tool(s)}  \\ 
			\midrule
			1 & tool supported & Timed automata controller models developed from UPPAAL templates & UPPAAL \\
			   &                         & Incomplete stochastic models of the controlled system and environment developed 
			                                  based on system specification and domain knowledge & PRISM \\ 
			2 & tool supported & Controller models verified to obtain controller assurance evidence & UPPAAL \\
			3 & manual & Partial assurance argument devised from GSN assurance argument pattern & -- \\
			4 & manual & Controller enacted by integrating the verified controller models and platform & --\\
			5 & manual &  Controlled system, controller and knowledge repository deployed & --\\
			6 & automated & MAPE control loop continually executed to ensure the system requirements & PRISM \& \approach\ controller platform \\
			7 & automated & GSN dynamic assurance argument generated & \approach\ controller platform \\
			\bottomrule
		\end{tabular}
	\end{small}
	
		\vspace*{-3mm}	
\end{table*}

The \emph{Probabilistic Verification Engine} consists of the verification libraries of the probabilistic model checker PRISM~\cite{KNP11} and is used by the analysis step of the MAPE control loop. As such, our \approach\ instance works with: 
\begin{enumerate}
\item Stochastic finite state transition models of the controlled system and the environment, defined in the PRISM high-level modelling language. Incomplete versions of these models are devised in Stage~1 of \approach, and have their unknowns resolved at runtime. All types of models that PRISM can analyse are supported, including discrete- and continuous-time Markov chains (DTMCs and CTMCs), Markov decision processes (MDPs) and probabilistic automata (PAs). 
\item Runtime-assured system requirements expressed in the appropriate variant of probabilistic temporal logic, i.e., probabilistic computation tree logic (PCTL) for DTMCs, MDPs and PAs, and continuous stochastic logic (CSL) for CTMCs.
\end{enumerate}
This makes our instantiation of the generic \approach\ methodology applicable to self-adaptive systems whose non-functional (e.g., reliability, performance, resource usage and cost-related) requirements can be specified in the above logics, and whose behaviour related to these requirements can be described using stochastic models. As shown by the recent work of multiple research groups (e.g., \cite{CalinescuGKMT2011,Calinescu2012:CACM,Camara16,Epifani2009:ICSE,FilieriGT12,FKP+12,DBLP:conf/fase/SuCFRT16,QuatmannD0JK16}), this represents a broad and important class of self-adaptive software that includes a wide range of service-based systems, web applications, resource management systems, and embedded systems.

Also developed in Stage~1 of \approach, the four controller models form an application-specific network of interacting timed automata~\cite{Alur94D}, and are expressed in the modelling language of the UPPAAL verification tool suite~\cite{behrmann2006uppaal}. 

Accordingly, UPPAAL is used in Stage~2 of \approach\ to verify the compliance of the controller models with the generic controller requirements and with any system requirements that can be assured at design time. These requirements are defined in computation tree logic (CTL)~\cite{Clarke:1986:AVF:5397.5399}.

In Stage~3 of our \approach\ instance, a partial assurance argument is devised starting from an assurance argument pattern represented in \emph{goal structuring notation} (GSN) \cite{Kelly04thegoal}. GSN is a community standard \cite{gsn-2011} that is widely used for assurance case development in industry \cite{spriggs-2012}.

The controller enactment from Stage~4 involves integrating the timed-automata controller models with our verified controller platform.

In Stage~5 of \approach, the controlled software system and its enacted controller are deployed, together with a \emph{Knowledge Repository} 
that supports the operation of the controller. Initially, this repository contains: (i)~the partial assurance argument from Stage~3; (ii)~the system requirements to be assured at runtime; and (iii)~the (incomplete) stochastic system and environment models from Stage~1.

During the execution of the MAPE loop in Stage~6 of \approach, the \emph{Monitor} obtains information about the system and its environment through \emph{Probes}. This information is used to resolve the unknowns from the stochastic models of the controlled system and its environment. Examples of such unknowns include probabilities of transition to `failure' states for a DTMC, MDP or PA, rates of transition to `success' states for a CTMC, and sets of states and transitions modelling certain system behaviours. After each update of the stochastic system and environment models, the \emph{Analyzer} reverifies the compliance of the self-adaptive system with its runtime-assured requirements. When the requirements are no longer met, the \emph{Analyzer} uses the verification results to identify a new system configuration that restores this compliance, or to find out that such a configuration does not exist and to select a predefined failsafe configuration. The step-by-step actions needed to achieve the new configuration are then established by the \emph{Planner} and implemented by the \emph{Executor} through the \emph{Effectors} of the controlled system.

Using the \emph{Probabilistic Verification Engine} enables the \emph{Analyzer} and \emph{Planner} to produce assurance evidence justifying their selection of new configurations and of plans for transitioning the system to these configurations, respectively. This adaptation assurance evidence is used to synthesise a fully-fledged, dynamic GSN assurance argument in Stage~7 of our \approach\ instance. As indicated in Figure~\ref{fig:architecture}, versions of the adaptation assurance evidence and of the dynamic assurance argument justifying each reconfiguration of the self-adaptive system are stored in the \emph{Knowledge Repository}.

The implementation of the \approach\ stages in our tool-supported instance of the methodology is summarised in Table~\ref{table:stage-summary} and described in further detail in Section~\ref{subsec:description}. The UUV system introduced in Section~\ref{sec:running} is used to support this description.

\subsection{Stage Descriptions \label{subsect:stage-description}}
\label{subsec:description}

\begin{figure*}
	\centering
	\includegraphics[width=15.5cm]{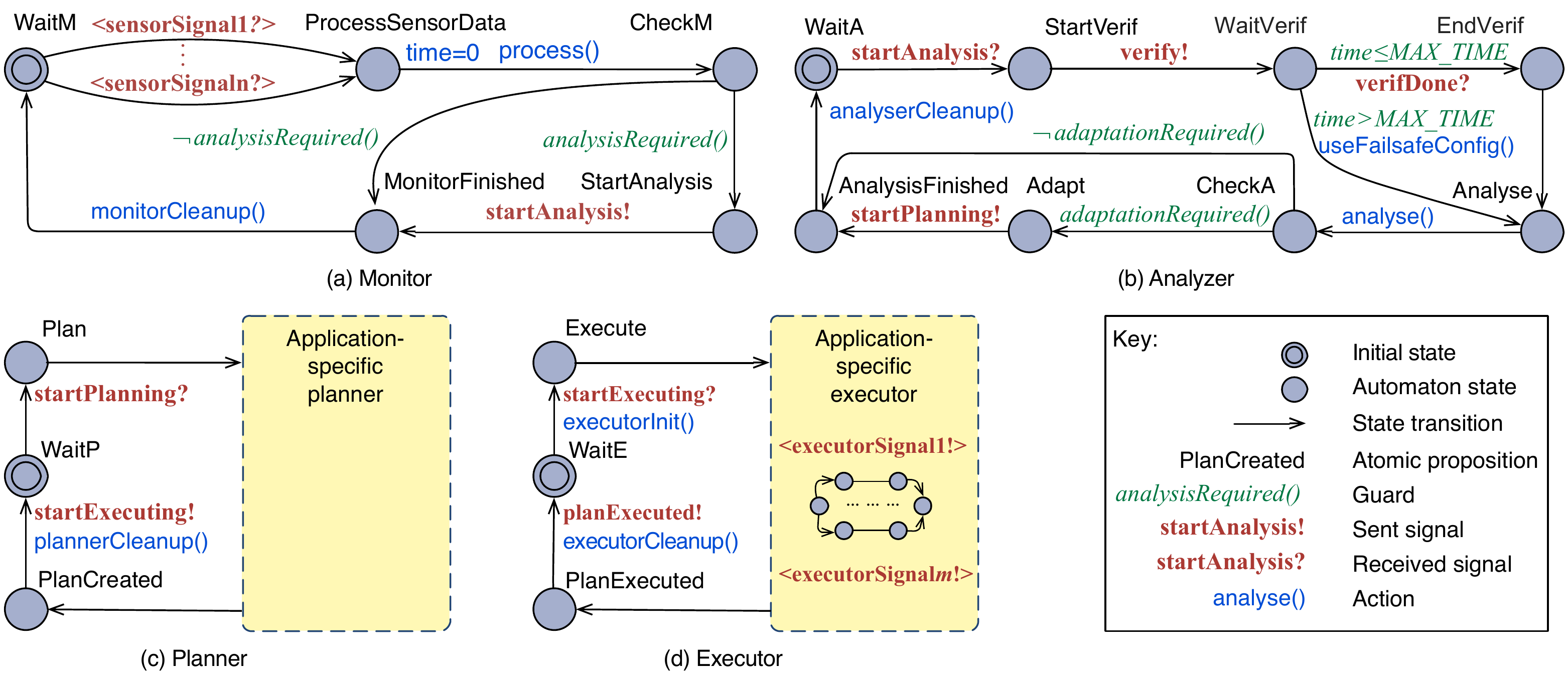}
	
	\vspace*{-2.5mm}
	\caption{Event-triggered MAPE model templates
		\label{fig:automatonTemplates}}
\end{figure*}

	\begin{figure*}
		\centering
		\includegraphics[width=15cm]{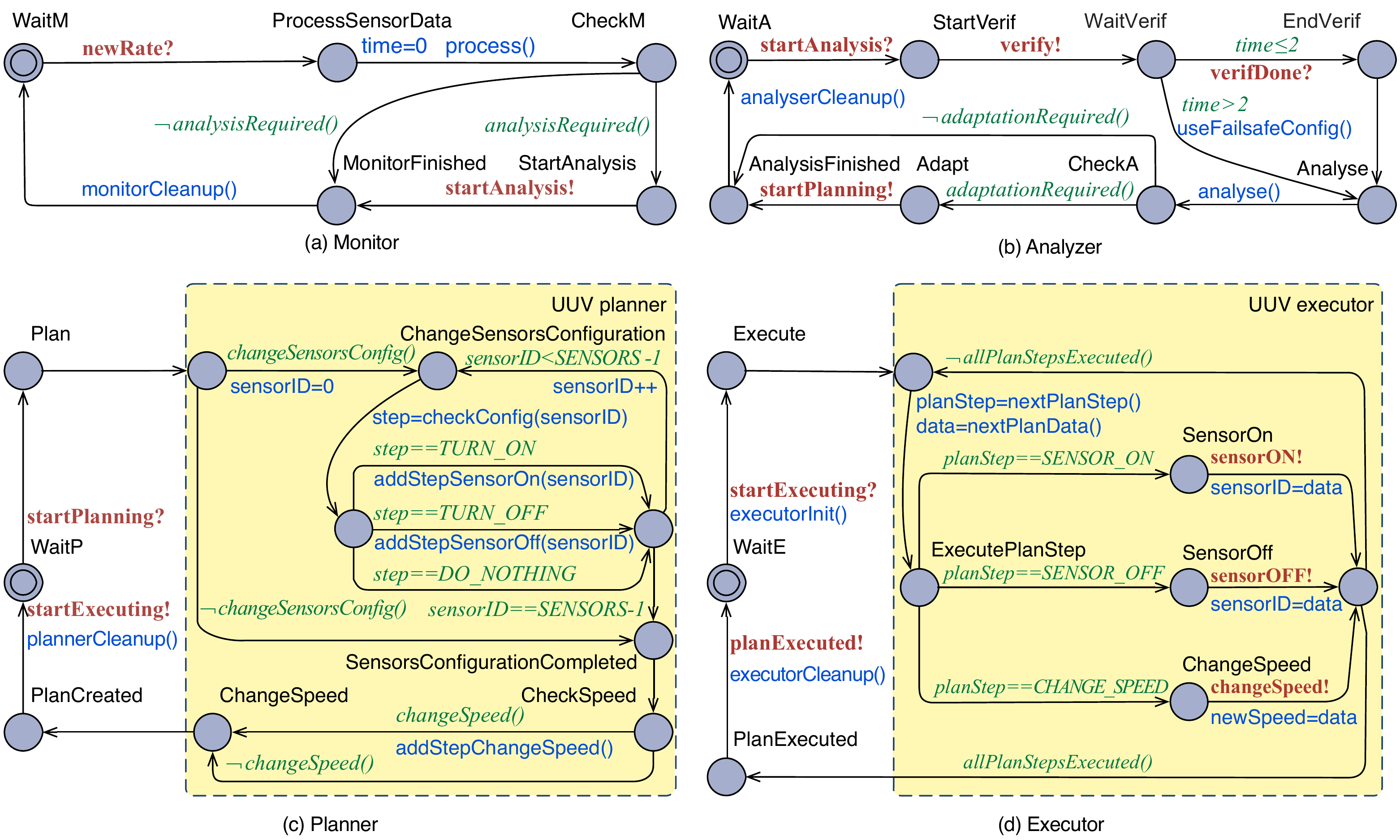}
		
		\vspace*{-2mm}
		\caption{UUV MAPE automata that instantiate the event-triggered \approach\ model templates \label{fig:uuvModels}}
		
		\vspace*{-3.5mm}
	\end{figure*}

\subsubsection{Development of Verifiable Models \label{subsec:modelsDevelopment}}

\noindent \textbf{Controller models.} We devised two types of templates for the four controller models from Fig.~\ref{fig:architecture}: (i)~\emph{event triggered}, in which the monitor automaton is activated by a sensor-generated signal indicating a change in the managed system or the environment; and (ii)~\emph{time triggered}, in which the monitor is activated periodically by an internal clock. The event-triggered automaton templates are shown in Fig.~\ref{fig:automatonTemplates} using the following font and text style conventions:
\squishlist
\item $\mathsf{Sans}\textsf{-}\mathsf{serif}\, \mathsf{font}$ is used to annotate states with the atomic propositions (i.e.\ boolean properties) that are true in those states, e.g.\ $\mathsf{PlanCreated}$ from the Planner automaton; 
\item \emph{Italics text} is used for the guards that annotate state transitions with the conditions which must hold for the transitions to occur, e.g.\ \emph{time$\leq$MAX\_TIME} from the Analyzer automaton;
\item State transitions are additionally annotated with the actions executed upon taking the transitions, and these actions are also shown in $\mathsf{sans}\textsf{-}\mathsf{serif}\, \mathsf{font}$, e.g. $\mathsf{time}\!\!=\!\!0$ to initialise a timer in the Monitor automaton; 
\item \textbf{Bold text} is used for the synchronisation channels between two automata---these channels are specified as pairs comprising a `!'-decorated sent signal and a `?'-decorated received signal with the same name, e.g., \textbf{startAnalysis!} and \textbf{startAnalysis?} from the monitor and analyzer automata, respectively. The two transitions associated with a synchronisation channel can only be taken at the same time. 
\squishend
Finally, signals in angle brackets `$\langle\rangle$' are placeholders for application-specific signal names, and guards and actions decorated with brackets `()' represent application-specific C-style functions. 

To specialise these model templates for a particular system and application, software engineers need: (a)~to replace the signal placeholders with real signal names; (b)~to define the guard and action functions;  and (c)~to devise the automaton regions shaded in Fig.~\ref{fig:automatonTemplates}.
For example, for the monitor automaton the engineers first need to replace the placeholders \textbf{$\langle$sensorSignal$_1$\hspace*{-0.4mm}?\hspace*{0.4mm}$\rangle$}, \ldots, \textbf{$\langle$sensorSignal$_n$\hspace*{-0.4mm}?\hspace*{0.4mm}$\rangle$} with sensor signals announcing relevant changes in the managed system. They must then implement the functions \textsf{process()},  \emph{analysisRequired()} and \textsf{monitorCleanup()}, whose roles are to process the sensor data, to decide if the change specified by this data requires the ``invocation'' of the analyzer through the \textbf{startAnalysis!} signal, and to carry out any cleanup that may be required, respectively. Details about the other automata from Fig.~\ref{fig:automatonTemplates} are available on our project website, which also provides implementations of these MAPE model templates in the modelling language of the UPPAAL verification tool suite \cite{behrmann2006uppaal}.
	
\vspace*{2mm}
\noindent
\textsc{Example 1.}
	We instantiated the \approach\ model templates for the UUV system from Section~\ref{sec:running}, obtaining the automata shown in Fig.~\ref{fig:uuvModels}. The signal \textbf{newRate?} is the only sensor signal that the monitor automaton needs to deal with, by reading a new UUV-sensor measurement rate (in \textsf{process()}) and checking  whether this rate has changed to such extent that a new analysis is required (in \emph{analysisRequired()}). If analysis is required, the analyzer automaton sends a \textbf{verify!} signal to invoke the runtime verification engine, and thus verifies which UUV configurations satisfy requirements R1 and R2 and with what $\mathit{cost}$. The function \textsf{analyse()} uses the verification results to select a configuration that satisfies R1 and R2 with minimum $cost$ (cf.\ requirement R3). If no such configuration exists or the verification does not complete within 2~seconds and the guard `\emph{time$>$2}' is triggered, a zero-speed configuration is selected (cf.\ requirement R4). If the selected configuration is not the one in use, \emph{adaptationRequired()} returns \textsf{true} and the \textbf{startPlanning!} signal is sent to initiate the execution of the planner automaton. The planner assembles a stepwise plan for changing to the new configuration by first switching on any UUV sensors that require activation, then switching off those that are no longer needed, and finally adjusting the UUV speed. These reconfiguration steps are  carried out by the executor automaton by means of \textbf{sensorON!}, \textbf{sensorOFF!} and \textbf{changeSpeed!} signals handled by the effectors from Fig.~\ref{fig:architecture}, as described in Section~\ref{sec:enactment}.

\vspace*{2mm}
\noindent \textbf{Parametric stochastic models.} These models used by  the \approach\ control loop at runtime are application specific, and need to be developed from scratch. Their parameters correspond to probabilities or rates of transition between model states, and are continually estimated at runtime, based on change information provided by the sensors of the controlled system. As such, the verification of these models at runtime enables the \approach\ analyzer to identify configurations it can use to meet the system requirements after unexpected changes, as described  in detail in \cite{Calinescu2012:CACM,CalinescuGKMT2011,Calinescu09Ka,Epifani2009:ICSE,Filieri2011:ICSE}.
The types of stochastic models supported by our \approach\ instance are shown in Table~\ref{table:modelTypes}. As illustrated by the research work cited in the table, the temporal logics used to express the properties of these models support the specification of numerous performance, reliability, safety, resource usage and other non-functional requirements that recent surveys propose for self-adaptive systems \cite{Cheng2014,Villegas:2011:FEQ:1988008.1988020}. 

\begin{table}
	\renewcommand{\arraystretch}{1.2}
	\centering

         \vspace*{-3mm}
	  \caption{Stochastic models supported by the \approach\ instance, with citations of representative research that uses them in self-adaptive systems}
\begin{small}

  \vspace*{-2.5mm}
  \begin{tabular}{p{4.6cm} p{3.5cm}} %
	\toprule
	\textbf{Type of stochastic model} & \textbf{Non-functional requirement specification logic}
 \\
\midrule
	Discrete-time Markov chains \cite{Epifani2009:ICSE,CalinescuGKMT2011,Calinescu:2014:AML:2568088.2568094,FilieriGT12,Filieri2011:ICSE,Ghezzi:2014:MBM:2568225.2568234} & PCTL$^a$, LTL$^b$, PCTL*$^c$\\
	Markov decision processes \cite{FKP+12} & PCTL$^a$, LTL$^b$, PCTL*$^c$\\
	Probabilistic automata \cite{Calinescu2012,JCK2013} & PCTL$^a$, LTL$^b$, PCTL*$^c$\\
	Continuous-time Markov chains \cite{Calinescu09Ka,CGB2015,Gerasimou2014:SEAMS} & CSL$^d$\\
	Stochastic games \cite{camara2015optimal,Camara16} & rPATL$^e$\\
\bottomrule
\multicolumn{2}{l}{\scriptsize $^a$Probabilistic Computation Tree Logic \cite{Hansson94J,Alfaro95}}\\[-0.6mm]
\multicolumn{2}{l}{\scriptsize $^b$Linear Temporal Logic \cite{4567924}}\\[-0.5mm]
\multicolumn{2}{l}{\scriptsize $^c$PCTL* is a superset of PCTL and LTL}\\[-0.5mm]
\multicolumn{2}{l}{\scriptsize $^d$Continuous Stochastic Logic \cite{Aziz00SSB,Baier03HHK}}\\[-0.5mm]
\multicolumn{2}{l}{\scriptsize $^e$reward-extended Probabilistic Alternating-time Temporal Logic \cite{CFK+12}}
  \end{tabular}
  \end{small}
  \label{table:modelTypes}
  
  \vspace*{-4mm}
\end{table}

To ensure the accuracy of the stochastic models described above, \approach\ can rely on recent advances in devising these models from logs \cite{Ghezzi:2014:MBM:2568225.2568234,Perez-PalacinCM2013} and UML activity diagrams \cite{CalinescuJR2013a,GallottiGMT08}, and in dynamically and accurately updating their parameters based on sensor-provided runtime observations of the controlled system \cite{Calinescu:2014:AML:2568088.2568094,CalinescuJR2011,Epifani2009:ICSE,DBLP:conf/icse/FilieriGL15}.

\begin{figure}
		\centering
		\includegraphics[width=\hsize]{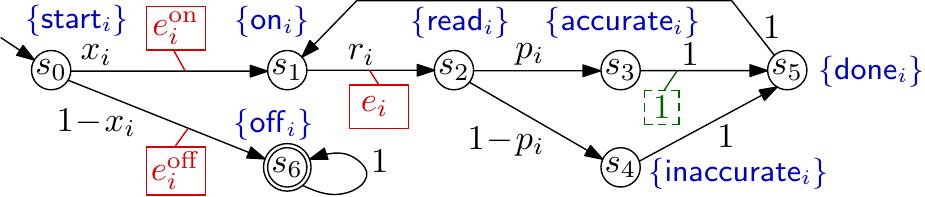}
		\caption{CTMC model $M_i$ of the $i$-th UUV sensor, adopted from \cite{Gerasimou2014:SEAMS} \label{fig:uuvCTMC}}
		
		\vspace*{-2mm}
	\end{figure}

\vspace*{2mm}
\noindent
\textsc{Example 2.}
	Fig.~\ref{fig:uuvCTMC} shows the CTMC model $M_i$ of the $i$-th UUV sensor from our running example. From the initial state $s_0$, the system transitions to state $s_1$ or $s_6$ if the sensor is switched on ($x_i=1$) or off ($x_i=0$), respectively. The sensor takes measurements with rate $r_i$, as indicated by the transition $s_1 \rightarrow s_2$. A measurement is accurate with probability $p_i$ as shown by the transition $s_2 \rightarrow s_3$; when inaccurate, the transition $s_2 \rightarrow s_4$ is taken. While the sensor is active this operation is repeated, as modelled by the transition $s_5 \rightarrow s_1$. 
	The model is augmented with two \emph{reward structures}. A ``measure" structure, shown in a dashed rectangular box, associates a reward of 1 to each accurate measurement taken. An ``energy" structure, shown in solid rectangular boxes,  associates the energy used to switch the sensor on ($e_i^\mathrm{on}$) and off ($e_i^\mathrm{off}$) and to perform a measurement ($e_i$) with the transitions modelling these events. 
	The model $M$ of the $n$-sensor UUV is given by the parallel composition of the $n$ sensor models: $M=M_1 || ... || M_n$; and the QoS system requirements are specified using CSL as follows:
	\squishlist
		\item [\textbf{R1}:] $R_{\geq 20}^\mathrm{\;measure}[C^{\leq 10/\mathit{sp}}]$
		\item [\textbf{R2}:] $R_{\leq 120}^\mathrm{\;energy}[C^{\leq 10/\mathit{sp}}]$
		\item [\textbf{R3}:] $\mathrm{minimise}\;\! (w_1 E \!+\! w_2 \mathit{sp}^{-1})$, where $E\!=\!R_{=?}^\mathrm{\;energy}[C^{\leq 10/\mathit{sp}}]$
	\squishend
	
	\vspace*{1mm}
	\noindent
	where $10/\mathit{sp}$ is the time taken to travel 10m at speed $\mathit{sp}$. As requirement R4 is a failsafe requirement, we verify it at design time as explained in the next section, so it is not encoded into CSL.
	
	 \begin{table*}
	\renewcommand{\arraystretch}{1.2}
	\centering
		\caption{Generic properties that should be satisfied by an \approach\ controller}

\vspace*{-2mm}
		\begin{small}
		
		\begin{tabular}{p{0.2cm} p{8cm} p{8.2cm}} %
			\toprule
			\textbf{ID} & \textbf{Informal description} & \textbf{Specification in computation tree logic (CTL) \cite{Clarke:1986:AVF:5397.5399}}  \\ 
			\midrule
			P1 & The \approach\ controller is deadlock free.  & A$\Box$ not deadlock \\
			P2 & Whenever analysis is required, the Analyser eventually carries out this action. & A$\Box$ (Monitor.StartAnalysis $\rightarrow$ A$\Diamond$ Analyzer.Analyse)\\
			P3 & Whenever the system requirements are violated, a stepwise reconfiguration plan is eventually assembled.  & A$\Box$ (Analyzer.Adapt $\rightarrow$ A$\Diamond$ Planner.PlanCreated)\\
			P4 & Whenever a stepwise plan is assembled, the Executor eventually implements it.  & A$\Box$ (Planner.PlanCreated $\rightarrow$ A$\Diamond$ Executor.PlanExecuted)\\
			P5 & Whenever the Monitor starts processing the received data, it eventually terminates its execution. & A$\Box$ (Monitor.ProcessSensorData $\rightarrow$ A$\Diamond$ Monitor.Finished)\\
			P6 & Whenever the Analyser begins the analysis, it eventually terminates its execution. & A$\Box$ (Analyzer.Analyse $\rightarrow$ A$\Diamond$ Analyzer.AnalaysisFinished)\\
			P7 & A plan is eventually created, each time the Planner starts planning. & A$\Box$ (Planner.Plan $\rightarrow$ A$\Diamond$ Planner.PlanCreated)\\
			P8 & Whenever the Executor starts executing a plan, the plan is eventually executed. & A$\Box$ (Executor.Execute $\rightarrow$ A$\Diamond$ Executor.PlanExecuted)\\
			P9 & Whenever adaptation is required, the current configuration and the best configuration differ. & A$\Box$ (Analyzer.Adapt $\rightarrow$ currentConfig != newConfig)\\
			\bottomrule
		\end{tabular}
	\end{small}
	\label{table:verifiedProperties}
\end{table*}


\subsubsection{Verification of Controller Models \label{subsec:controllerVerification}}

During this \approach\ stage, a trusted model checker is used to verify the network of MAPE automata devised in the previous section. This verification yields evidence that the MAPE models satisfies a set of key safety and liveness properties that may include both generic and application-specific properties. Table~\ref{table:verifiedProperties} shows a non-exhaustive list of generic properties that we assembled for the current version of \approach. Although these properties are application-independent, verifying that an \approach\ controller satisfies them is possible only after its application-specific MAPE models were devised. This involves completing the application-specific parts of the planner and executor automata, and implementing the functions for the \emph{guards} and \textsf{actions} from all the model templates. 

Additionally, automata that simulate the controller sensors, runtime probabilistic verification engine and effectors from Fig.~\ref{fig:architecture} need to be defined to enable this verification. The sensors, verification engine and effectors automata have to synchronise with the relevant monitor, analyzer and executor signals, respectively. The sensors automaton and verification automaton also have to exercise the possible paths through the monitor, analyzer and planner automata (and indirectly the executor automaton). To this end, they can nondeterministically populate the knowledge repository with data that satisfies all the different guard combinations. Alternatively, a finite collection of the two automata can be used to verify subsets of all possible MAPE paths, as long as the union of all such subsets covers the entire behaviour space of the MAPE network of automata. 

Note that these application-specific elements of the MAPE automata are much larger than the application-independent elements from the MAPE model templates. Therefore, we do not use compositional model checking \cite{ClarkeLM89,JCK2013} to verify the two parts of the MAPE automata separately, with the application-independent elements verified once and for all.  Such an approach would increase the complexity of the verification task (e.g.\ by requiring the identification and verification of less intuitive ``assumptions'' \cite{CobleighGP03} that the application-specific parts of the automata need to ``guarantee'') without any noticeable reduction in the verification time, almost all of which would be required to verify the application-specific automata elements.

\vspace*{3mm}
\noindent
\textsc{Example 3.} We used the UPPAAL model checker~\cite{behrmann2006uppaal} to verify that the network of MAPE  automata from Fig.~\ref{fig:uuvModels} (which we made available on our project website) satisfies all the generic correctness properties from Table~\ref{table:verifiedProperties}, as well as the application-specific property
\squishlist
  \item [\textbf{R4}:] A$\Box$ (Analyzer.Analyse $\wedge$ Analyzer.time$>$2 $\rightarrow$\\ 
                               \hspace*{1.75cm}A$\Diamond$ Planner.Plan $\wedge$ newConfig.speed==0),
\squishend
which represents the CTL encoding of requirement R4. To carry out this verification, we defined simple sensors, verification engine and effectors automata as described above. We used a simple one-state effectors automaton with transitions returning to its single state for each of the received signals \textbf{sensorON?}, \textbf{sensorOFF?} \textbf{changeSpeed?} and \textbf{planExecuted?}; and a finite collection of sensor--verification engine automata pairs that together exercised all possible paths of the MAPE automata from Fig.~\ref{fig:uuvModels}. These auxiliary UPPAAL automata are available on the project website.


\subsubsection{Partial Instantiation of Assurance Argument Pattern}

We used the \emph{Goal Structuring Notation} (GSN) introduced in Section~\ref{subsect:gsn} to devise a reusable \emph{assurance argument pattern} (cf.\ Section~\ref{subsect:assurance-paterns}) for self-adaptive software. Unlike all existing assurance argument patterns \cite{hawkins2011using}, our new pattern captures the fact that for self-adaptive software the assurance process 
cannot be completed at design time. Instead, it is a continual process where some design features and code elements are dynamically reconfigured and executed during self-adaptation. 
As such, the detailed claims and evidence for meeting the system requirements must vary with self-adaption, and thus \approach\ assurance cases must evolve dynamically at runtime.

\begin{figure}
\centering
\includegraphics[width=\hsize]{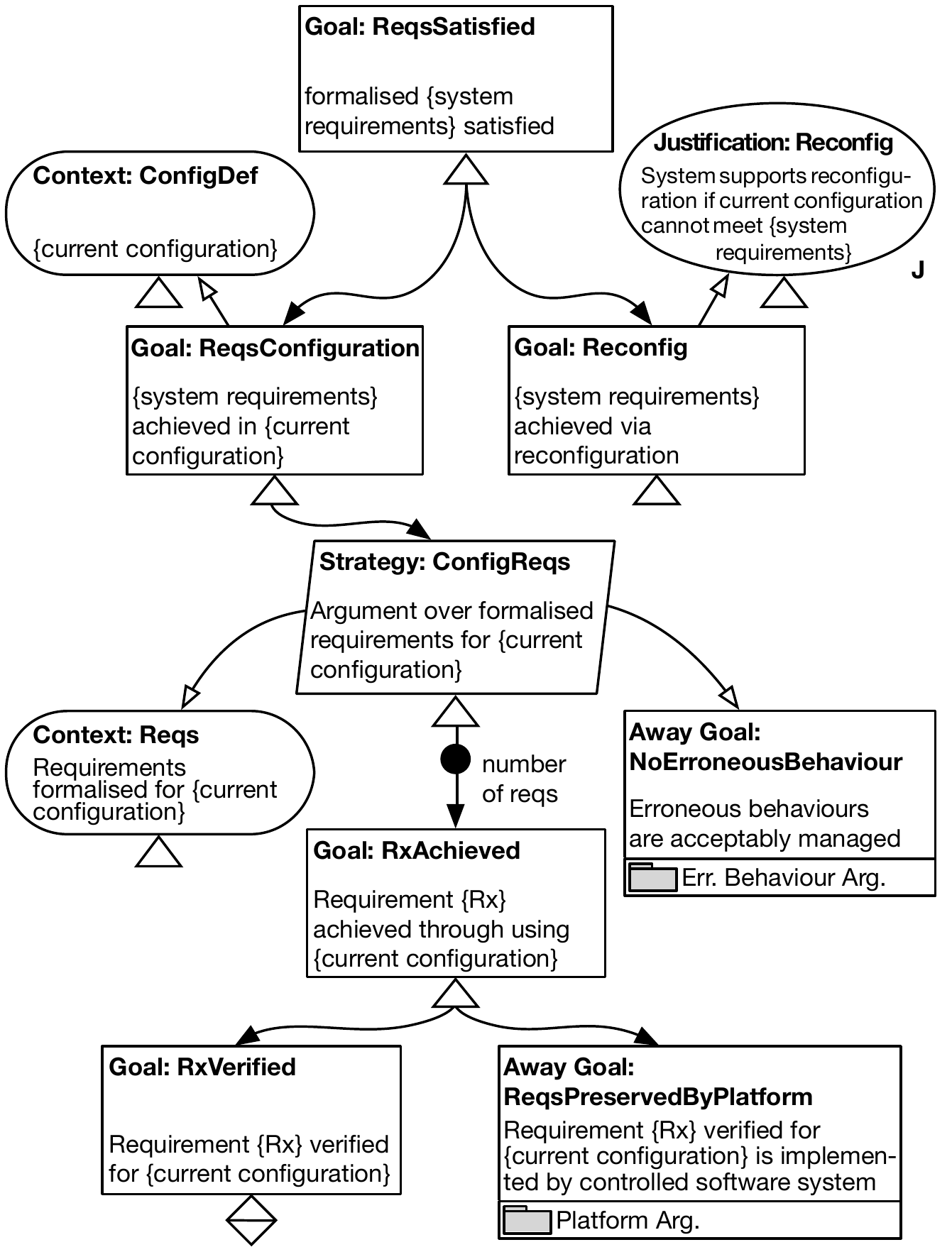}

\vspace*{-2mm}
\caption{ \approach\ assurance argument pattern. 
\label{fig:pattern}}

\vspace*{-5mm}
\end{figure}

\begin{figure*}
\centering
\includegraphics[width=0.7\hsize]{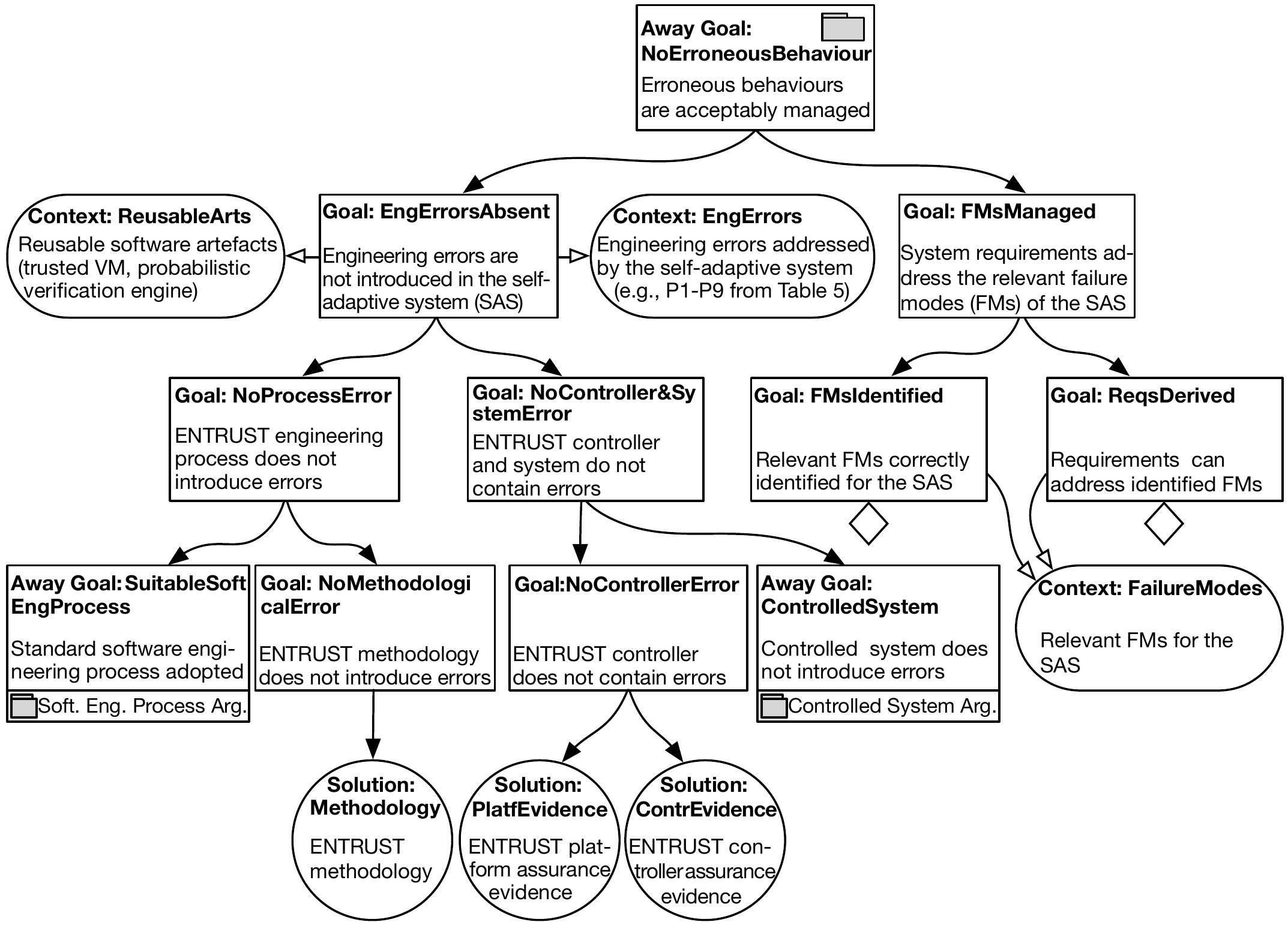}

\vspace*{-2mm}
\caption{ Away goal \textbf{NoErroneousBehaviour}, which justifies the absence of errors due to reconfiguration and is based on the existing GSN pattern Hazardous Contribution Software Safety Argument from the existing GSN catalogue~\cite{hawkins2011using} \label{fig:pattern2}}
\end{figure*}

The \approach\ assurance argument pattern is shown in Fig.~\ref{fig:pattern}. Its root goal, \textbf{ReqsSatisfied}, states that the system requirements are satisfied at all times. These requirements are typically allocated to the software from the higher-level system analysis process, so the justifications of their derivation, validity and completeness are addressed as part of the overall system assurance case (which is outside the scope of the software assurance case). \textbf{ReqsSatisfied} is supported  by a sub-claim based on (i.e. in the context of) the current configuration  (\textbf{ReqsConfiguration}) and by a reconfiguration sub-claim (\textbf{Reconfig}). That is, the pattern shows that we are guaranteeing that the current configuration satisfies the requirements (in the absence of changes) and that the \approach\ controller will plan and execute a reconfiguration that will satisfy these requirements (should a change occur). 

The pattern justifies how the system requirements are achieved for each configuration by using a sub-goal \textbf{RxAchieved} for each requirement Rx. Further, a new configuration has the potential to introduce erroneous behaviours (e.g., deadlocks). The justification for the absence of these errors is provided via the away goal \textbf{NoErroneousBehaviour} (described below).
The pattern concludes with the goals \textbf{RxVerified} and \textbf{ReqsPreservedByPlatform}, which justify the verification and the implementation of the formalised requirements, respectively. The away goal \textbf{ReqsPreservedByPlatform} confirms that the controlled system handles correctly the reconfiguration commands received through effectors. This away goal is obtained using standard assurance processes, which are outside the scope of this paper. 

As shown Fig.~\ref{fig:pattern2}, the \textbf{NoErroneousBehaviour} away goal is supported by two sub-claims. The \textbf{FMsManaged} sub-claim uses the goals \textbf{FMsIdentified} and \textbf{ResDerived} to state that the relevant ``failure modes'' for the self-adaptive system have been identified and that the system requirements fully address these failure modes. We leave the two goals undeveloped, as they are achieved using standard requirements engineering and assurance practices. The \textbf{EngErrorsAbsent} sub-claim states that the engineering of the self-adaptive system does not introduce errors in the context of the \approach\ reusable artefacts (i.e., of our trusted virtual machine and probabilistic verification engine) and of the generic properties that an \approach\ controller has to satisfy. \textbf{EngErrorsAbsent} is in turn supported by two sub-goals, \textbf{NoProcessError} and \textbf{NoController\&SystemError}. The former sub-goal is obtained through using suitable software engineering processes (via the away goal \textbf{SuitableSoftEngProcess}, which also covers the use of the methods mentioned in Section~\ref{subsec:modelsDevelopment} to ensure the accuracy of the \approach\ stochastic models) and through avoiding methodological errors by using the \approach\ methodology. The latter sub-goal, \textbf{NoController\&SystemError}, is achieved by claims about: 
\squishlist
\item[1)] The absence of controller errors. This is supported (i)~by the controller verification evidence from Stage~2 of \approach\ (cf.\ Fig.~\ref{fig:methodology}); and (ii)~by the reusable platform assurance evidence, which includes (testing) evidence about the correct operation of the model checkers UPPAAL and PRISM, based on their long track record of successful adoption across multiple domains and on our own experience of using them to develop self-adaptive systems. 
\item[2)] The absence of controlled system errors, covered by the \textbf{ControlledSystem} away goal. 
\squishend
The away goals \textbf{SuitableSoftEngProcess} and \textbf{ControlledSystem} are obtained following existing software assurances processes, and thus we do not describe them here.

The partial instantiation of the assurance argument pattern in the last design-time stage of \approach\ produces a \emph{partially-developed and partially-instantiated}  assurance argument \cite{DHP2015}. This includes placeholders for items of evidence that can only be instantiated and developed based on operational data, i.e., the runtime verification evidence that is generated by the analysis and planning steps of the \approach\ controller. 

\begin{figure}
\centering
\includegraphics[width=\hsize]{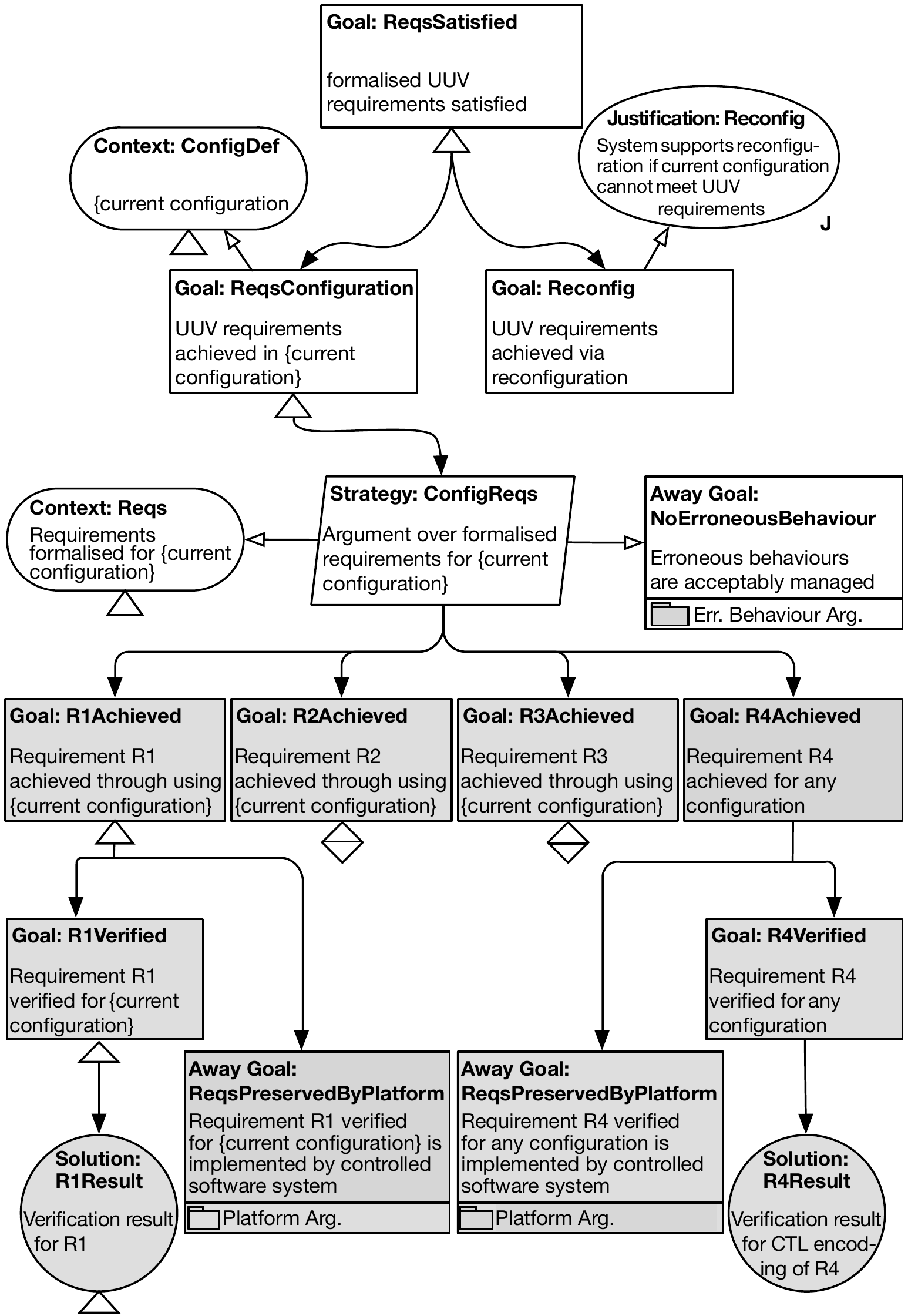}

\vspace*{-2mm}
\caption{Partially-instantiated assurance argument for the UUV system \label{fig:template}}

\vspace*{-3mm}
\end{figure}

\vspace*{3mm}
\noindent
\textsc{Example 4.}
 Fig.~\ref{fig:template} shows the partially-instantiated assurance argument pattern for the self-adaptive UUV system, in which we shaded the (partially) instantiated GSN elements. To keep the diagram clear, we only show the expansion for requirements R1 and R4, leaving R2 and R3 undeveloped. The goal \textbf{R1Achieved} (which needs to be further instantiated when the system configuration is dynamically selected) is supported by: (a)~sub-claim \textbf{R1Verified}, whose associated solution placeholder \textbf{R1Result} remains uninstantiated and should constantly be updated by the \approach\ controller at runtime; and (b)~the away goal \textbf{ReqsPreservedByPlatform} described earlier in this section. The undeveloped and partially instantiated goals \textbf{R2Achieved} and \textbf{R3Achieved} have the same structure as \textbf{R1Achieved}. In contrast, the (failsafe) goal \textbf{R4Achieved} is fully instantiated because the solution \textbf{R4Result}, comprising UPPAAL verification evidence that R4 is achieved irrespective of the configuration of the self-adaptive system, was obtained in the second \approach\ stage (verification of controller models), cf.~Example~3.


\subsubsection{Enactment of the Controller \label{sec:enactment}}

In this stage, the controller from Fig.~\ref{fig:architecture} is assembled by integrating the MAPE controller models discussed in Section~\ref{subsec:modelsDevelopment}, the \approach\ verified controller platform and application-specific sensor, effector and stochastic model management components. The application-specific components include generic functionality such as the signals through which these components synchronise with the MAPE automata (e.g., \textbf{verify?} and \textbf{planExecuted?}). Accordingly, our current version of \approach\ includes abstract Java classes that provide this common functionality. These abstract classes, which we made available on the project website, need to be specialised for each application. Thus, the specialised sensors and effectors must use the APIs of the managed software system to observe its state and environment, and to modify its configuration, respectively. The stochastic model management component must specialise the probabilistic verification engine so that it instantiates the parametric stochastic models using the actual values of the managed system and environment parameters (provided by sensors) and analyses the application-specific requirements.

\vspace*{3mm}
\noindent
\textsc{Example 5.} To assemble an \approach\ controller for the UUV system from our running example, we implemented Java classes that extend the functionality of the abstract \textsf{Sensors}, \textsf{Effectors} and \textsf{VerificationEngine} classes from the \approach\ distribution. In addition to synchronising with the relevant application-specific signals from the MAPE automata (e.g., \textbf{newRate?}), the specialised sensors and effectors invoke the relevant API methods of our UUV simulator. The specialised verification engine instantiates the parametric sensor models $M_i$ from Fig.~\ref{fig:uuvCTMC}, $1\!\leq\! i\!\leq\! n$, and verifies the CSL-encoded requirements from Example~2.


\subsubsection{Deployment of the Self-Adaptive System}

As explained in Section~\ref{sect:deployment-stage}, the role of this stage is to integrate the \approach\ controller and the controlled software system into a self-adaptive software system that is then installed, preconfigured and set running. In particular, the pre-configuration must select initial values for all the parameters of the controlled system. Immediately after it starts running and until the first execution of the MAPE control loop, the system functions as a traditional, non-adaptive software system. As such, a separate assurance argument (which is outside the scope of this paper) must be developed using traditional assurance methods, to confirm that the initial system configuration is suitable. 

The newly running software starts to behave like a self-adaptive system with the first execution of the MAPE control loop, as described in the next two sections.

\begin{figure*}
	\centering
	\includegraphics[width=15.5cm]{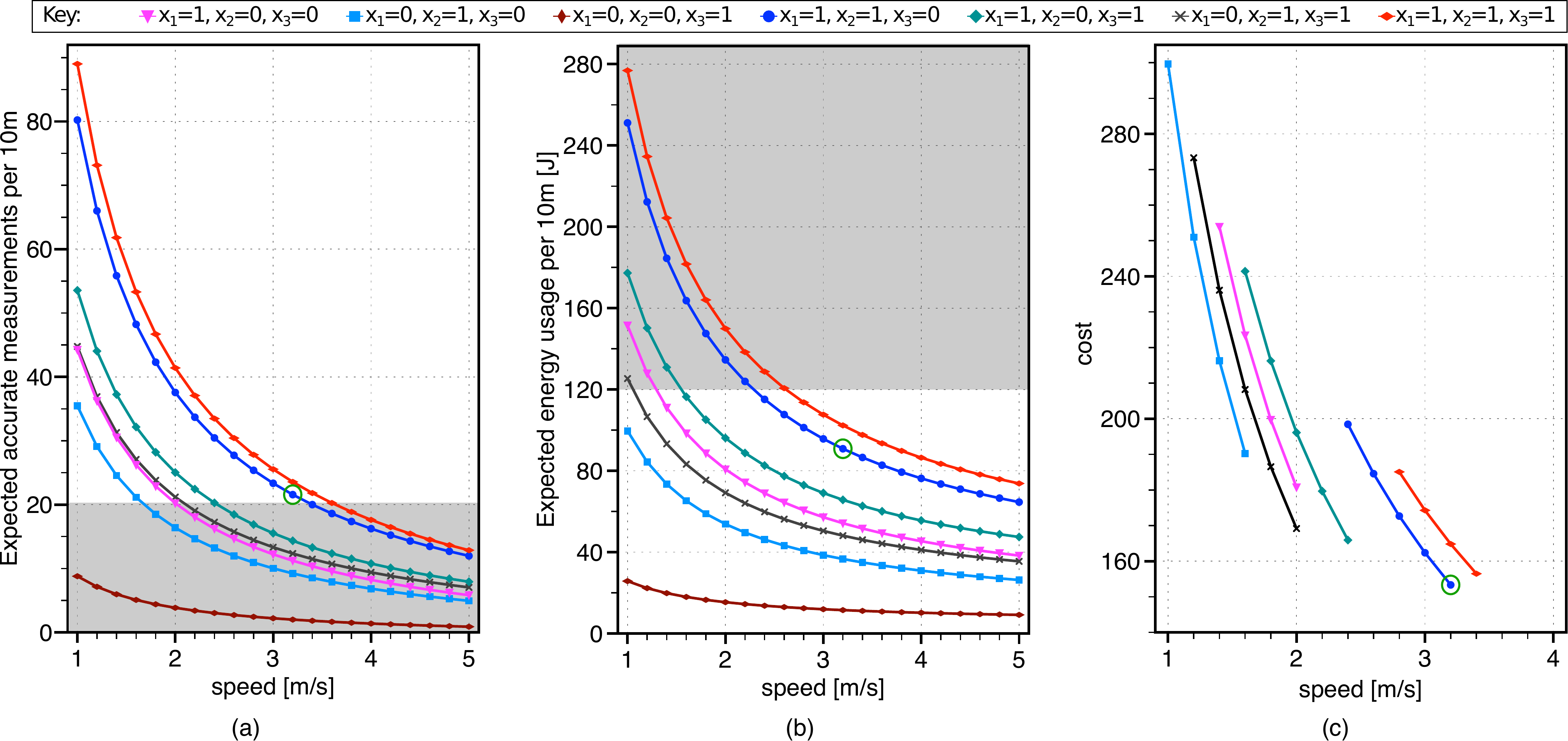}
	
	\vspace*{-3mm}
	\caption{Verification results for requirement (a) R1, (b) R2, and (c) cost of the feasible configurations; 21 speed values between 1m/s and 5m/s are considered for each of the seven combinations of active sensors, corresponding to $21\times 7 = 147$ alternative configurations. The best configuration (circled) corresponds to $x_1=x_2=1$, $x_3=0$ (i.e.\ UUV using only its first two sensors) and $\mathit{sp}=3.2$m/s, and the shaded regions correspond to requirement violations.
		\label{fig:rqvExample}}
		
		\vspace*{-3mm}
\end{figure*}

\noindent
\textsc{Example 6.} For the system from our running example, we used the open-source MOOS-IvP\footnote{Mission-Oriented Operating Suite -- Interval Programming} platform (\url{oceanai.mit.edu/moos-ivp}) for the implementation of autonomous applications on unmanned marine vehicles \cite{Benjamin2013:MRO}, and we developed a fully-fledged three-sensor UUV simulator that is available on the \approach\ website. We then exploited the publish-subscribe architecture of MOOS-IvP to interface the \approach\ sensors and effectors (and thus the controller from Example~5) with the UUV simulator, we installed the controller and the controlled system on a computer with a similar spec to that of the payload computer of a mid-range UUV, and we preconfigured the system to start with zero speed and all its sensors switched off. We chose this configuration, corresponding to initial UUV parameter values $(x_1,x_2,x_3,\mathit{sp})=(0,0,0,0)$, to ensure that the system started with a configuration satisfying its failsafe requirement R4 (cf.\ Section~\ref{sec:running}).\footnote{The use of a failsafe initial configuration is our recommended approach for \approach\ self-adaptive systems. When this is not possible, an execution of the MAPE loop must be initiated as part of the system start-up, to ensure that an initial  configuration meeting the system requirements is selected.}

\subsubsection{Self-Adaptation}

In this \approach\ stage, the deployed self-adaptive system is dynamically adjusting its configuration in line with the observed internal and environmental changes. The use of continual verification within the \approach\ control loop produces assurance evidence that underpins the dynamic generation of assurance cases in the next stage of our \approach\ instance.

\vspace*{3mm}
\noindent
\textsc{Example 7.} Consider the scenario in which the UUV system from our running example comprises $n\!=\!3$ sensors with: initial measurement rates $r_1\!=\!5$s$^{-1}, r_2\!=\!4$s$^{-1}, r_3\!=\!4$s$^{-1}$; energy consumed per measurement $e_1\!=\!3$J, $e_2\!=\!2.4$J, $e_3\!=\!2.1$J; and energy used for switching a sensor on and off $e_1^{\mathrm{on}}\!=\!10$J, $e_2^{\mathrm{on}}\!=\!8$J, $e_3^{on}\!=\!5$J and $e_1^{\mathrm{off}}\!=\!2$J, $e_2^{\mathrm{off}}\!=\!1.5$J, $e_3^{\mathrm{off}}\!=\!1$J, respectively. 
Also, suppose that the current UUV configuration is $(x_1,x_2,$ $x_3,\mathit{sp}) \!=\! (0, 1, 1, 2.8)$, and that sensor 3 experiences a degradation such that $r_3^{\mathrm{new}}\!=\!1$s$^{-1}$.
The \approach~ controller gets this new measurement rate through the monitor. As the sensor rates differ from those in the knowledge repository, the guard \emph{analysisRequired()} returns true and the  \textbf{startAnalysis!} signal is sent. Upon receiving the signal, the \emph{analyser model} invokes the probabilistic verification engine, whose analysis results for requirements \textbf{R1--R3} are depicted in Fig.~\ref{fig:rqvExample}. The \emph{analyse()} action filters the results as follows: configurations that violate requirements \textbf{R1} or \textbf{R2}, i.e., the shaded areas from Fig.~\ref{fig:rqvExample}a and Fig.~\ref{fig:rqvExample}b, respectively, are discarded.\footnote{
Note that R1 and R2 are ``conflicting'' requirements, in the sense that the configurations that satisfy R1 by the widest margin violate R2, and the other way around. In such scenarios, \approach\ supports the selection of configurations based on trade-offs between the conflicting requirements, as specified by a cost (or utility) function. If either requirement became much stricter (e.g.\ if R1 required over 50 measurements per every 10m), no configuration would satisfy both R1 and R2. In this case, \approach\  would choose the configuration specified by the failsafe requirement R4, i.e.\ would reduce the UUV speed to 0m/s, and would record the probabilistic model checking evidence showing the lack of a suitable non-failsafe configuration.} 
The remaining configurations are feasible, so their cost~(\ref{eq:cost1}) is computed for  $w_1=1$ and $w_2=200$. The configuration minimising the cost (i.e., $(x_1,x_2,x_3,\mathit{sp}) \!=\! (1, 1, 0, 3.2)$ -- circled in Fig.~\ref{fig:rqvExample}a-c) is selected as the best configuration. Since the best and the current configurations differ, the analyzer invokes the planner to assemble a stepwise reconfiguration plan with which i)~sensor 1 is switched on; ii)~next, sensor 3 is switched off; and iii) finally the speed is adjusted to 3.2m/s. Once the plan is assembled, the executor is enforcing this plan to the UUV system. 
The adaptation results from Fig.~\ref{fig:rqvExample} provide the evidence required for the generation of the assurance case as described next.


\subsubsection{Synthesis of Dynamic Assurance Argument \label{subsec:last-stage}}

The \approach\ assurance case evolves in response to the results of the MAPE process, e.g., \textit{time-triggered} and \textit{event-triggered} outputs of the monitor, the outcomes of the analyzer, the mitigation actions developed by the planner and their realisation by the executor. This offers a dynamic approach to assurance because the full instantiation of the \approach\ assurance argument pattern is left to runtime, i.e. the only stage when the evidence required to complete the argument becomes available. 
As such, the assurance case resulting from this stage captures the full argument and evidence for the justification of the current configuration of the self-adaptive system. 

\begin{figure}

\vspace*{-5mm}
\centering
\includegraphics[trim= 20mm 39mm 18mm 9mm, clip, width=\hsize]{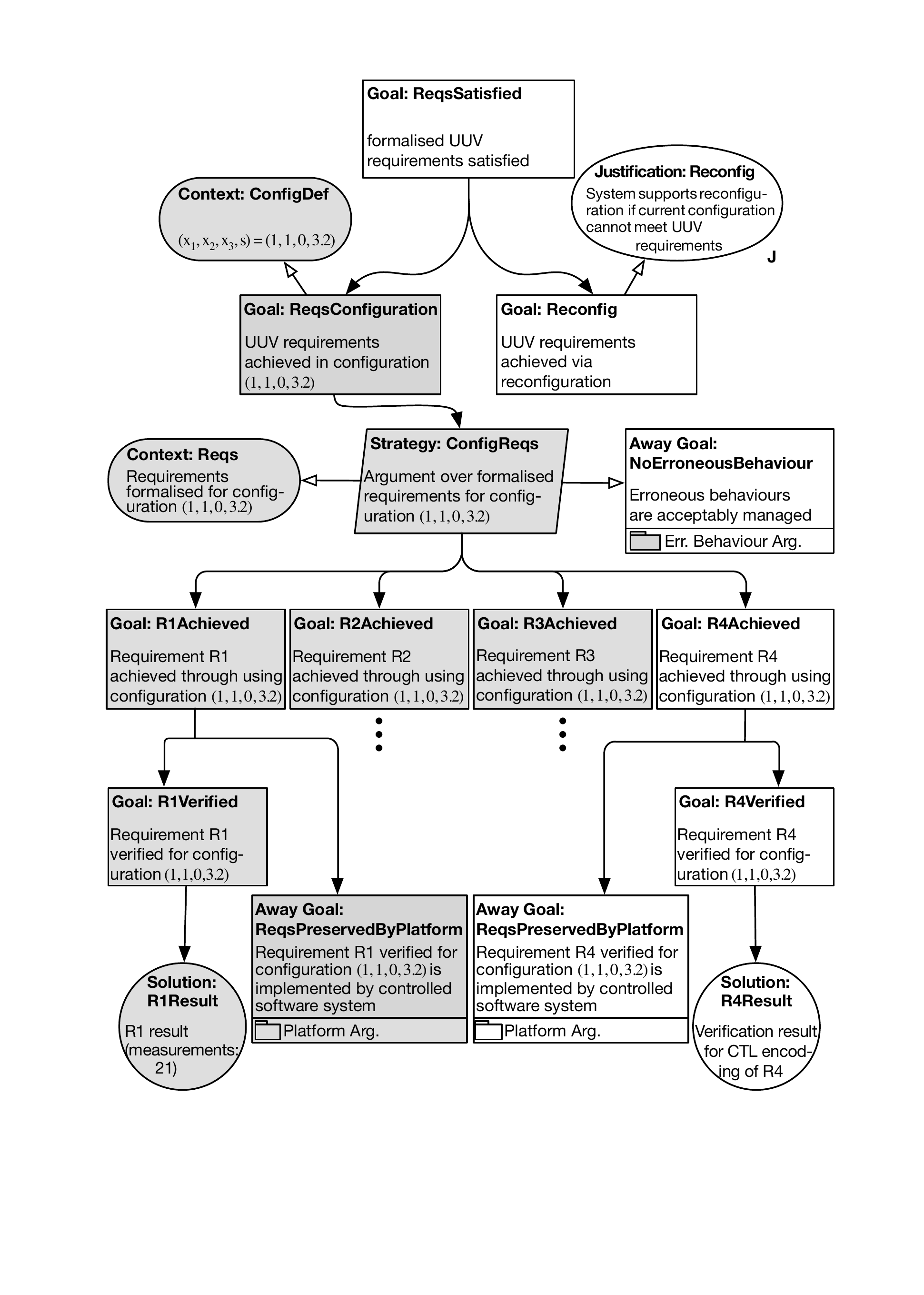}

\vspace*{-3mm}
\caption{Fully-instantiated assurance argument for the UUV system; the subgoals for \textbf{R2Achieved} and \textbf{R3Achieved} (not included due to space constraints) are similar to those for \textbf{R1Achieved}, and shading is used to show the elements instantiated at runtime\label{fig:full_argument}}

\vspace*{-4mm}
\end{figure}

\vspace*{3mm}
\noindent
\textsc{Example 8.}
Consider again the partially-instantiated assurance argument pattern for our UUV system (Fig.~\ref{fig:template}). After the \approach\ controller activities described in Example~7 conclude with the selection of the UUV configuration $(x_1,x_2,x_3,\mathit{sp}) \!=\! (1, 1, 0, 3.2)$ and the generation of runtime verification evidence that this configuration satisfies requirements R1--R3, this partially-instantiated assurance argument pattern is fully instantiated as shown in Fig.~\ref{fig:full_argument}. 

	\begin{figure*}
		\centering
		\includegraphics[width=15cm]{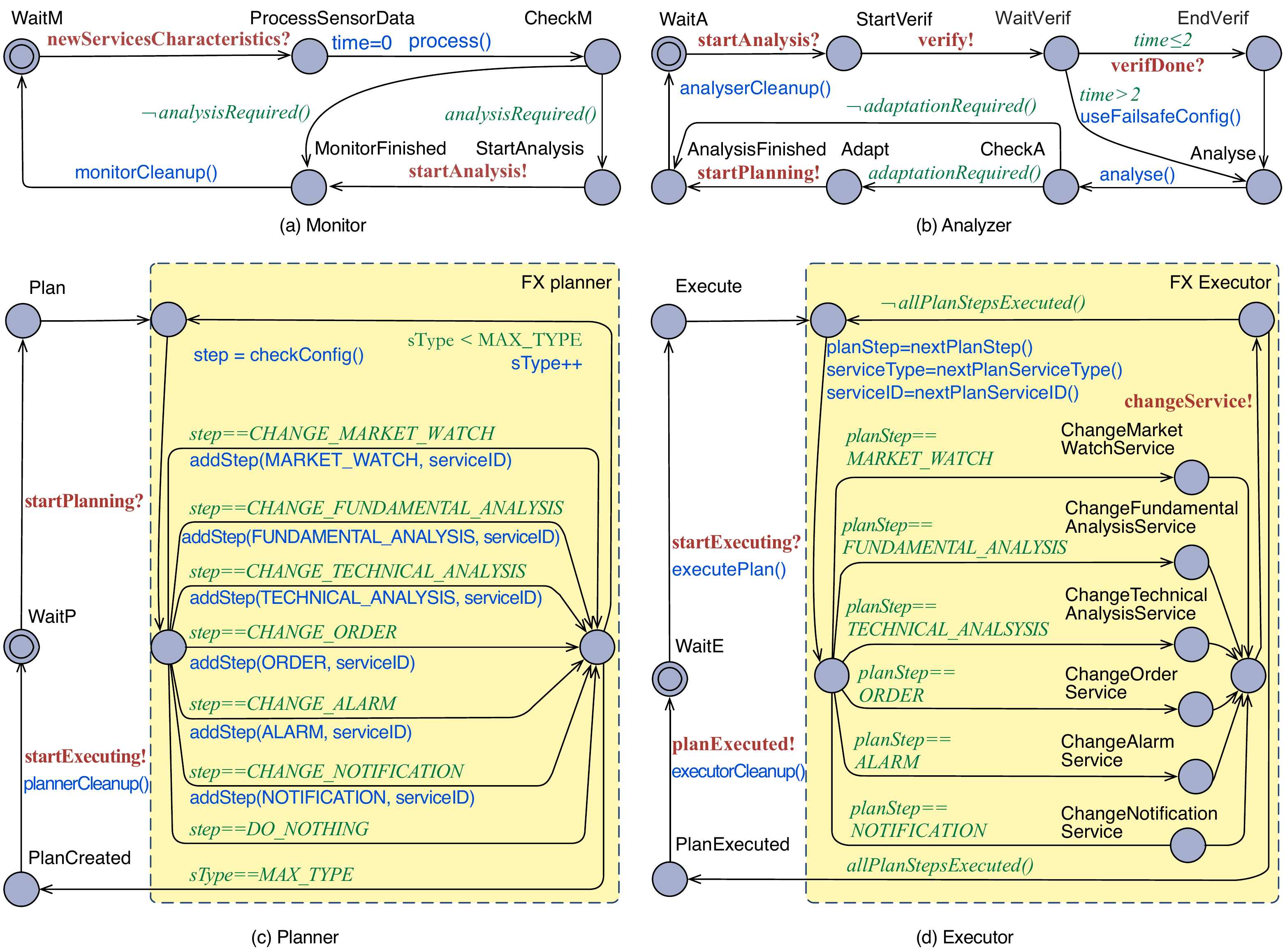}
		
		\vspace*{-3mm}
		\caption{FX MAPE automata that instantiate the event-triggered \approach\ model templates \label{fig:fxModels}}
		
		\vspace*{-3mm}
	\end{figure*}

\subsection{Self-Adaptive Service-Based System \label{sec:sbs}}

We complete the presentation of the tool-supported instance of \approach\ with a description of its use to engineer of the second self-adaptive system introduced in Section~\ref{sec:systems}.

\vspace*{2mm}
\noindent
\emph{Stage 1 (Development of Verifiable Models)} 
We specialised our event-triggered MAPE model templates for the FX system. The resulting MAPE models are shown in Fig.~\ref{fig:fxModels}, where the shaded areas in Planner and Executor automata indicate the FX-specific steps for assembling a plan and executing the adaptation, respectively. The implementations of all \emph{guards} and \textsf{actions} decorated with brackets `()' (which represent application-specific C-style functions, as explained in Section~\ref{subsec:modelsDevelopment}) are available on our project website. 

	\begin{figure}
	
		\vspace*{-2mm}
		\centering
		\includegraphics[trim= 7mm 4mm 4mm 4mm, clip, width=\hsize]{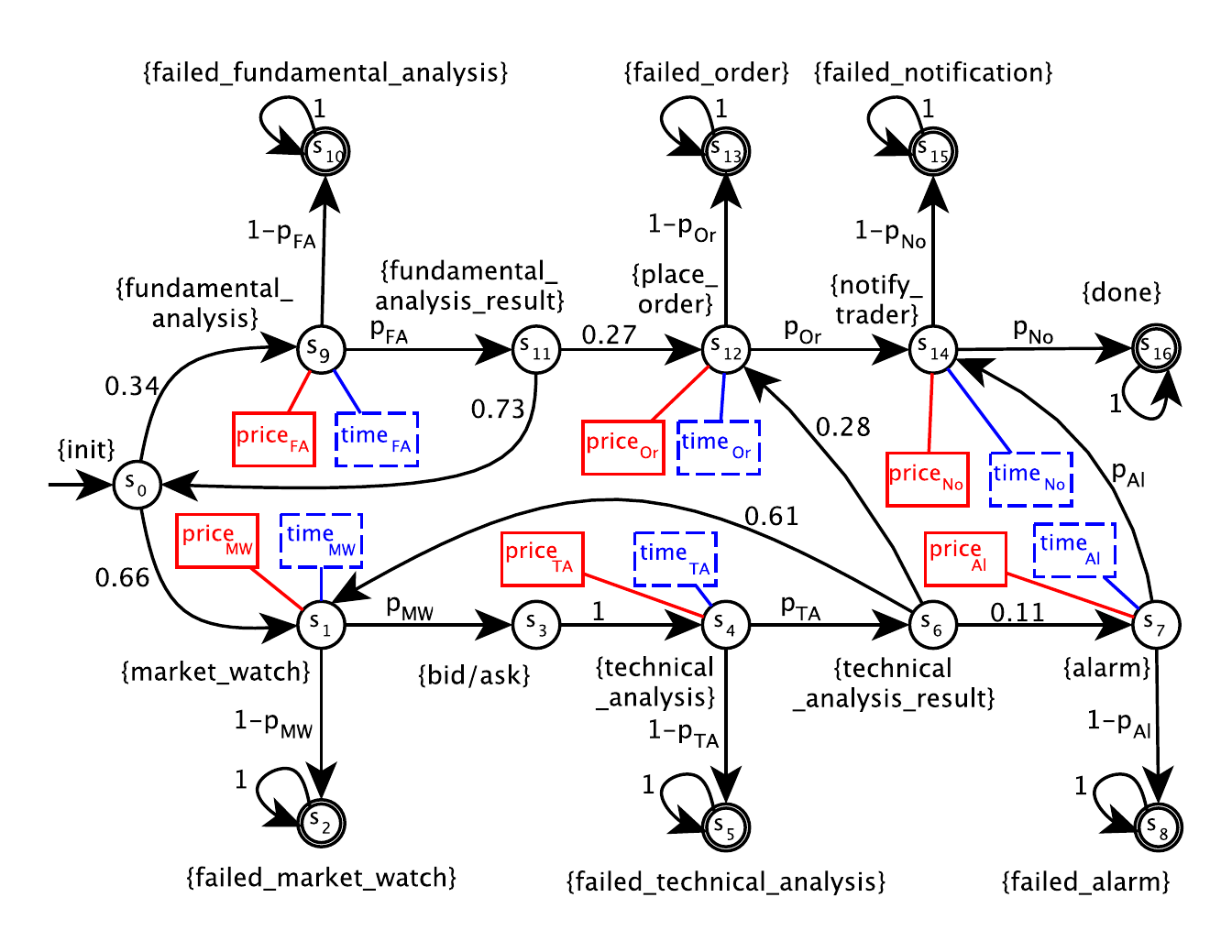}
		
		\vspace*{-2mm}
		\caption{Parametric DTMC model of the FX system; $p_\textsf{MW}$, $p_\textsf{TA}$, \ldots, $\mathit{time}_\textsf{MW}$, $\mathit{time}_\textsf{TA}$, \ldots, and $\mathit{price}_\textsf{MW}$, $\mathit{price}_\textsf{TA}$, \ldots, represent the \emph{reliability} (i.e. success probability), the \emph{response time} and the \emph{price}, respectively, of the implementations used for the \textsf{MW}, \textsf{TA}, \ldots system services. \label{fig:fxDTMC}}
		
		\vspace*{-5mm}
	\end{figure}

To model the runtime behaviour of the FX system, we used the parametric discrete-time Markov chain (DTMC) depicted in Fig.~\ref{fig:fxDTMC}. In this DTMC, constant transition probabilities derived from system logs are associated with the branches of the FX workflow from Fig.~\ref{fig:workflow}. In contrast, state transitions that model the success or failure of service invocations are associated with parameterised probabilities, which are unknown until the runtime selection of the FX services. Likewise, the ``price'' and (response) ``time'' reward structures (shown in solid and dashed boxes, respectively) are parametric and depend on the combination of FX services dynamically selected by the \approach\ controller.

Finally, we formalised requirements R1--R3 in rewards-augmented probabilistic computational tree logic (PCTL), and the failsafe requirement R4 in CTL as follows:
	\squishlist
		\item [\textbf{R1}:] $P_{\geq 0.9}[F\;\! \textsf{done}]$  
		\item [\textbf{R2}:] $R_{\leq 5}^\mathrm{\;time}[F\;\! \textsf{done}]$
		\item [\textbf{R3}:] $\mathrm{minimise}\;\!(w_1 \mathit{price} + w_2 \mathit{time})$, where\\ 
				\hspace*{0.3cm}$\mathit{price}=R_{=?}^\mathrm{\;price}[F\;\! \mathsf{done}]$ and $\mathit{time}=R_{=?}^\mathrm{\;time}[F\;\! \mathsf{done}]$
		\item [\textbf{R4}:] A$\Box$ (Analyzer.Analyse $\wedge$ Analyzer.time$>$2 $\rightarrow$\\ 
                               \hspace*{0.7cm}A$\Diamond$ Planner.Plan $\wedge$ newConfig.Order==\textsf{NoSvc})
	\squishend
where `newConfig.Order==\textsf{NoSvc}' signifies that no service is used to implement the \emph{Order} operation (i.e., the operation is skipped).

\vspace*{2mm}
\noindent
\emph{Stage 2 (Verification of Controller Models)} We used the model checker UPPAAL to verify that the MAPE automata network from Fig.~\ref{fig:fxModels} satisfies the generic controller correctness properties in Table~\ref{table:verifiedProperties}, and the FX-specific CSL property R4.

\begin{table*}
	\renewcommand{\arraystretch}{1.1}
	\centering
	\caption{Initial characteristics of the service instances used by the FX system \label{table:fx-services}}	
	\begin{footnotesize}
		\begin{tabular}{p{2.4cm} | 
				p{0.8cm} p{0.8cm}  | 
				p{0.8cm} p{0.8cm}  | 
				p{0.8cm} p{0.8cm} | 
				p{0.8cm} p{0.8cm} | 
				p{0.8cm} p{0.8cm} | 
				p{0.8cm} p{0.8cm}} %
			\toprule
			\textbf{Operation:} 
				& \multicolumn{2}{c|}{\emph{Market Watch}}  
				& \multicolumn{2}{c|}{\emph{Technical Analysis}}  
				& \multicolumn{2}{c|}{\emph{Fundam.\ Analysis}}
				& \multicolumn{2}{c|}{\emph{Alarm}} 
				& \multicolumn{2}{c|}{\emph{Order}} 
				& \multicolumn{2}{c}{\emph{Notification}}      \\
			\textbf{Service ID:} 
				& \textsf{MW$_0$} & \textsf{MW$_1$} 
				& \textsf{TA$_0$} & \textsf{TA$_1$}  
				& \textsf{FA$_0$} & \textsf{FA$_1$}
				& \textsf{Al$_0$} & \textsf{Al$_1$} 
				& \textsf{Or$_0$} & \textsf{Or$_1$} 
				& \textsf{No$_0$} & \textsf{No$_1$} \\ 
			\midrule
			\mbox{response time [s]} 
				& .5 & .5  
				& .6 & 1.0 
				& 1.6 & .7 
				& .6 & .9 
				& .6 & 1.3 
				& 1.8 & .5 \\
			reliability 
				& .976 & .995
				& .998 & .985 
				&.998 & .99
				& .995 & .99  
				& .995 & .95 
				& .99 & .99  \\
			price 
				&5 & 10
				&6 & 4 
				& 23 & 25
				&15 & 9  
				&25 & 20 
				& 5 & 8 \\
			\bottomrule
		\end{tabular}
	\end{footnotesize}
\end{table*}

	\begin{figure}
		\centering
\includegraphics[width=\hsize]{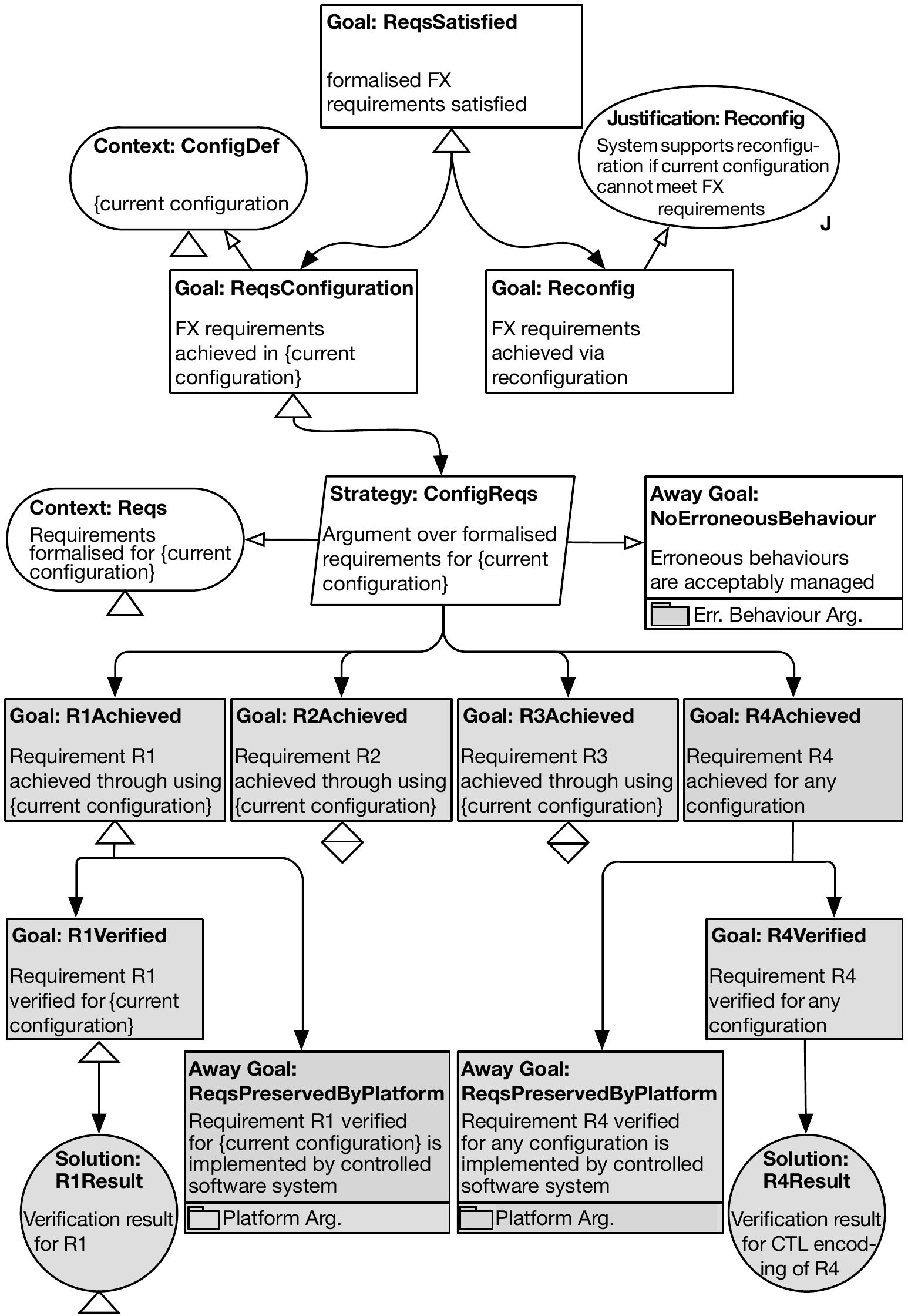}
		\caption{Partially-instantiated assurance argument for the FX system; the elements (partially) instantiated in Stage~3 of \approach\ are shaded. \label{fig:FX-partialGSN}}
		
		\vspace*{-3mm}
	\end{figure}

\vspace*{2mm}
\noindent
\emph{Stage 3 (Partial Instantiation of Assurance Argument Pattern)} We partially instantiated the \approach\ assurance argument pattern for our self-adaptive FX system, as shown in Fig.~\ref{fig:FX-partialGSN}. 

\vspace*{2mm}
\noindent
\emph{Stage 4 (Enactment of the Controller)} To assemble the \approach\ controller for the FX system, we combined the controller and stochastic models from Stage~1 with our generic controller platform, and with FX-specific Java classes that we implemented to specialise the abstract \textsf{Sensors}, \textsf{Effectors} and \textsf{VerificationEngine} abstract classes of \approach. The \textsf{Sensors} class synchronises with the Monitor automaton from Fig.~\ref{fig:fxModels} through the \textbf{newServicesCharacteristics!} signal (issued after changes in the properties of the FX services are detected). In addition, the \textsf{Sensors} and \textsf{Effectors} classes use the relevant API methods of an FX implementation that we developed as explained below. The specialised \textsf{VerificationEngine} instantiates the parametric DTMC model from Fig.~\ref{fig:fxDTMC} at runtime, and verifies the PCTL formulae devised in Stage~1 for requirements R1--R3.

\begin{figure*}
	\centering
	\includegraphics[width=15.5cm]{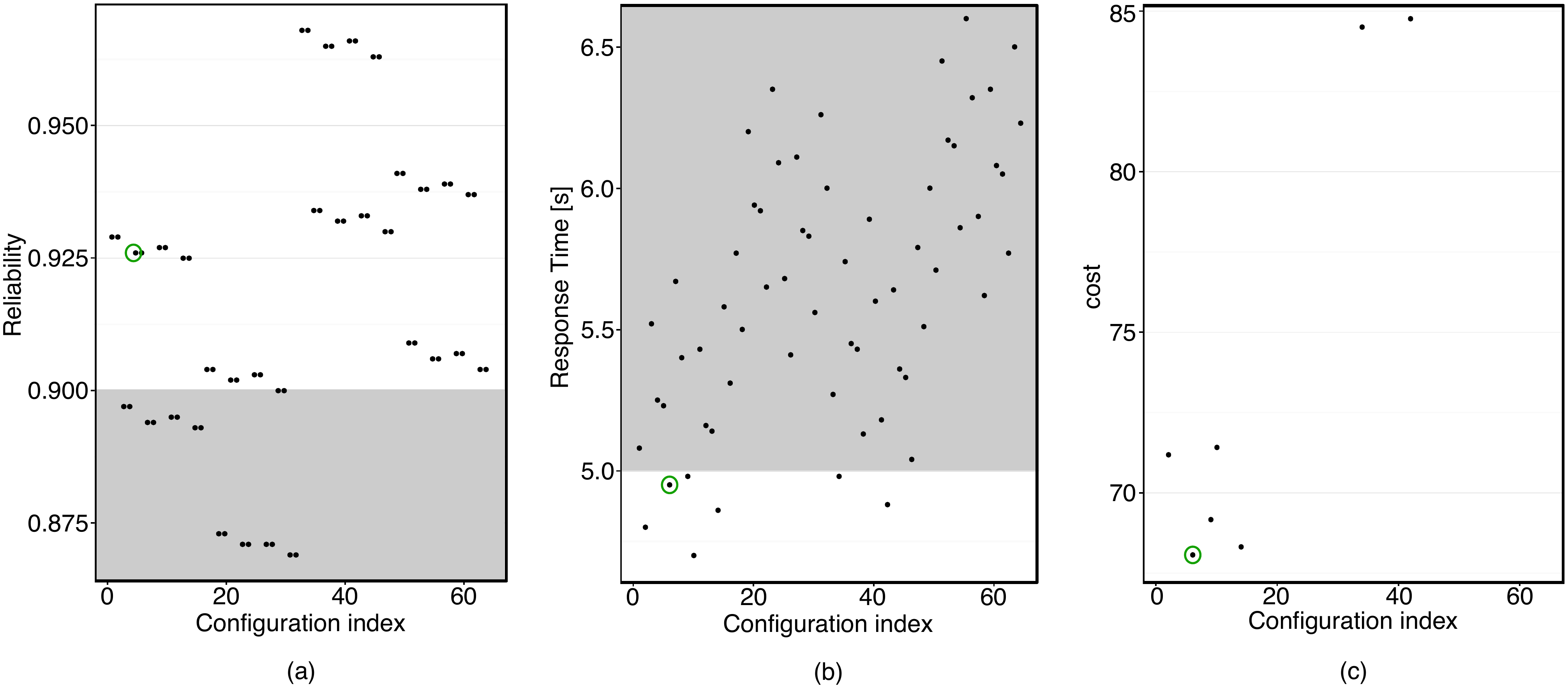}
	
	\vspace*{-2mm}
	\caption{Runtime verification results for FX requirement (a) R1, (b) R2, and (c) R3---cost of the feasible configurations, where the configuration index $i_1i_2i_3i_4i_5i_6$ in number base 2 corresponds to the FX configuration that uses services \textsf{MW$_{i_1}$, TA$_{i_2}$, FA$_{i_3}$, Al$_{i_4}$, Or$_{i_5}$} and \textsf{No$_{i_6}$}. The best configuration (circled) has index $5_{(10)}=000101_{(2)}$, corresponding to \textsf{MW$_0$, TA$_0$, FA$_0$, Al$_1$, Or$_0$} and \textsf{No$_1$}. Shaded regions correspond to requirement violations. 
		\label{fig:rqvFX}}
		
		\vspace*{-2mm}
\end{figure*}

\vspace*{2mm}
\noindent
\emph{Stage 5 (Deployment of the Self-Adaptive System)} We implemented a prototype version of the FX system using Java web services deployed in Tomcat/Axis, and a Java FX workflow that we integrated with the \approach\ controller from Stage~4. Our self-adaptive FX system (whose code is available on our project website) could select from two functionally equivalent web service implementations for each of the six FX services from Fig.~\ref{fig:workflow}, i.e. from 12 web services with the initial characteristics shown in Table~\ref{table:fx-services}. For simplicity and without loss of generality, we installed the components of the self-adaptive FX system on a single computer with the characteristics detailed in Section~\ref{sec:RQ1}, and we preconfigured the system to start by using the first web service implementation available for each service (i.e. \textsf{MW$_0$, TA$_0$,} etc.), except for the \emph{Order} service. For  \emph{Order}, \textsf{NoSvc} was selected initially, to ensure that the failsafe requirement R4 was satisfied until a configuration meeting requirements R1--R3 was automatically selected by the first execution of the MAPE loop, shortly after the system started.

\vspace*{2mm}
The remainder two stages of \approach, presented next, were continually performed by the self-adaptive FX system as part of its operation.

\vspace*{2mm}
\noindent
\emph{Stage 6 (Self-Adaptation)} In this stage, the self-adaptive FX system dynamically reconfigures in response to observed changes in the characteristics of the web services it uses. Several such reconfigurations are described later in the paper, in Section~\ref{sec:RQ1} and in Fig.~\ref{fig:FX-scenarios}. To illustrate this process in detail, consider the system configuration immediately after change \textsf{C} from Fig.~\ref{fig:FX-scenarios}, where the FX workflow uses the services \textsf{MW$_1$, TA$_0$, FA$_0$, Al$_0$, Or$_0$} and \textsf{No$_1$}. This configuration is reached after the FX services, initially operating with the characteristics from Table~\ref{table:fx-services}, experience degradations in the reliability of \textsf{MW}$_0$ ($p_{\mathsf{MW}_0}^\mathsf{new}=0.9$, change \textsf{B} in Fig.~\ref{fig:FX-scenarios}) and in the response time of \textsf{FA}$_1$ ($\mathit{time}_{\mathsf{FA}_1}^\mathsf{new}=1.2$s, change \textsf{C} in Fig.~\ref{fig:FX-scenarios}). With the FX system in this configuration, suppose that the \emph{Market Watch} service \textsf{MW$_0$} recovers, i.e., $p_{\mathsf{MW}_0}^\mathsf{new}=0.976$ as in Table~\ref{table:fx-services}. Under these circumstances, which correspond to change \textsf{D} from Fig.~\ref{fig:FX-scenarios}, the \approach\ controller receives the updated characteristics of \textsf{MW$_0$} via its monitor. As the new service characteristics differ from those in the knowledge repository, the guard \emph{analysisRequired()} holds and the \textbf{startAnalysis!} signal is sent. The analyser model receives the signal and invokes the runtime probabilistic verification engine, whose analysis of the FX requirements R1--R3 over the $2^6\!=\!64$ possible system configurations (corresponding to six services each provided by two implementations) is shown in Fig.~\ref{fig:rqvFX}. As part of this analysis, configurations that violate requirements R1 or R2 (i.e., those from the shaded areas in Fig.~\ref{fig:rqvFX}a and Fig.~\ref{fig:rqvFX}b, respectively) are discarded. The remaining configurations are feasible, so their cost is calculated (for $w_1\!=\!1$ and $w_2\!=\!2$) as shown in Fig.~\ref{fig:rqvFX}c. The feasible configuration using services \textsf{MW$_0$, TA$_0$, FA$_0$, Al$_1$, Or$_0$} and \textsf{No$_1$} has the lowest cost and is thus selected as the best system configuration. Since the best and the current configurations differ, the guard \emph{adaptationRequired()} holds and the analyser invokes the planner through the \textbf{startPlanning!} signal to assemble a stepwise reconfiguration plan through which: (i)~\textsf{MW$_0$} replaces \textsf{MW$_1$}; and (ii)~\textsf{Al$_1$} replaces \textsf{Al$_0$}. Once the plan is ready, the executor automaton receives the \textbf{startExecuting?} signal and is ensuring the implementation of this plan by sending the signal \textbf{changeService!} to the system effectors.

\begin{figure}
\centering
\includegraphics[trim= 22mm 39mm 19mm 9mm, clip, width=\hsize]{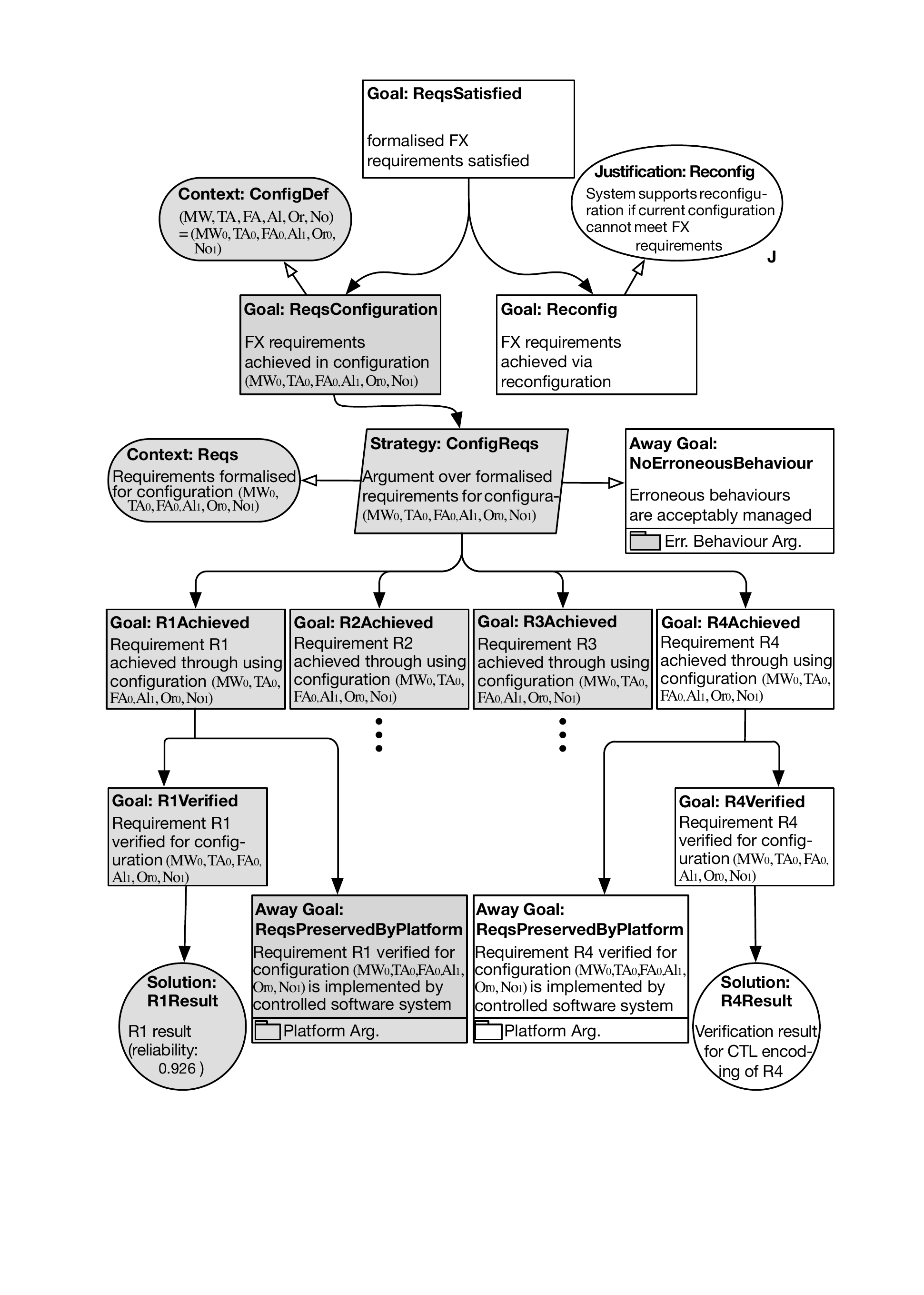}

\vspace*{-2mm}
\caption{Fully-instantiated assurance argument for the FX system; the subgoals for \textbf{R2Achieved} and \textbf{R3Achieved} (not included due to space constraints) are similar to those for \textbf{R1achieved}, and shading is used to show the elements instantiated at runtime \label{fig:full_argument_FX}}
\end{figure}

\vspace*{2mm}
\noindent
\emph{Stage 7 (Synthesis of Dynamic Assurance Argument)} The partially instantiated FX assurance pattern from Fig.~\ref{fig:FX-partialGSN} is updated into a full assurance argument after each selection of a new configuration by the \approach\ controller. This involves using the new evidence generated by the runtime probabilistic verification engine to complete the instantiation of the assurance pattern. As an example, Fig.~\ref{fig:full_argument_FX} shows the complete assurance pattern synthesised as part of the configuration change that we have just used to illustrate the previous stage of \approach.

\section{Evaluation}
\label{sec:evaluation}

This section presents our evaluation of \approach. We start with a description of our evaluation methodology in Section~\ref{subsec:eval-methodology}. Next, we detail our experimental results, and discuss the findings of the \approach\ evaluation in Section~\ref{sect:experiments}. Finally, we assess the main threats to validity in Section~\ref{subsec:threats}.

\subsection{Evaluation Methodology \label{subsec:eval-methodology}}

To evaluate the effectiveness and generality of \approach, we used our methodology and its tool-supported instance to engineer the two self-adaptive software systems from Section~\ref{sec:systems}. In each case, we first implemented a simple version of the managed software system using an established development platform for its domain (cf.~Example~6 and Section~\ref{sec:sbs} -- Stage~5). We then used our methodology to develop an \approach\ controller and a partially-instantiated assurance argument pattern for the system. Next, we deployed the \approach\ self-adaptive system in a realistic environment seeded with simulated changes specific to the application domain. Finally, we examined the correctness and efficiency of the adaptation and of the assurance cases produced by \approach\ in response to each of these unexpected environmental changes. The experimental results are discussed in Section~\ref{sect:experiments}. 
The aim of our evaluation was to answer the following research questions.
\squishlist
\item[\textbf{RQ1 (Correctness)}:] Are \approach\ self-adaptive systems making the right adaptation decisions and generating valid assurance cases?
\item[\textbf{RQ2 (Efficiency):}] Does \approach\ provide design-time and runtime assurance evidence with acceptable overheads for realistic system sizes?
\item[\textbf{RQ3 (Generality)}:] Does \approach\ support the development of self-adaptive software systems and dynamic assurance cases across application domains?
\squishend
As the focus of our evaluation was the \approach\ methodology and its tool-supported instance, we necessarily made a number of assumptions. In particular, we assumed that established assurance processes could be used to construct assurance arguments for all aspects of the controlled systems from our case studies, including their correct design, development, operation, ability to respond to effector requests, and any real-time considerations associated with achieving the new configurations decided by the \approach\ controller. As such, these aspects are outside the scope of \approach\ and are not covered in our evaluation. We further assumed that the derivation, validity, completeness and formalisation of the self-adaptive system requirements are addressed as part of the overall system assurance cases for the two case studies, and therefore also outside the scope of our evaluation of \approach.

\subsection{Experimental Results and Discussion \label{sect:experiments}}

\subsubsection{RQ1 (Correctness) \label{sec:RQ1}} 

To answer the first research question, we carried out experiments that involved running the UUV and FX systems in realistic environments comprising (simulated) unexpected changes specific to their domains. For the UUV system, the experiments were seeded with failures including sudden degradation in the measurement rates of sensors and complete failures of sensors, and with recoveries from these problems. For the FX system, we considered variations in the response time and the probability of successful completion of third-party service invocation.
All the experiments were run on a MacBook Pro with 2.5 GHz Intel Core i7 processor, and 16 GB 1600 MHz DDR3 RAM. 

\begin{figure}
\centering
\includegraphics[width=\hsize]{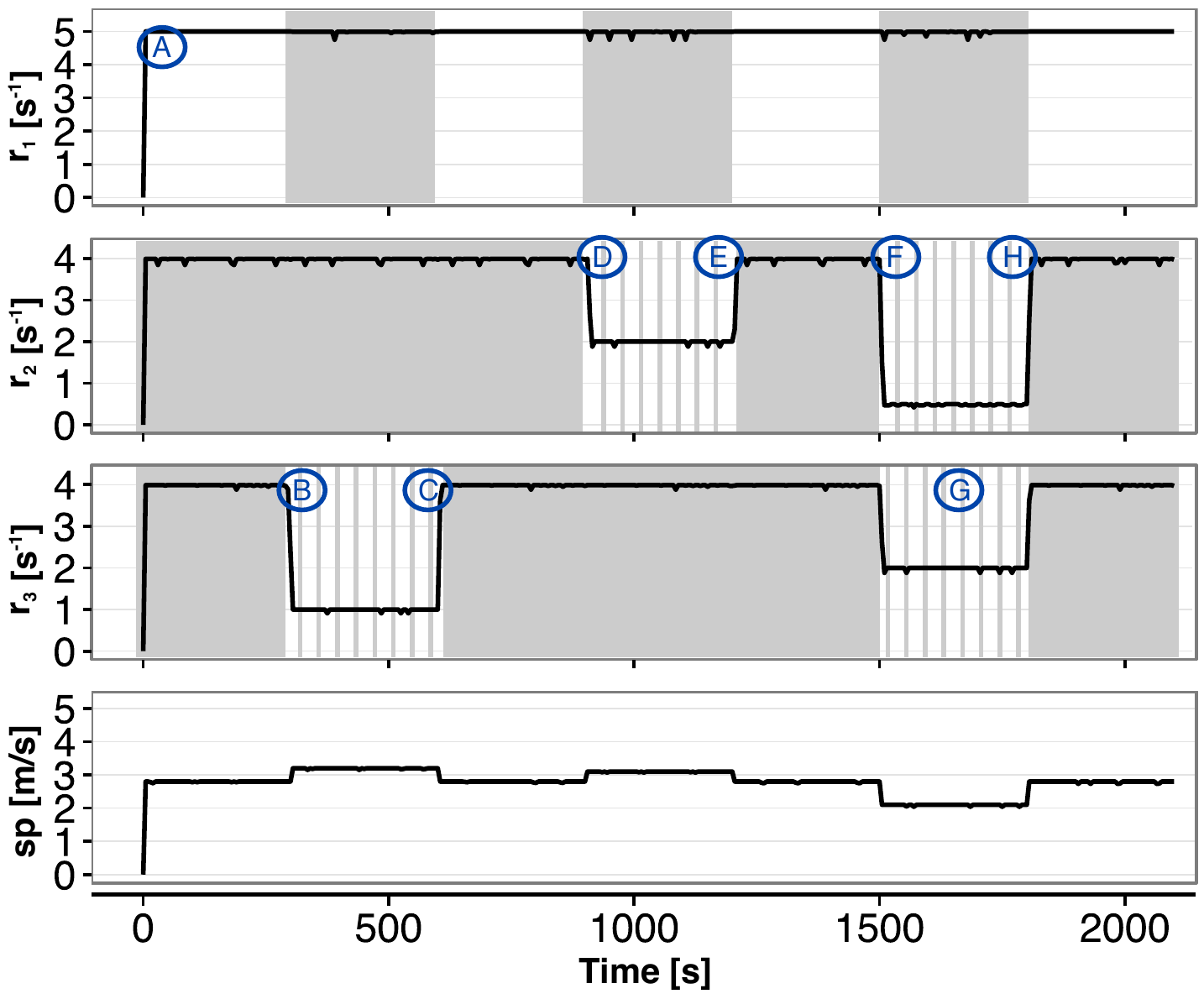}

\vspace*{-2mm}
\caption{\textcolor{black}{Change scenarios for the self-adaptive UUV system over 2100 seconds of simulated time. Extended shaded regions indicate the sensors switched on at each point in time, and narrow shaded areas show the periodical testing of sensors switched off due to degradation (to detect their recovery).
\label{fig:UUV-scenarios}}}

\vspace*{-4mm}
\end{figure}

For the UUV system, we described a concrete change scenario and the resulting self-adaptation process and generation of an assurance case in Examples~7 and~8, earlier in the paper. The complete set of change scenarios we used in this experiment is summarised in Fig.~\ref{fig:UUV-scenarios}, which depicts the changes in the sensor rates and the new UUV configurations selected by the \approach\ controller. The labels \textsf{A--H} from Fig.~\ref{fig:UUV-scenarios} correspond to following key events: 
\squishlist
\item[\textsf{A)}] The UUV starts with the initial state and configuration from Example~7;
\item[\textsf{B)}] Sensor~3 experiences the degradation described in Example~7 ($r_3^\mathsf{new}\!=\!1$), so the higher-rate but less energy efficient sensor~1 is switched on (allowing a slight increase in speed to $\mathit{sp}\!=\!3.2$m/s) and sensor~3 is switched off; 
\item[\textsf{C)}] Sensor~3 recovers and the initial configuration is resumed; 
\item[\textsf{D)}] Sensor~2 experiences a degradation, and is replaced by sensor~1, with the speed increased to $\mathit{sp}\!=\!3.1$m/s; 
\item[\textsf{E)}] Sensor~2 recovers and the initial configuration is resumed;
\item[\textsf{F)}] Both sensor~2 and sensor~3 experience degradations, so sensor~1 alone is used, with the UUV travelling at a lower speed $\mathit{sp}\!=\!2.1$m/s; 
\item[\textsf{G)}] Periodic tests (which involve switching sensors~2 and 3 on for short periods of time) are carried out to detect a potential recovery of the degraded sensors; 
\item[\textsf{H)}] Sensors~2 and 3 resume operation at nominal rates and the initial UUV configuration is reinstated.
\squishend

If the UUV system was not self-adaptive, it would have to operate with a fixed 
configuration, which would lead to requirement violations for extended periods 
of time. To understand this drawback of a non-adaptive UUV, consider that its 
fixed configuration is chosen to coincide with the initial UUV configuration 
from Fig.~\ref{fig:UUV-scenarios} (i.e.\ 
$(x_1,x_2,x_3,\mathit{sp})=(0,1,1,2.8)$) -- a natural choice because manual 
analysis can be used to find that this configuration satisfies the UUV 
requirements at deployment time. However, with this fixed configuration, the 
UUV will violate its throughput requirement R1 whenever one or both of UUV 
sensors~1 and~2 experience a non-trivial degradation, i.e.\ in the time 
intervals B--C (only 13 measurements per 10m instead of the 
required 20 measurements, according to additional analysis we carried out), 
D--E (only 15 measurements per 10m) and F--H 
(only 7 measurements per 10m) from 
Fig.~\ref{fig:UUV-scenarios}. Although a different fixed configuration may 
always meet requirement R1, such a configuration would violate other 
requirement(s), e.g.\  having all three UUV sensors switched on meets R1 but 
violates the resource usage requirement R2 at all times. 

	Finally, we performed experiments to assess how the adaptation decisions may be affected by changes in the weights $w_1,w_2$ from the UUV cost (\ref{eq:cost1}) and the energy usage of the $n$ UUV sensors. We considered UUVs with $n\!\in\!\{3,4,5,6\}$ sensors, and for each value of $n$ we carried out 30 independent experiments with the weights $w_1, w_2$ randomly drawn from the interval $[1,500]$, and the energy consumption for taking a measurement and switching on and off a sensor (i.e., $e_i, e^\mathrm{on}_i$ and $e^\mathrm{off}_i$, $1\leq i\leq n$) randomly drawn from the interval $[0.1J,10J]$. 
	The experimental results (available, together with the PRISM-generated assurance evidence, on the project website) show that \approach\ successfully reconfigured the system irrespective of the weight and energy usage values. In particular, if a 
	configuration satisfying requirements R1--R3 existed for a 
	specific change and system characteristics combination, \approach\ 
	reconfigured the UUV system to use this configuration. As expected, the 
	configuration minimising the cost (\ref{eq:cost1}) depended both on the 
	values of the weights $w_1,w_2$ and on the sensor energy usage. 
	When no configuration satisfying requirements R1--R3 was available, 
	\approach\ employed the zero-speed failsafe configuration from 
	requirement R4 until configurations satisfying requirements R1--R3 were again possible
	after a sensor recovery.

\begin{figure}
\centering
\includegraphics[width=\hsize]{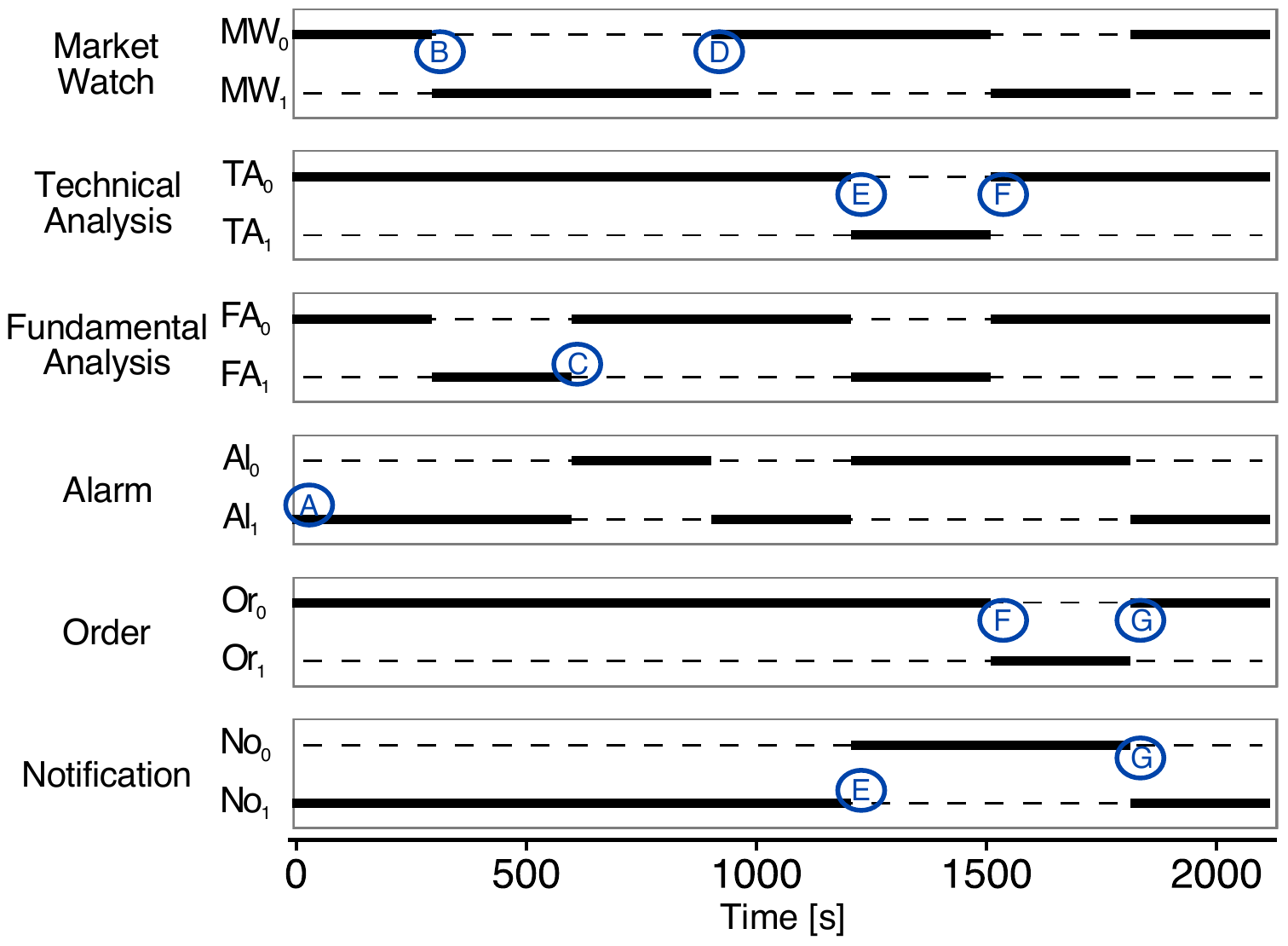}

\vspace*{-2mm}
\caption{\textcolor{black}{Change scenarios for the self-adaptive FX system, with the initial services characteristics shown in Table~\ref{table:fx-services}.
The thick continuous lines depict the services selected at each point in time.}
\label{fig:FX-scenarios}}

\vspace*{-2mm}
\end{figure}

For the FX system, a concrete change scenario is detailed in Section~\ref{sect:FX-case-study}, and the complete set of change scenarios used in our experiments is summarised in Fig.~\ref{fig:FX-scenarios}, where labels \textsf{A--G} correspond to the following events:
\squishlist
\item[\textsf{A)}] The FX starts with the initial services characteristics from Table~\ref{table:fx-services} and uses a configuration comprising the services \textsf{MW$_0$, TA$_0$, FA$_0$, Al$_1$, Or$_0$} and \textsf{No$_1$}, which satisfies requirements R1 and R2 and optimises R3;
\item[\textsf{B)}] The \emph{Market Watch} service \textsf{MW$_0$} experiences a significant reliability degradation ($p^\mathsf{new}_\mathsf{MW_0}=0.9$), so FX starts using the significantly more reliable \textsf{MW$_1$}, and thus ``affords'' to also switch to the slightly less reliable but faster  \emph{Fundamental Analysis} service \textsf{FA$_1$} in order to minimise the $\mathit{cost}$ defined in requirement R3;
\item[\textsf{C)}] Due to an increase in response time of \emph{Fundamental Analysis} service \textsf{FA$_1$} ($\mathit{time}_\mathsf{FA_1}^\mathsf{new}=1.2$s), the FX switches to using \textsf{FA$_0$} and also replaces the \emph{Alarm} service \textsf{Al$_1$} with the faster but more expensive service \textsf{Al$_0$} (to meet the timing requirement R2); 
\item[\textsf{D)}] The \emph{Market Watch} service \textsf{MW$_0$} recovers, so FX switches back to this services and also resumes using the less reliable \emph{Alarm} service \textsf{Al$_1$};
\item[\textsf{E)}] The \emph{Technical Analysis} service \textsf{TA$_0$} and the \emph{Notification } service \textsf{No$_1$} exhibit unexpected degradations in reliability ($p_\mathsf{TA_0}^\mathsf{new}=0.98$) and in response time ($\mathit{time}_\mathsf{No_1}^\mathsf{new}=1$s), respectively, so the FX system self reconfigures to use \textsf{MW$_0$, TA$_1$, FA$_1$, Al$_0$, Or$_0$} and \textsf{No$_0$};
\item[\textsf{F)}] As a result of a reliability degradation in the \emph{Order} service \textsf{Or$_0$} ($p_\mathsf{Or_0}^\mathsf{new}=0.91$) and recovery of the  \emph{Technical Analysis} service \textsf{TA$_0$}, the FX system replaces services \textsf{MW$_0$}, \textsf{TA$_1$}, \textsf{FA$_1$} and \textsf{Or$_0$} with \textsf{MW$_1$}, \textsf{TA$_0$}, \textsf{FA$_0$} and \textsf{Or$_1$}, respectively; 
\item[\textsf{G)}] All the degraded services recover, so the initial configuration \textsf{MW$_0$, TA$_0$, FA$_0$, Al$_1$, Or$_0$} and \textsf{No$_1$} is reinstated.
\squishend

As in the case of the UUV system, a non-adaptive FX version will fail to meet the system requirements for extended periods of time. For example, choosing to always use the initial FX configuration from Fig.~\ref{fig:FX-scenarios} would lead to a violation of the reliability requirement R1 while service \textsf{MW$_0$} experiences a significant reliability degradation in the time interval B--D. While using service \textsf{MW$_1$} instead of \textsf{MW$_0$} would avoid this violation, \textsf{MW$_1$} is more expensive but no faster than \textsf{MW$_0$} (cf.~Table~\ref{table:fx-services}) so its choice would increase the cost~(\ref{eq:cost2}), thus violating the cost requirement R3 in the time interval A--B.

For each change scenario from our experiments within the two case studies (cf. Figs.~\ref{fig:UUV-scenarios} and~\ref{fig:FX-scenarios}), we performed two checks. 
For the former check, we confirmed that the \approach\ controller operated correctly. To this end, we established that the change was accurately reported by the sensors and correctly processed by the monitor, leading the analyzer to select the right new configuration, for which a correct plan was built by the planner and implemented by the executor. 

For the latter check, we determined the suitability of the \approach\ assurance cases.  We started from the guidelines set by safety and assurance standards, which highlight the importance of demonstrating, using available evidence, that an assurance argument is \emph{compelling}, \emph{structured} and \emph{valid} \cite{cc2012,DefenceStandard00-56,LittlewoodW07}. Also, we considered the fact that ENTRUST has been examined experimentally but has not been tested in real-world scenarios to generate the industrial evidence necessary before approaching the relevant regulator. However, our preliminary results show, based on formal design-time and runtime evidence, that the primary claim of \approach\ assurance cases is supported by a direct and robust argument. Firstly, the argument assures the achievement of the requirements either based on a particular active configuration or through reconfiguration, while maintaining a failsafe mechanism. Secondly, the argument and patterns are well-structured and conform to the GSN community standard \cite{gsn-2011}. 
Thirdly, \approach\ provides rigorous assessments of validity not only at design time but also through-life, by means of monitoring and continuous verification that assess and challenge the validity of the assurance case based on actual operational data. This continuous assessment of validity is a core requirement for safety standards, as highlighted recently for medical devices \cite{RAEng2013}. As such, our approach satisfies five key principles of dynamic assurance cases \cite{DHP2015}: 
\squishlist
\item \emph{continuity} and \emph{updatability}, as evidence is generated and updated at runtime to ensure the continuous validity of the assurance argument (e.g. the formal evidence for solution \textbf{R1Result} from the UUV argument in Fig.~\ref{fig:full_argument}, which satisfies a system requirement given the current configuration); 
\item \emph{proactivity}, since the assurance factors that provide the basis for the evidence in the assurance argument are proactively identified (e.g.\ the \textbf{ConfigDef} context from the UUV argument in Fig.~\ref{fig:full_argument}, which captures the parameters of the current configuration);
\item \emph{automation}, because the runtime evidence is dynamically synthesised by the MAPE controller;
\item \emph{formality}, as the assurance arguments are formalised using the GSN standard.
\squishend

In conclusion, subject to the limitations described above, our experiments provide strong empirical evidence that \approach\ self-adaptive systems make the right adaptation decisions and generate valid assurance cases.

\begin{figure}
	\centering
	\includegraphics[width=\hsize]{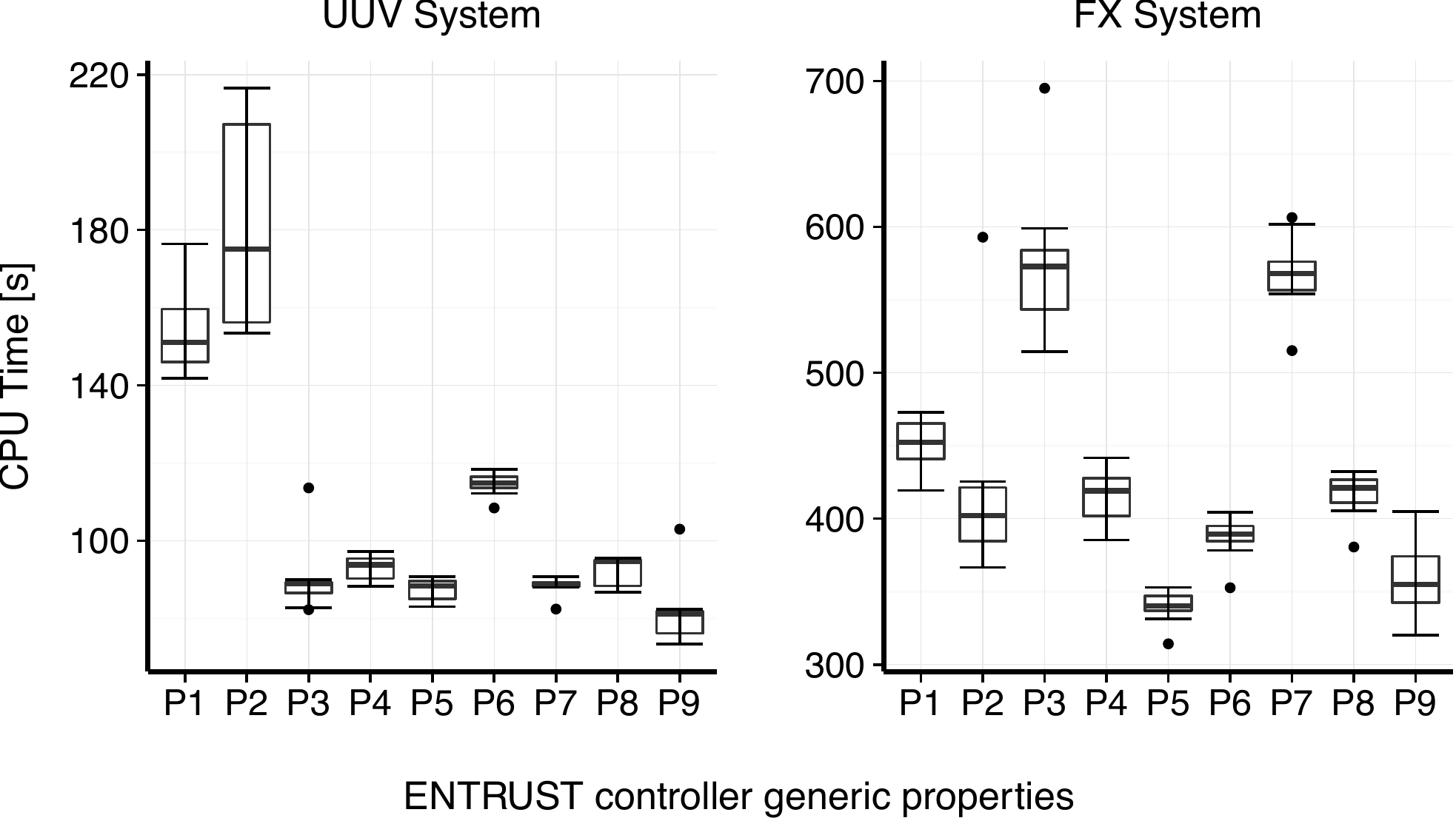}
	
	\vspace*{-2mm}
	\caption{CPU time for the UPPAAL verification of the generic controller properties in Table~\ref{table:verifiedProperties} (box plots of 10 independent measurements)}
	\label{fig:Stage2CPU}
	
	\vspace*{-3mm}
\end{figure}

\subsubsection{RQ2 (Efficiency)} 

To assess the efficiency of the \approach\ generation of assurance evidence, we measured the CPU time taken by (i)~the design-time UPPAAL model checking of the generic controller properties from Table~\ref{table:verifiedProperties}; and (ii)~the runtime probabilistic model checking performed by the \approach\ analyzer. Fig.~\ref{fig:Stage2CPU} shows the time taken to verify the generic controller properties from Table~\ref{table:verifiedProperties} for a three-sensor UUV system, and for an FX system comprising two third-party implementations for each workflow service. With typical CPU times of several minutes per property and a maximum below 12 minutes, the overheads for the verification of all controller properties are entirely acceptable. 

\begin{figure}[t]
	\centering
	\includegraphics[width=\hsize]{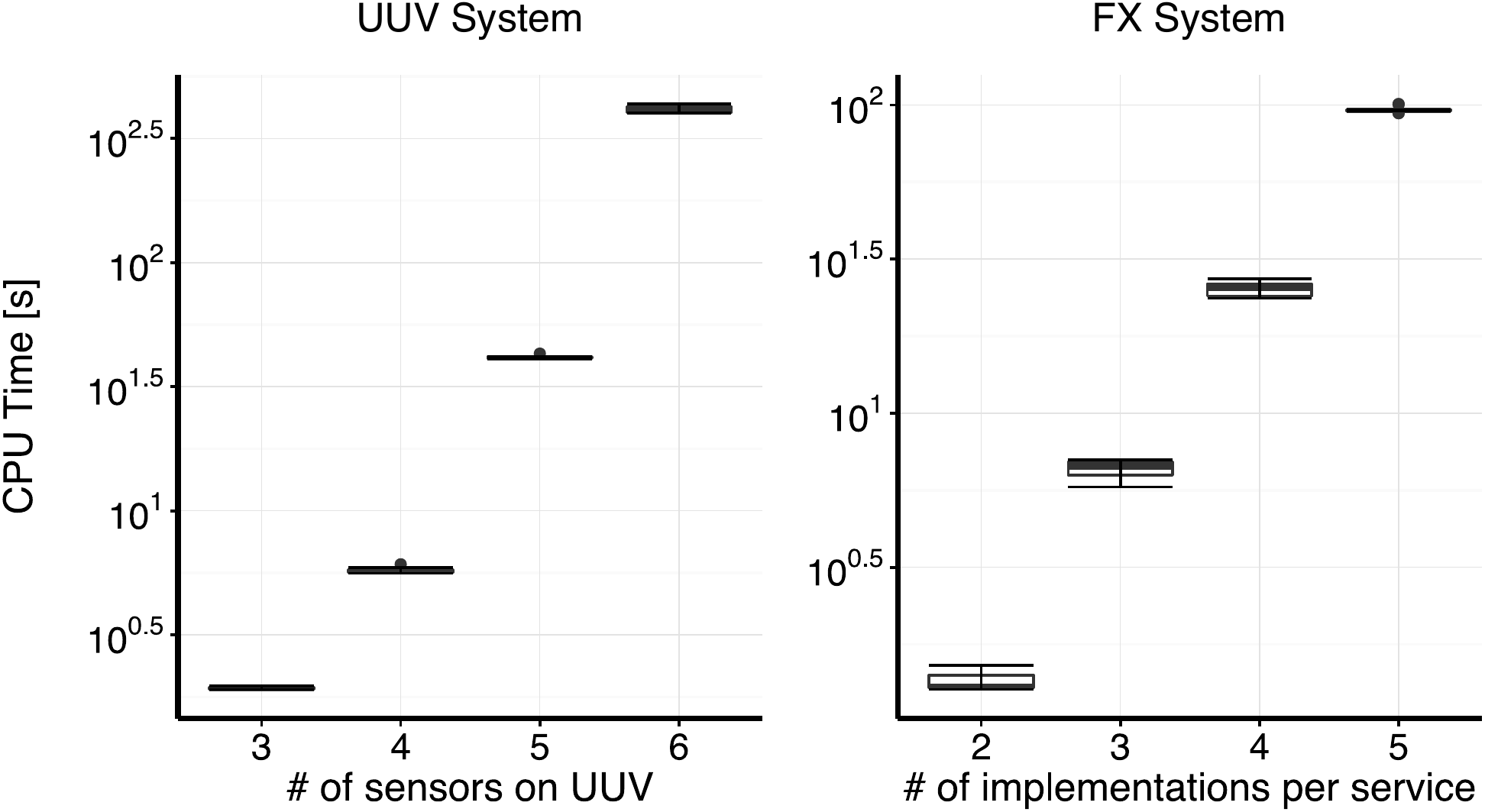}
	
	\vspace*{-2mm}
	\caption{CPU time for the runtime probabilistic model checking of the QoS requirements after changes (box plots based on 10 system runs comprising seven changes each---70 measurements in total)}
	\label{fig:CPU_UUV}
	
	\vspace*{-2mm}
\end{figure}

The CPU times required for the runtime probabilistic model checking of the QoS requirements for alternative configurations of the two systems (Fig.~\ref{fig:CPU_UUV}) have values below 1.5s and 2s, respectively. These runtime overheads, which correspond to under 10ms for the verification of a UUV configuration and under 30ms for the verification of an FX configuration, are acceptable for two reasons. First, failures and other changes requiring system reconfigurations are infrequent in the 
systems for which \approach\ is intended. Second, these systems have failsafe 
configurations that they can temporarily assume if needed during the infrequent 
reverifications of the \approach\ stochastic models.

As shown in Fig.~\ref{fig:CPU_UUV}, we also ran experiments to assess the 
increase in runtime overhead with the system size and number of alternative 
configurations, by considering UUVs with up to six sensors, and FX system 
variants with up to five implementations per service. Typical for model 
checking, the CPU time increases exponentially with these system 
characteristics. This makes the current implementation of our \approach\ 
instance suitable for self-adaptive systems with up to hundreds of 
configurations to analyse and select from at runtime. However, our recent work 
on compositional \cite{Calinescu2012}, incremental \cite{JCK2013}, 
caching-lookahead \cite{Gerasimou2014:SEAMS} and distributed \cite{CGB2015}  
approaches to probabilistic model checking and on metaheuristics for 
probabilistic model synthesis \cite{gerasimouCB2015} suggests that these more 
efficient model checking approaches could be used to extend the applicability 
of our \approach\ instance to much larger configuration space sizes. As an 
example, in \cite{Gerasimou2014:SEAMS} we used \emph{caching} of recent runtime 
probabilistic model checking results and anticipatory verification of likely 
future configurations (i.e. \emph{lookahead}) to significantly reduce the mean 
time required to select new configurations for a variant of our self-adaptive 
UUV system (by over one order of magnitude in many scenarios). Integrating 
\approach\ with these approaches is complementary to the purpose of this paper 
and represents future work.

\begin{table}
	\renewcommand{\arraystretch}{1.}
	\centering
	\caption{Comparison of self-adaptive systems used to evaluate \approach \label{table:uuv-fx-comparison}}
	
	\vspace*{-2mm}	
	\begin{footnotesize}
		\begin{tabular}{p{1.5cm}  p{3cm}  p{3cm}} %
			\toprule
			& \textbf{UUV} & \textbf{FX}      \\
			\midrule
			Type & embedded system & service-based system \\
			\midrule
			Domain & oceanic monitoring & exchange trade \\
			\midrule 
			Requirements & \mbox{throughput, resource use,} cost, safety & \mbox{reliability, response time,} cost, safety \\
			\midrule
			Sensor data & UUV sensor measure- & service response time \\
			& ment rate & and reliability \\
			\midrule
			Adaptation & switch sensors on/off, & change service instance \\
			actions & change speed  \\
			\midrule
			Uncertainty modelling & CTMC models of UUV sensors & DTMC model of system \\
			\midrule
			Assurance  & \multicolumn{2}{l}{testing evidence for correct operation of trusted} \\
			evidence  & \multicolumn{2}{l}{virtual machine; model checking evidence for} \\
			before & \multicolumn{2}{l}{correctness of MAPE controller and for UUV/FX} \\
			deployment & safety requirement \\
			\midrule
			Assurance  & probabilistic model 
			  & probabilistic model  \\
			evidence & checking evidence for 
			  & checking evidence for  \\
			obtained & \mbox{throughput, resource use}
			  & \mbox{reliability, response time} \\
			at runtime & and cost requirements
			  & and cost requirements \\
			
			\bottomrule
		\end{tabular}
	\end{footnotesize}
\end{table}

\subsubsection{RQ3 (Generality)} 

As shown in Table~\ref{table:uuv-fx-comparison}, we used \approach\ to develop an embedded 
system from the oceanic monitoring domain, and a service-based 
system from the exchange trade domain. Self-adaptation within these systems was underpinned by the verification of continuous- and discrete-time Markov chains, respectively; and the requirements and types of changes for the two systems differed. Finally, the \approach\ assurance arguments for the two systems were based on evidence obtained through testing, model checking, and probabilistic model checking.
Although evaluation in additional areas is needed, these results indicate that our \approach\ instance can be used across application domains.

To assess the overall generality of \approach, we note that probabilistic model checking can effortlessly be replaced with simulation in our experiments, because the probabilistic model checker PRISM can be configured to use discrete-event simulation instead of model checking techniques. Using this PRISM configuration requires no change to the Markov models or probabilistic temporal logic properties we analysed at runtime. As for any simulation, the analysis results would be approximate, but would be obtained with lower overheads than those from Fig.~\ref{fig:CPU_UUV}.

The uncertainties that affect self-adaptive systems are often of a stochastic nature, and thus the use of stochastic models and probabilistic model checking to analyse the behaviour of these systems is very common (e.g.\ \cite{CalinescuGKMT2011,Calinescu2012:CACM,Camara16,Epifani2009:ICSE,FilieriGT12,FKP+12,DBLP:conf/fase/SuCFRT16,QuatmannD0JK16}). As such, our \approach\ instance is applicable to a broad class of self-adaptive systems. 

Nevertheless, other methods have been used to synthesise MAPE controllers and to support their operation. Many such methods (e.g.\ based on formal proof, traditional model checking, other simulation techniques and testing) are described in Section~\ref{sec:related}. Given the generality of \approach, these methods could potentially be employed at design time and/or at runtime by alternative instantiations of \approach, supported by different modelling paradigms, requirement specification formalisms, and tools. For example, the use of the (non-probabilistic) graph transformation models or dynamic tests proposed in \cite{Becker06} and \cite{Fredericks14}, respectively, in the self-adaptation \approach\ stage is not precluded by any of our assumptions (cf.\ Section~\ref{sssec:self-adaptation}), although the method chosen for this stage will clearly constrain the types of requirements for which assurance evidence can be provided at runtime.

\subsection{Threats to Validity \label{subsec:threats}}

\textbf{Construct validity} threats may be due to the assumptions made when implementing our simple versions of the UUV and FX systems, and in the development of the stochastic models and requirements for these systems. To mitigate these threats, we implemented the two systems using the  well-established UUV software platform MOOS-IvP and (for FX) standard Java web services deployed in Tomcat/Axis. The model and requirements for the UUV system are based on a validated case study that we are familiar with from previous work \cite{Gerasimou2014:SEAMS}, and those for the FX system were developed in close collaboration with a foreign exchange expert.

\textbf{Internal validity} threats can originate from how the experiments were performed, and from bias in the interpretation of the results due to researcher subjectivity. To address these threats, we reported results over multiple independent runs; we worked with a team comprising experts in all the key areas of ENTRUST (self-adaptation, formal verification and assurance cases); and we made all experimental data and results publicly available to enable replication. 

\textbf{External validity} threats may be due to the use of only two systems in our evaluation, and to the experimental evaluation having been done by only the authors' three research groups. To reduce the first threat, we selected systems from different domains with different requirements. The evaluation results show that ENTRUST supports the development of trustworthy self-adaptive solutions with assurance cases for the two different settings. To reduce the second threat, we based \approach\ on input from, and needs identified by, the research community \cite{CamaraLGL2013,Cheng2014,delemos_et_al:DR:2014:4508,seamsRoadmap2013}. In addition, we fine tuned \approach\ based on feedback from industrial partners involved in the development of mission-critical self-adaptive systems, and these partners are now using our methodology in planning future engineering activities. Nevertheless, additional evaluation is required to confirm  generality for domains with characteristics that differ from those in our evaluation (e.g., different timing patterns and types of requirements and disturbances) and usability by a larger number of users.

\section{Related Work}
\label{sec:related}

\newcolumntype{P}[1]{>{\centering\arraybackslash}p{#1}}

Given the uncertain operating conditions of self-adaptive systems, a central aspect of providing assurances for such systems is to collect and integrate evidence that the requirements are satisfied during the entire lifetime. To this end, researchers from the area of self-adaptive systems have actively studied a wide variety of assurance methods and techniques applicable at design time and/or at runtime\cite{Cheng2014,delemos_et_al:DR:2014:4508,Magee2006,Tamura2013,Weyns2012,Weyns:2016,Zoghi:2016:DAA:2872308.2822896}.  Tables \ref{table:related-1} and \ref{table:related-2} summarise the state of the art, partitioned into categories based on the main method used to provide assurances, e.g. formal proof, model checking or simulation. We consider as the main method of a study from our analysis the method that the study primarily focuses on; the approaches from these studies may implicitly use additional methods, such as testing of their platforms and tools, but this is not emphasised by their authors.   
We summarise the representative approaches included in each category according to their:
\squishlist
\item[1)] \emph{Assurances evidence}, comprising separate parts for the methods used to provide assurance evidence for: (i)~the correctness of the platform used to execute the controller, (ii)~the correctness of the controller functions, and (iii)~the correctness of the runtime adaptation decisions; 
\item[2)] \emph{Methodology}, comprising three parts: the engineering process (i.e. a methodical series of steps to provide the assurances), tool support (i.e., tools used by engineers to provide evidence at design time and tools used at runtime by the controller, e.g. during analysis or planning), and other reusable components (i.e. third-party libraries and purpose-built software components used as part of the controller, and other artefacts that can be used at design time or at runtime, including models, templates, patterns, algorithms).
\squishend

 \begin{table*}
	\caption{Overview of related research on assurances for self-adaptive systems - part I \label{table:related-1}}
	\centering
	\begin{scriptsize}		
	\renewcommand{\arraystretch}{1.2}
	\setlength{\tabcolsep}{0.4em}
	\begin{tabular}{| P{1.7cm} | P{2cm} | P{3.5cm} | P{3.5cm} | P{1.8cm} | P{2.1cm} | P{2.1cm} | }
			\hline
			 \multirow{3}{*}{\textbf{Approach}}&  \multicolumn{3}{c|}{\textbf{Assurance evidence}}  &  \multicolumn{3}{c|}{\textbf{Methodology}}  \\	
			\cline{2-7}
			 & \textbf{Controller} & \textbf{Controller}  & \textbf{Adaptation} & \textbf{Engineering} & \textbf{Tool} & \textbf{Other reusable} \\
			& \textbf{platform} & \textbf{functions}  & \textbf{decisions} & \textbf{process} & \textbf{support} & \textbf{components} \\ 
			\hline
			\multicolumn{7}{|c|}{\cellcolor{gray!25}\textbf{Formal proof}}\\
			\hline
			Adaptation semantics \cite{Zhang2006b} & & Proof of safety and liveness \mbox{properties of adaptive programs} and program compositions & & & & Model checking algorithm\\
			\hline
			 Synthesis of behavioral models \cite{Dippolito10} &  &  Proof of completeness and soundness of synthesized behavioral models  &  &   &  &  Controller synthesis technique\\	
			 \hline			 Controller synthesis \cite{Khakpour16} &  &  &  Proof that controller synthesis algorithm generates controllers that guarantee correct and deadlock free adaptations &   Controller synthesis process only &  Tool to generate controller offline &  Controller synthesis algorithm\\	
			\hline	 			 
			 Correctness adaptation effects \cite{Brukman08} &  &  & Proof of safety, no deadlock, and no starvation of system processes as a result of adaptation &  &  & Verified middle- ware that ensures safety and liveness \mbox{$\!$of monitored system} \\
			 \hline
			 Guaranteed qualities \cite{Almeida06} &  &  &  Proof of optimizing resource \mbox{$\!$allocation under QoS constraints} &  &   &  \mbox{$\!$Ad-hoc solver of op-} \mbox{timisation problem}\\		
			 \hline	
			\multicolumn{7}{|c|}{\cellcolor{gray!25}\textbf{Model checking}}\\
			\hline
			Correct adaptation functions \cite{Iftikhar14} & Thoroughly tested virtual machine used to interpret and run controller models & UPPAAL model checking of interacting timed automata to ensure controller deadlock freeness, liveness, etc. and functional system requirements & & UPPAAL used to verify controller models at design time & & Tested reusable virtual machine; controller model templates\\
			\hline					 			 
			 Controller synthesis and enactment
\cite{Braberman13} &  &  Synthesised controller that is guaranteed not to be anomalous &  &  & Tool used for controller synthesis & Reusable inter- preter and config- uration manager for controller enactment \\
			 \hline			
			 Safe adaptation configurations \cite{Becker06} &  &  &  Verification of safety properties of system transitions using a graph transformation model &  &  & Symbolic verification procedure \\	
			 \hline
			 Guaranteed qualities \cite{Calinescu09} &  &  &  Probabilistic model checking of continually updated stochastic models of the controlled system and the environment to ensure non-functional requirements &  &  & PRISM verification library for analysis of stochastic system and envir- onment models \\
			 \hline
			 Resilience to controller failures \cite{Camara15} &  &  &  Probabilistic model checking of resilience properties of synthesized Markov models of the managed system &  Procedure to check resilience to controller failures &  &  Reusable operational profiles to check resilience \\
			\hline
		\end{tabular}
	\end{scriptsize}
\end{table*}

\begin{table*}
	\caption{Overview of related research on assurances for self-adaptive systems - part II \label{table:related-2}}
	\centering
	\begin{scriptsize}		
	\renewcommand{\arraystretch}{1.2}
	\setlength{\tabcolsep}{0.4em}
	\begin{tabular}{| P{1.7cm} | P{2cm} | P{3.5cm} | P{3.5cm} | P{1.8cm} | P{2.1cm} | P{2.1cm} | }
			\hline
			 \multirow{3}{*}{\textbf{Approach}}&  \multicolumn{3}{c|}{\textbf{Assurance evidence}}  &  \multicolumn{3}{c|}{\textbf{Methodology}}  \\	
			\cline{2-7}
			 & \textbf{Controller} & \textbf{Controller}  & \textbf{Adaptation} & \textbf{Engineering} & \textbf{Tool} & \textbf{Other reusable} \\
			& \textbf{platform} & \textbf{functions}  & \textbf{decisions} & \textbf{process} & \textbf{support} & \textbf{components} \\ 
			\hline
			\multicolumn{7}{|c|}{\cellcolor{gray!25}\textbf{Simulation}}\\
			\hline
			Evaluation novel approach \cite{Sykes11} & & & Offline simulation to ensure the scalability and robustness to node failures and message loss & & & \\
			\hline
			  Support for design \cite{Camara16} &  &  & Offline simulations to check if the performance of a latency-aware adaptation algorithm falls within predicted bounds &  &  OMNeT++ simulator for checking algorithm performance &  \\	
			\hline
			 Runtime analysis \cite{Weyns16} &  &  & Runtime simulation of stochastic models of managed system and environment to ensure non-functional requirements with certain level of confidence  &  &  UPPAAL-SMC used for online  simulation of stochastic system and environment models &  \\		
			 \hline	
			\multicolumn{7}{|c|}{\cellcolor{gray!25}\textbf{Testing}}\\
			\hline				
			 Test effectiveness of adaptation framework \cite{GarlanSASS2004} &  &  &  Offline stress testing in client-server system, showing that self-repair significantly improves system performance &  &  & Rainbow framework to realise self-adaptation \\
			 \hline				
			 Test controller robustness \cite{Camara15} &  &  Robustness testing of controller by injecting invalid inputs at the controller's interface and employ responses to classify robustness & &  Robustness testing procedure only &  &  Probabilistic response specification patterns for robustness testing\\	
			\hline
			 Runtime testing  \cite{Fredericks14} &  &  &  Dynamic tests to validate safe and correct adaptation of system using test cases adapted to changes in the system and environment &  One-stage process for test case adaptation &  &  \\
			\hline
			\multicolumn{7}{|c|}{\cellcolor{gray!25}\textbf{Other approaches}}\\
			\hline				
			 Control-theoretic approaches, e.g., \cite{Filieri14} &  &  Control-theoretic guarantees for one goal (setpoint) using automatically synthesised controller at runtime &  Controller guarantees for stability, overshoot, setting time and robustness of system operating under disturbances & &  ARPE tool to build online a first-order model of the system & Kalman filter and change point detection procedure for model updates \\ 
			 \hline				
			 Runtime verification \cite{Stoller11} &  &  &  Online verification of the probability that a temporal property is satisfied given a sample execution trace &  &  TRACE-CONTRACT tool used for trace analysis &  \\
			 \hline				
			 Sanity checks \cite{Uttamchandani05} &  &  &  Sanity checks evaluate the correctness of resource sharing decisions made by a reasoning engine &  &  CHAMELEON tool providing performance guarantees 
			 &  \\
			\hline
		\end{tabular}
	\end{scriptsize}
\end{table*}

\noindent
Providing assurances for self-adaptive systems with strict requirements requires covering all these aspects, as well as an \emph{assurance argument} that integrates the assurance evidence into a compelling, comprehensible and valid case that the system requirements are satisfied. Unlike \approach\ (Table~\ref{table:related-3}), the current research disregards this need for an assurance argument. We discuss below the different approaches and point out limitations that we overcome with ENTRUST.

\textit{Formal proof} establishes theorems to prove properties of the controller or the system under adaptation. Proof was used to provide evidence for safety and liveness properties of self-adaptive systems with different semantics (one-point adaptation, overlap adaptation, and guided adaptation) \cite{Zhang2006b}. Formal proof was also used to provide evidence for properties of automatically synthesised controllers, e.g. the completeness and soundness of synthesised behavioral models that satisfy an expressive subset of liveness properties \cite{Dippolito10} and correctness and deadlock free adaptations performed by automatically synthesised controllers \cite{Khakpour16}. Finally, formal proof was used to demonstrate the correctness of adaptation effects, e.g. proofs for safety, no deadlock, and no starvation of system processes as a result of adaptation \cite{Brukman08}, and guarantees for the required qualities of adaptations, e.g. proofs for optimised resource allocation, while satisfying quality of service constraints \cite{Almeida06}. The focus of all these approaches is on providing assurance evidence for particular aspects of adaptation. All of them offer reusable components, however, these solutions require complete specifications of the system and its environment, and---unlike \approach---cannot handle aspects of the managed system and its environment that are unknown until runtime. 

\textit{Model checking} enables verifying that a property holds for all reachable states of a system, either offline by engineers and/or online by the controller software. Model checking was used to ensure correctness of the adaptation functions that are modeled as interacting automata, with the verified models directly interpreted during execution by a thoroughly tested virtual machine \cite{Iftikhar14}. Model checking was also used to provide guarantees for automatic controller synthesis and enactment, e.g. to assure that a synthesised controller and reusable model interpreter have no anomalies \cite{Braberman13}. Model checking has extensively been used to provide guarantees for the effects of adaptation actions on the managed system, e.g. for safety properties of the transitions of a managed system that is modeled as a graph transformation system \cite{Becker06}, to ensure non-functional requirements by runtime verification of continually updated stochastic models of the controlled system and the environment \cite{Calinescu09}, and to provide evidence for resilience properties of synthesized Markov models of the managed system  \cite{Camara15}. Again, the focus of all the approaches is on providing assurance evidence for particular aspects of adaptation. The \approach\ instance presented in Section~\ref{sec:tool-supported} uses two of these techniques (i.e., \cite{Iftikhar14} and \cite{Calinescu09}) to verify the correctness of the MAPE logic at design time and to obtain evidence that adaptation decisions are correct at runtime, respectively. In addition, \approach\ offers a process for the systematic engineering of all components of the self-adaptive system, which includes employing an industry-adopted standard for the formalization of assurance arguments. 

 \begin{table*}
	\caption{Comparison of \approach\ to related research on assurances for self-adaptive systems \label{table:related-3}}
	\centering
	\begin{scriptsize}		
	\renewcommand{\arraystretch}{1.2}
	\setlength{\tabcolsep}{0.4em}
	\begin{tabular}{| P{1.5cm} | P{1.85cm} | P{1.85cm} | P{1.85cm} | P{1.85cm} | P{2.5cm} | P{2.5cm} | P{2.5cm} | }
			\hline
			 \multirow{3}{*}{\textbf{Approach}}&  \multicolumn{3}{c|}{\textbf{Assurance evidence}}  & \cellcolor{gray!25} & \multicolumn{3}{c|}{\textbf{Methodology}}  \\	
			\cline{2-4}\cline{6-8}
			 & \textbf{Controller} & \textbf{Controller}  & \textbf{Adaptation} & \cellcolor{gray!25}\textbf{Assurance} & \textbf{Engineering} & \textbf{Tool} & \textbf{Other reusable} \\
			& \textbf{platform} & \textbf{functions}  & \textbf{decisions} & \cellcolor{gray!25}\textbf{argument}& \textbf{process} & \textbf{support} & \textbf{components} \\ 
			\hline
			 Generic ENTRUST methodology & 
			 Reuse of verified application-independent controller functionality & 
			 Verification of controller models to ensure generic controller requirements and some system requirements & 
			 Automated synthesis of adaptation assurance evidence during the analysis and planning steps of the MAPE control loop & 
			 \cellcolor{gray!25} Development of partial assurance argument at design time, and synthesis of dynamic assurance argument during self-adaptation & 
			 Seven-stage process for the systematic engineering of all components of the self-adaptive system, and of an assurance case arguing its suitability for the intended application & 
			 Tools specific to each ENTRUST instance & 
			 Reusable software artefacts: controller platform, controller model templates; Reusable assurance artefacts: platform assurance evidence, generic controller requirements, assurance argument pattern \\
			 \hline				
			 Tool-supported ENTRUST instance & 
			 Reuse of thoroughly tested  virtual machine to directly interpret and run controller models, and of established probabilistic model checking engine & 
			 UPPAAL model checking of interacting timed automata models to ensure controller deadlock-freeness, liveness, etc. and functional system requirements & 
			 PRISM probabilistic model checking of continually \mbox{$\!\!$updated\hspace*{-0.2mm} stochastic} models of the controlled system and the environment to ensure non-functional requirements & 
			 \cellcolor{gray!25} Assurance argument synthesised using the industry-adopted Goal Structuring Notation (GSN) standard & 
			 Seven-stage process for the systematic engineering of all components of the self-adaptive system, and of an assurance case arguing its suitability for the intended application & 
			 UPPAAL used to verify controller models; PRISM used to verify stochastic system and environment models & 
			 Reusable controller platform (virtual machine, probabilistic verification engine), timed automata controller model templates;
Reusable platform assurance evidence, CTL generic controller require- ments, GSN assurance argument pattern \\
			\hline
		\end{tabular}
	\end{scriptsize}
 \end{table*}

\textit{Simulation approaches} provide evidence by analysing the output of the execution of a model of the system. Simulation was used to evaluate novel self-adaptation approaches, e.g. to ensure the scalability and robustness to node failures and message loss of a self-assembly algorithm \cite{Sykes11}, and to support the design of self-adaptive systems, e.g., to check if the performance of a latency-aware adaptation algorithm falls within predicted bounds \cite{Camara16}. Recently some efforts have been made to let the controller exploit simulation at runtime to support analysis, e.g. runtime simulation of stochastic models of managed system and environment has been used to ensure non-functional requirements with certain level of confidence \cite{Weyns16}. The primary focus of simulation approaches has been on providing assurance evidence for the adaptation actions (either as a means to check the controller effects or to make a prediction of the expected effects of different adaptation options). The approaches typically rely on established simulators. 

\textit{Testing} is a standard method for assessing if a software system performs as expected in a finite number of scenarios. Testing was used to test the effectiveness of adaptation frameworks, e.g. checking whether a self-repair framework applied to a client-server system keeps the latencies of clients within certain bounds when the network is overloaded \cite{GarlanSASS2004}. Testing was used to provide evidence for the robustness of controllers by injecting invalid inputs at the controller's interface and use the responses to classify robustness \cite{Camara15}. Several studies have applied testing at runtime, e.g. to validate safe and correct adaptations of the managed system based on adapt test cases generated in response to changes in the system and environment  \cite{Fredericks14}. While simulation and testing approaches can be employed within the generic \approach\ methodology to obtain assurance evidence for particular aspects of self-adaptive systems, they need to be complemented by assurances for other components of a self-adaptive system and integrated in a systematic process as provided by \approach. 

\textit{Other approaches}.  To conclude, we highlight some other related approaches that have been used to provide assurances for self-adaptive systems. Recently, there has been a growing interest in applying control theory to build ``correct by construction'' controllers.  The approach was used to automatically synthesise controllers at runtime, providing control-theoretic guarantees for stability, overshoot, setting time and robustness of system operating under disturbances \cite{Filieri14}. This research is at an early stage, and its potential to deliver solutions for real-world systems and scenarios has yet to be confirmed. In contrast, \approach\ relies on proven software engineering techniques for modelling and analysing software systems and assuring their required properties. Runtime verification is a well-studied lightweight verification technique based on extracting information from a running system to detect whether certain properties are violated. For example, sequences of events can be modeled as observation sequences of a Hidden Markov Model allowing to verify the probability that a temporal property is satisfied by a run of a system given a sampled execution trace \cite{Stoller11}. Sanity checks are another approach to check the conformance of requirements of adaptive systems. Sanity checks have been used to evaluate the correctness of resource sharing decisions made by a reasoning engine \cite{Uttamchandani05}.  Approaches such as runtime verification and sanity checks are often supported by established tools. However, these approaches provide only one piece of evidence. Such approaches can also be used by our generic \approach\ methodology, which supports the integration of assurance evidence from multiple sources in order to continuously generate an assurance case. 

Another line of related research (not specifically targetting self-adaptation and thus not included in Table~\ref{table:related-2}) is \textit{runtime certification}, proposed in \cite{Rushby:Cases15} and further developed in \cite{Knight15,Schneider:2013:CSC:2491465.2491467}. Runtime certification involves the proactive runtime monitoring of the assumptions made in the assurance case, thereby providing early warnings for potential failures. \approach\ goes beyond the mere monitoring of assumptions, to evolving the arguments and evidence dynamically based on the runtime verification data, particularly for self-adaptive software assurance. ENTRUST also extends existing work on assurance argument patterns \cite{Denney2013} by enabling runtime instantiation. 

The ENTRUST methodology and the other research summarised in this section also build on results from the areas of configurable software, configuration optimisation, and performance tuning. For instance, symbolic evaluation has been used to understand the behaviour of configurable software systems~\cite{Reisner:2010}, dedicated support to automatically verify the correctness of dynamic updates of client-server systems has been proposed~\cite{Hayden:2012}, and specification languages have been devised to help program library developers expose multiple variations of the same API using different algorithms~\cite{Tapus:2002}. However, none of these results could be directly applied to self-adaptive software systems, which need to reconfigure dynamically in response to runtime environmental uncertainties and goal changes.

The sparsity of Tables~\ref{table:related-1} and~\ref{table:related-2} makes clear that existing approaches are confined to providing correctness evidence for specific aspects of the self-adaptive software. In contrast to existing work on assurances for self-adaptive systems, Table~\ref{table:related-3} shows that \approach\ offers an end-to-end methodology for the development of trustworthy self-adaptive software systems. Unique to our approach, this includes the development of assurance arguments. The upper part of Table~\ref{table:related-3} shows how the generic \approach\ methodology covers the whole spectrum of aspects that are required to provide assurances for self-adaptive systems with strict requirements. The lower part of Table~\ref{table:related-3} shows a concrete tool-supported instantiation of \approach\ and summarises how the various assurances aspects are covered for this instance. Details about 
the information summarised in the table are provided in Sections~\ref{sec:methodology} and~\ref{sec:tool-supported}.

\section{Conclusion}
\label{sec:conclusion}

We introduced \approach, the first end-to-end methodology for the engineering of trustworthy self-adaptive software systems and the dynamic generation of their assurance cases. \approach\ and its tool-supported instance presented in the paper include methods for the development of verifiable controllers for self-adaptive systems, for the generation of design-time and runtime assurance evidence,  and for the runtime instantiation of an assurance argument pattern that we devised specifically for these systems.

The future research directions for our project include evaluating the usability of \approach\ in a controlled experiment,  extending the runtime model checking of system requirements to functional requirements, and reducing the runtime overheads by exploiting recent advances in probabilistic model checking at runtime \cite{CGB2015,Calinescu2012,Filieri2011:ICSE,Gerasimou2014:SEAMS,JCK2013}. In addition, we are planning to explore the applicability of \approach\ to other systems and application domains.

\bibliographystyle{IEEEtranS}
\bibliography{ms}

\end{document}